\documentclass[%
reprint,
amsmath,amssymb,
aps,
]{revtex4-1}

\usepackage{graphicx}% Include figure files
\usepackage{dcolumn}% Align table columns on decimal point
\usepackage{bm}% bold math

\usepackage[utf8]{inputenc}
\usepackage[T1]{fontenc}
\usepackage{mathptmx}

\usepackage[caption=false]{subfig}

\begin{document}

\preprint{APS/123-QED}

\title{Modeling and Simulation of Non-equilibrium Flows with Uncertainty Quantification}
% Force line breaks with \\

\author{Tianbai Xiao}%
 \email{tianbai.xiao@kit.edu}
\affiliation{ 
Karlsruhe Institute of Technology, Karlsruhe 76131, Germany%\\This line break forced with \textbackslash\textbackslash
}%

\date{\today}% It is always \today, today,

\begin{abstract}

In the study of gas dynamics, theoretical modeling and numerical simulation are mostly set up with deterministic settings.
Given the coarse-grained modeling in theories of fluids, considerable uncertainties may exist between flow-field solutions and real-world physics. 
To study the emergence, propagation and evolution of uncertainties from molecular to hydrodynamic level poses great opportunities and challenges to develop both sound theories and reliable multi-scale algorithms. 
In this paper, we study the stochastic behavior of multi-scale gas dynamic systems, especially focusing on the non-equilibrium effects.
The theoretical analysis is presented on the basis of kinetic model equation and its upscaling macroscopic system, with the reformulation from the stochastic Galerkin method.
A newly developed stochastic kinetic scheme is employed to conduct numerical simulation of homogeneous relaxation, normal shock structure, shear layer and lid-driven cavity problems.
Different kinds of uncertainties are involved in conjunction with the gas evolutionary processes.
New physical observations, such as the synergistic propagation pattern between mean fields and uncertainties, sensitivity of different orders of uncertainties, and the influence of boundary effects from continuum to
rarefied regimes, will be identified and analyzed theoretically.
The paper serves as a heuristic study of quantifying the uncertainties within multi-scale flow dynamics.

\end{abstract}

\maketitle

\section{\label{sec:intro}Introduction}

The study of fluid dynamics is profoundly based on a hierarchy of governing equations from different scales.
Counting a great many of fluid elements, the Navier-Stokes equations are the first-principle modeling of conservation laws with a macroscopic perspective.
On the other hand, the Boltzmann equation describes the gas dynamic system by tracking the evolution of probabilistic distribution function of single particle.
Hilbert’s 6th problem \cite{hilbert1902mathematical} served as an intriguing beginning of trying to describe the behavior of interacting many-particle systems, including the gas dynamic equations, across different scales. 
It has been shown since then that some hydrodynamic equations can be derived as the asymptotic limits of kinetic solutions \cite{chapman1970mathematical,grad1949kinetic}.

The theories of fluids can be regarded as coarse-grained approximation of flow physics in the real world.
Therefore, considerable uncertainties may be introduced due to the lack of comprehensive knowledge or reduced degrees of freedom in the simplified models.
For instance, for the evaluation of collision kernel in the kinetic equations, the phenomenological model parameters often need to be calibrated by experiments, e.g. the Lennard-Jones molecule model \cite{lennard1924determination}.
As a result, the errors inherited from experiments will inevitably influence the numerical evolution of particle interactions at kinetic scale that ought to be deterministic, as well as the reproduced constitutive relationships for the macroscopic moment system.

To evaluate the quality of reduced model and assess the effects of uncertainties on it falls into the topic of uncertainty quantification (UQ).
Two fundamental tasks can be related to UQ problems, i.e. the forward and inverse problems.
The former pertains to uncertainty propagation from model inputs to outputs where input uncertainties have been characterized stochastically.
The latter aims at the parameterization of uncertainties based on existing data sets.
In this paper, we will focus on the forward problem, on which there has been an increasing research interest in computational fluid dynamics (CFD) in recent literature.
For example,
Caucuci \cite{cacuci2003sensitivity} used local sensitivity analysis and Saltelli \cite{saltelli1999quantitative} adopted moment methods to solve flow problems with relatively small uncertainties.
Besides, the spectral methods have been progressively used given the development of polynomial chaos (PC) methods for the probabilistic representation of uncertainty.
Xiu and Karniadakis \cite{xiu2003modeling} discussed the application of generalized polynomial chaos (gPC) to deal with uncertainties with moderate magnitude.
Walters and Huyse \cite{walters2002uncertainty} analyzed these different UQ strategies, and Najm \cite{najm2009uncertainty} addressed the usage of polynomial chaos with the application in compressible flows.

The existing UQ applications of CFD mainly focus on macroscopic fluid dynamic equations with standard stochastic settings. 
In the recent decades, rapid development has been made in multi-scale algorithms, e.g. the continuum-rarefied hybrid methods \cite{bourgat1995coupling,sun2004hybrid,wijesinghe2004three,degond2007moving} and the asymptotic-preserving (AP) schemes \cite{lemou2008new,filbet2010class,xu2010unified,xiao2017well}, which has been proved to be an efficient choice to seek the discrete Hilbert’s 6th path.
However, limited work has been conducted either on the
Boltzmann equation in a stochastic sense or on the evolutionary process of uncertainties in multi-scale physics \cite{hu2017uncertainty,dimarco2019multi}.
Given the nonlinear system including intermolecular collisions, initial inputs, fluid-surface interactions and geometric complexities, uncertainties may emerge from molecular-level nature, develop upwards, affect macroscopic collective behaviors, and vice versa. 
To study the emergence, propagation and evolution of uncertainty poses great opportunities and challenges to develop both sound theories and reliable multi-scale numerical algorithms.

In this paper, theoretical analysis and numerical experiments will be carried out to study the uncertainty propagation in multi-scale and non-equilibrium flows quantitatively.
For the theoretical analysis, the kinetic theory of gases is reformulated with the stochastic Galerkin (SG) method, which is an intrusive methodology based on the generalized polynomial chaos (gPC).
The numerical experiments are produced by the newly developed stochastic kinetic scheme (SKS) \cite{xiao2020stochasticflow}, which is a hybrid method of intrusive stochastic Galerkin and collocation schemes and is able to conduct accurate and efficient simulations.
Several CFD applications, including homogeneous relaxation, normal shock structure, and lid-driven cavity, under the different uncertainties from initial flow fields, boundary conditions and intermolecular collision kernels, will be studied in detail.

The rest of this paper is organized as follows. 
Sec. \ref{sec:theory} is a brief introduction of stochastic kinetic theory and its asymptotic analysis. 
Sec. \ref{sec:algorithm} presents a brief introduction of the solution algorithm employed for numerical simulations.
Sec. \ref{sec:experiments} includes numerical experiments to present and analyze some new physical observations related to uncertainty propagation in fluid dynamics.
The last section is the conclusion.

\section{\label{sec:theory}Stochastic Kinetic theory of gases}

\subsection{Boltzmann equation with uncertainty}

The Boltzmann equation depicts the time-space evolution of particle probability distribution function.
In the absence of external force, it can be written as
\begin{equation}
    f_t + \mathbf u \cdot \nabla_\mathbf x f=Q(f),
    \label{eqn:boltzmann}
\end{equation}
where $\mathbf u \in \mathbb R^3$ is particle velocity, and $Q(f)$ is the collision term.
Considering the possible uncertainties in intermolecular collisions, initial and boundary conditions, we can extend the Boltzmann equation with stochastic settings and reformulate the gas kinetic system, i.e.,
\begin{equation}
\begin{aligned}
    &f_t + \mathbf u \cdot \nabla_\mathbf x f=Q(f)(t,\mathbf{x}, \mathbf{u}, \mathbf{z}), \quad t,\mathbf{x}, \mathbf{u}, \mathbf{z} \in [0,T] \times D \times \mathbb{R}^{3} \times I, \\
    &\mathcal B(f)(t, \mathbf{x}, \mathbf{u}, \mathbf{z})=0, \quad t,\mathbf{x}, \mathbf{u}, \mathbf{z} \in [0,T] \times \partial D \times \mathbb{R}^{3} \times I,\\
    &f(0, \mathbf{x}, \mathbf{u}, \mathbf{z})=f_0(\mathbf{x}, \mathbf{u}, \mathbf{z}), \quad \mathbf{x}, \mathbf{u}, \mathbf{z} \in D \times \mathbb{R}^{3} \times I, \\
\end{aligned}
\label{eqn:sbe system}
\end{equation}
where $\mathbf z \in I$ is the random variable.
For brevity, the following analysis will be conducted on basis of the Bhatnagar-Gross-Krook (BGK) model,
\begin{equation}
    Q(f)=\nu (\mathcal M - f),\ \mathcal M = \rho \left(\frac{\lambda}{\pi}\right)^{3/2}e^{-\lambda(\mathbf u - \mathbf U)^2},
\end{equation}
where $\mathcal M$ is the Maxwellian distribution function, and $\nu$ is the collision frequency, and $\lambda=\rho/(2p)$.
The full Boltzmann collision integral can be also implemented in the numerical simulation \cite{wu2013deterministic}.

\subsection{Stochastic Galerkin formulation}

The methods of uncertainty quantification can be roughly divided into two subsets, i.e. the intrusive and non-intrusive.
The non-intrusive method is sampling-based technique.
Many realizations of random inputs are generated based on the prescribed probability distribution, for which a deterministic problem is solved.
The intrusive methods work in a way such that we reformulate the original deterministic system. 
It promises an intuitive physical insight and higher-order convergence in the random space.
In this part, we introduces the intrusive stochastic Galerkin (SG) method on basis of the generalized polynomial chaos (gPC).

Let us consider a gPC expansion of particle distribution function with degree $N$ in random space, i.e.,
\begin{equation}
    f(t,\mathbf x,\mathbf u,\mathbf z) \simeq f_{N} = \sum_{|i|=0}^N \hat f_{i} (t,\mathbf x,\mathbf u) \Phi_i (\mathbf z) = \hat{\boldsymbol f}^T \boldsymbol \Phi,
    \label{eqn:polynomial chaos}
\end{equation}
where $i$ could be a scalar or a $K$-dimensional vector $i=(i_1,i_2,\cdots,i_K)$ with $|i|=i_1+i_2+\cdots+i_K$.
The $\hat f_{i}$ is the coefficient of $i$-th polynomial chaos expansion,
and the basis functions used are orthogonal polynomials \{$\Phi_ i(\mathbf z)$\} satisfying the following constraints,
\begin{equation}
\begin{aligned}
    \mathbb{E}[\Phi_ j (\mathbf z) \Phi_ k (\mathbf z)] 
    &= \int_{I_{\mathbf z}} \Phi_{ j}(\mathbf z) \Phi_{ k}(\mathbf z) \varrho(\mathbf z) d \mathbf z \\
    &= \gamma_ k \delta_{ j  k}, \quad 0 \leq | j|, | k| \leq N,
\end{aligned}
\end{equation}
where $\varrho$ is the probability density function and $\gamma_k$ is the normalization factor,
\begin{equation}
\gamma_ k=\mathbb{E}[\Phi_ k^2 (\mathbf z)], \quad 0 \leq |k| \leq N.
\end{equation}
For brevity, we use the notation $\langle \rangle$ to denote taking moments along random space henceforth.
The expectation value and variance can be evaluated through
\begin{equation}
\begin{aligned}
    &\mathbb{E}(f_N) = \left\langle \sum_i^N \hat f_i \Phi_i \right\rangle = \hat f_0, \\
    &\mathrm{var}(f_N) = \left\langle \left(\sum_i^N \hat f_i \Phi_i - \mathbb{E}(f_N) \right)^2 \right\rangle \simeq \sum_{|i|>0}^N \left( \gamma_i \hat f_i^2 \right)
\end{aligned}
\label{eqn:gpc mean and variance}
\end{equation}

After substituting the Eq.(\ref{eqn:polynomial chaos}) into the kinetic equation (\ref{eqn:sbe system}) and performing a Galerkin projection, we then obtain
\begin{equation}
\frac{\partial \hat{\boldsymbol f}}{\partial t}+\mathbf{u} \cdot \nabla_{\mathbf{x}} \hat {\boldsymbol f} = \hat {\boldsymbol Q},
\label{eqn:stochastic bgk equation}
\end{equation}
where $\hat{\boldsymbol Q}$ is the gPC coefficient vector of the projection from collision operator to the polynomial basis.
With the assumption of collision frequency,
\begin{equation}
    \nu\simeq\nu_N=\sum_i^N \nu_i \Phi_i,
\end{equation}
the collision term in gPC expansion can be written as
\begin{equation}
\begin{aligned}
    & {f}_{N}=\hat{\boldsymbol f}^{T} \boldsymbol \Phi=\sum_{i}^{N} \hat{f}_{i} \Phi_{i},\\
    & \hat{f}_{i}\left(f_{N}\right)=\frac{\sum_{j}^{N} \sum_{k}^{N} \hat{\nu}_{j} \hat{m}_{k}\left\langle\Phi_{j} \Phi_{k}, \Phi_{i}\right\rangle-\sum_{j}^{N} \sum_{k}^{N} \hat{\nu}_{j} \hat{f}_{k}\left\langle\Phi_{j} \Phi_{k}, \Phi_{i}\right\rangle}{\gamma_i},
\end{aligned}
\end{equation}
where $\hat {m}_{k}$ is the $k$-th coefficient in gPC expansion of equilibrium distribution and can be determined by $\hat {\boldsymbol f}$.

\subsection{\label{sec:asymptotic}Asymptotic analysis}

The kinetic theory of gases indicates the correspondence between macroscopic and microscopic variables.
In the stochastic sense, the macroscopic flow system can also be derived by taking moments along phase space,
\begin{equation}
\begin{aligned}
    \mathbf W  \simeq \mathbf W_{N} &= \int f_{N} \varpi d\mathbf u =\int \sum_{ i}^N \hat f_{i} (t,\mathbf x,\mathbf u) \Phi_ i (\mathbf z) \varpi d\mathbf u \\
    &= \sum_{i}^N \left( \int \hat f_{i} \varpi d\mathbf u \right) \Phi_ i = \sum_{ i}^N \hat w_{i} \Phi_ i,
\end{aligned}
\end{equation}
where $\varpi$ is a vector of velocity moments factors.
For conservative flow variables, it holds $\varpi=(1,\mathbf u,\mathbf u^2/2)^T$.
In the following, we are going to analyze the current stochastic Galerkin BGK equation, especially targeting its asymptotic limiting cases.

\subsubsection{\label{sec:asymptotic homogeneous}Homogeneous case}

Let us begin with spatially homogeneous case.
In this case, the BGK equation reduces to
\begin{equation}
    f_t=\nu(\mathcal M - f).
\end{equation}
It holds the following analytical solution,
\begin{equation}
    f=f_0 e^{-\nu t} + \mathcal M(1-e^{-\nu t}),
    \label{eqn:bgk homogeneous solution}
\end{equation}
where $f_0$ is the particle distribution at initial time instant.
Given a certain degree of uncertainty, we can reformulate the solution above into the stochastic Galerkin form, i.e.,
\begin{equation}
\begin{aligned}
    & f\simeq {f}_{N}=\hat{\boldsymbol f}^{T} \boldsymbol \Phi=\sum_{i}^{N} \hat{f}_{i} \Phi_{i},\\
    & \hat{f}_{i}=\frac{\left\langle \sum_{j}^{N} (\hat{f}_{0j}-\hat{m}_{j}) \Phi_{j} \exp(-\sum_{k}^{N}  \hat{\nu}_{k} \Phi_{k}t), \Phi_{i}\right\rangle}
    {\gamma_{i}}
    +\hat{m}_{i},
\end{aligned}
\label{eqn:sgbgk homogeneous solution}
\end{equation}

Different kinds of uncertainties can be considered.
For example, we suppose the uncertainty comes from stochastic collision frequency.
The initial distribution $f_0$ is deterministic, so are the macroscopic variables and the Maxwellian.
Threfore, Eq.(\ref{eqn:sgbgk homogeneous solution}) reduces to
\begin{equation}
    \hat{f}_{i}=\frac{\left\langle ({f}_{0}-\mathcal{M}) \exp(-\sum_{j}^{N}  \hat{\nu}_{j} \Phi_{j}t) + \mathcal M, \Phi_{i}\right\rangle}
    {\gamma_{i}}.
    \label{eqn:sgbgk homogeneous solution 1}
\end{equation}
It is obvious that the variance of solution is zero at either $t=0$ or $t\rightarrow \infty$, with an extremum existing in between.
Besides, since the collision frequency is independent of particle velocity, it will not affect the shape of particle distribution in velocity space, but plays as a scalar multiplier and affects thermodynamic properties only.

On the other hand, if the uncertainty is imprinted within initial distribution $f_0$ and the collision kernel is deterministic, Eq.(\ref{eqn:sgbgk homogeneous solution}) becomes,
\begin{equation}
    \hat{f}_{i}=\hat{f}_{0i} e^{-\nu t} +\hat{m}_{i} (1-e^{-\nu t}).
\end{equation}
Notice the initial and Maxwllian distributions correspond to the same macroscopic conservative variables, which keeps constant in the absence of particle transport.
Therefore, the following equation holds,
\begin{equation}
\begin{aligned}
    \hat w_{i} = \left( \int \hat f_{0i} \varpi d\mathbf u \right) = \left( \int \hat m_{i} \varpi d\mathbf u \right) = \left( \int \hat f_{i} \varpi d\mathbf u \right) ,
\end{aligned}
\end{equation}
and the expected macroscopic solutions and variances should keep constant during the evolution.

\subsubsection{Inhomogeneous case}

The BGK equation (\ref{eqn:boltzmann}) can be rewritten into the successive form, 
\begin{equation}
    f = \mathcal M - \tau D f = \mathcal M - \tau D(\mathcal M - \tau Df) = \cdots,
\end{equation}
where $D$ denotes the full derivatives along particle trajectories, and $\tau=1/\nu$ is the mean relaxation time.
Truncating the right hand side at a certain order yields concrete kinetic solution as well as its upscaling moment system, e.g. the Euler with $O(\tau)$ truncation, Navier-Stokes with $O(\tau^2)$ truncation, etc.

Now we look into the stochastic Galerkin system.
Truncate the solution with zeroth order, i.e.,
\begin{equation}
    \hat{\boldsymbol f} \simeq \hat{\boldsymbol m},
\end{equation}
and insert the above solution into Eq.(\ref{eqn:stochastic bgk equation}).
The compatibility condition of collision term leads to 
\begin{equation}
    \int\left(\begin{array}{c}
    1 \\
    \mathbf u \\
    \frac{1}{2} \mathbf u^{2}
    \end{array}\right)\left( \hat{\boldsymbol m}_{t}+\mathbf u \cdot \nabla_\mathbf x \hat{\boldsymbol m} \right) d \mathbf u=0,
\end{equation}
and the corresponding Euler equations yields 
\begin{equation}
    \frac{\partial}{\partial t}\left(\begin{array}{c}
    \hat{\boldsymbol \rho} \\
    \hat{(\boldsymbol{\rho U})} \\
    \hat{(\boldsymbol{\rho E})}
    \end{array}\right)+\nabla_{\mathbf x} \cdot \left(\begin{array}{c}
    \hat{\boldsymbol{F}}_{\rho} \\
    \hat{\boldsymbol{F}}_{m} \\
    \hat{\boldsymbol{F}}_{e}
    \end{array}\right)=0,
\end{equation}
where $\{\hat{\boldsymbol{F}}_{\rho},\hat{\boldsymbol{F}}_{m},\hat{\boldsymbol{F}}_{e}\}$ are the gPC coefficients vector of fluxes for density, momentum and energy.

For the second-order truncation, the particle distribution in gPC expansion becomes
\begin{equation}
\begin{aligned}
    & f\simeq f_N = \sum_i^N \hat f_i \Phi_i, \\
    & \hat f_i = \hat m_i - \frac{\sum_{j}^{N} \sum_{k}^{N} \hat{\tau}_{j} (\hat{m}_{kt} + \mathbf u \cdot \nabla_\mathbf x m_k)\left\langle\Phi_{j} \Phi_{k}, \Phi_{i}\right\rangle}{\gamma_i}.
\end{aligned}
\end{equation}
Substituting the above solution into Eq.(\ref{eqn:stochastic bgk equation}), we come to
\begin{equation}
    \frac{\partial}{\partial t}\left(\begin{array}{c}
    \hat{\boldsymbol \rho} \\
    \hat{(\boldsymbol{\rho U})} \\
    \hat{(\boldsymbol{\rho E})}
    \end{array}\right)+\nabla_{\mathbf x} \cdot \left(\begin{array}{c}
    \hat{\boldsymbol{F}}_{\rho} \\
    \hat{\boldsymbol{F}}_{m} \\
    \hat{\boldsymbol{F}}_{e}
    \end{array}\right)= 
    \left(\begin{array}{c}
    0 \\
    \hat{\boldsymbol{S}}_{m}\\
    \hat{\boldsymbol{S}}_{e}
    \end{array}\right),
\label{eqn:sgns}
\end{equation}
where
\begin{equation}
\begin{aligned}
    \hat{\boldsymbol S}_m = \int &\mathbf u \sum_{j}^{N} \sum_{k}^{N} \hat{\tau}_{j} (\hat{m}_{ktt} + 2\mathbf u \cdot \nabla_\mathbf x m_{kt} + \mathbf u \cdot \nabla_\mathbf x (\mathbf u \cdot \nabla_\mathbf x m_k))\\ &\left\langle\Phi_{j} \Phi_{k}, \Phi_{i}\right\rangle \varpi d\mathbf u, \\
    \hat{\boldsymbol S}_e = \int & \frac{1}{2}\mathbf u^2 \sum_{j}^{N} \sum_{k}^{N} \hat{\tau}_{j} (\hat{m}_{ktt} + 2\mathbf u \cdot \nabla_\mathbf x m_{kt} + \mathbf u \cdot \nabla_\mathbf x (\mathbf u \cdot \nabla_\mathbf x m_k))\\ &\left\langle\Phi_{j} \Phi_{k}, \Phi_{i}\right\rangle \varpi d\mathbf u.
\end{aligned}
\end{equation}
Notice in the deterministic limit, the derivatives of Maxwellian can be evaluated with
\begin{equation}
\begin{aligned}
    \frac{\partial \hat m_0}{\partial t}=& \frac{1}{\hat \rho_0} \frac{\partial \hat \rho_0}{\partial t} \hat m_0 +\frac{3}{2 \hat \lambda_0} \frac{\partial \hat \lambda_0}{\partial t} \hat m_0 \\
    & +\left(-\mathbf u^{2}+2 \mathbf u \cdot \hat {\mathbf U}_0-\hat {\mathbf U}_0^{2}\right) \frac{\partial \hat \lambda_0}{\partial t} \hat m_0 \\
    &+\left(2 \mathbf u \lambda-2 \hat{\mathbf U}_0 \lambda\right) \cdot \frac{\partial \hat{\mathbf U}_0}{\partial t} \hat m_0, \\
    \nabla_\mathbf x \hat m_0=& \frac{1}{\hat \rho_0} \nabla_\mathbf x \hat \rho_0 \hat m_0+\frac{3}{2 \hat \lambda_0} \nabla_\mathbf x \hat \lambda_0 \hat m_0 \\
    & +\left(-\mathbf u^{2} +2 \mathbf u \cdot \hat{\mathbf U}_0-\hat {\mathbf U}_0^{2}\right) \nabla_\mathbf x \hat{\lambda}_0 \hat m_0 \\
    &+\left(2 \mathbf u \hat{\lambda}_0 -2 \hat{\mathbf U}_0 \hat \lambda_0\right) \cdot \nabla_\mathbf x \hat{\mathbf U}_0 \hat m_0,
\end{aligned}
\end{equation}
and Eq.(\ref{eqn:sgns}) reduces to deterministic Navier-Stokes equations,
\begin{equation}
\begin{aligned}
    &\frac{\partial}{\partial t}\left(\begin{array}{c}
    \hat{\rho}_0 \\
    \hat{\rho}_0 \hat{\mathbf U}_0 \\
    \hat{\rho}_0 \hat E_0
    \end{array}\right)+\nabla_{\mathbf x} \cdot \left(\begin{array}{c}
    \hat{\rho}_0 \hat{\mathbf U}_0 \\
    \hat{\rho}_0 \hat{\mathbf U}_0 \hat{\mathbf U}_0 \\
    \hat{\rho}_0 \hat{\mathbf U}_0 \hat E_0
    \end{array}\right) \\
    &= \nabla_\mathbf x \cdot 
    \left(\begin{array}{c}
    0 \\
    \hat{\mathbf P}_0\\
    \hat{\mathbf U}_0 \cdot \hat{\mathbf P}_0 - \hat{\mathbf q}_0
    \end{array}\right).
\end{aligned}
\end{equation}
The stress tensor $\hat{\mathbf P}_0$ and heat flux $\hat{\mathbf q}_0$ are related to particle transport phenomena with non-vanishing mean free path, i.e.,
\begin{equation}
\begin{aligned}
    &\hat{\mathbf P}_0=-\hat p_0 \mathbf I+\mu\left(\nabla_\mathbf x \hat{\mathbf U}_0+\nabla_\mathbf x \hat{\mathbf U}_0^T-\frac{2}{3} (\nabla_\mathbf x \cdot \hat{\mathbf U}_0) \mathbf I \right), \\
    &\hat{\mathbf q}_{0}=-\kappa \nabla_\mathbf x \hat{T}_0, \\
    & \mu=\tau_0 \hat p_0, \ \kappa=\frac{5}{2} \frac{k}{m} \hat \tau_0 \hat p_0,
\end{aligned}
\end{equation}
where $p$ is the thermodynamic pressure, $\mathbf I$ is the identity tensor, and $k$ is the Boltzmann constant.

\section{\label{sec:algorithm}Solution Algorithm}

Sec. \ref{sec:theory} corroborates the current stochastic kinetic model in the hydrodynamic limit.
However, as we enter the deep end of Knudsen regimes with looser particle interactions, we can't expect the validity of asymptotic analysis.
The direct numerical modeling and simulation should be employed to investigate the non-equilibrium flow dynamics in conjunction with uncertainty propagation.

In this paper, a newly developed stochastic kinetic scheme (SKS) \cite{xiao2020stochasticflow} is employed to conduct numerical experiments.
In the following let us briefly go through the solution algorithm of the scheme.
To avoid tedious repetition, we suggest that the interested readers refer to the literature with detailed numerical implementation \cite{xiao2020stochasticflow,xiao2020stochasticplasma}.

Within the finite volume framework, the gPC coefficients of particle distribution function in the control volume can be expressed as,
\begin{equation}
    \hat {\boldsymbol f}(t^n,\mathbf x_i,\mathbf u_j)=\hat {\boldsymbol f}_{i,j}^n=\frac{1}{\Omega_{i}(\mathbf x)\Omega_{j}(\mathbf u)} \int_{\Omega_{i}} \int_{\Omega_{j}}\hat {\boldsymbol f} (t^n,\mathbf x,\mathbf u) d\mathbf xd\mathbf u,
\end{equation}
where $\Omega_i(\mathbf x) \Omega_j(\mathbf u)$ are the cell area in the discrete physical and velocity space.
The update of particle distribution function is as follows,
\begin{equation}
\begin{aligned}
    \hat {\boldsymbol f}_{i,j}^{n+1}
    =&\hat {\boldsymbol f}_{i,j}^n+\frac{1}{\Omega_{i}}\int_{t^n}^{t^{n+1}} \sum_{S_{r} \in \partial \Omega_{i}} S_r \hat {\boldsymbol F}_{r,j}^f dt + \int_{t^n}^{t^{n+1}} \hat {\boldsymbol Q}_{i,j}^f dt .
\end{aligned}
\label{eqn:galerkin micro update}
\end{equation}
where $\hat{\boldsymbol{F}}^f_r$ is the time-dependent fluxes for distribution function at interface $r$ in physical space, $S_r$ is the interface area, and $\hat{\boldsymbol Q}^f$ is the collision term.
Taking velocity moments of Eq.(\ref{eqn:galerkin micro update}), we obtain the corresponding macroscopic system,
\begin{equation}
\begin{aligned}
    \hat {\boldsymbol{W}}_{i}^{n+1}
    =\hat {\boldsymbol{W}}_{i}^n+\frac{1}{\Omega_{i}}\int_{t^n}^{t^{n+1}}\sum_{\mathbf S_{r} \in \partial \Omega_{i}} \mathbf{S}_{r} \cdot \hat{\boldsymbol{F}}_{r}^{W} dt,
\end{aligned}
\label{eqn:galerkin macro update}
\end{equation}
where $\hat{\boldsymbol{F}}^W_r$ is the fluxes for conservative variables.

The evaluation of interface flux functions is modeled by the evolving solution of kinetic equation.
With a simplified notation of the interface location $\mathbf x_{i+1/2}=0$ and the initial time instant within a time step $t^n=0$, if we assume the collision frequency as a local constant along physical, velocity and random space, the integral solution of Eq.(\ref{eqn:stochastic bgk equation}) holds along the characteristics, 
\begin{equation}
\begin{aligned}
    \hat{\boldsymbol f}(t,0,\mathbf u_j)=&\nu \int_{0}^{t} \hat{\boldsymbol m}(t',\mathbf x',\mathbf u_j)e^{-\nu (t-t')}dt' \\
    &+ e^{-\nu t}\hat{\boldsymbol f}(0,-\mathbf u_j t,\mathbf u_j),
\end{aligned}
\label{eqn:interface integral solution}
\end{equation}
where $\mathbf x'=\mathbf x - \mathbf u t$ is the particle trajectory.

The initial solution of particle distribution $\hat{\boldsymbol f}(0,-\mathbf u_j t,\mathbf u_j)$ can be obtained through reconstruction technique, e.g.
\begin{equation}
\hat{\boldsymbol f}(0,\mathbf x,\mathbf u_j)=\left\{
\begin{aligned}
&\hat{\boldsymbol f}_{i+1/2,j}^L, \quad x < 0, \\
&\hat{\boldsymbol f}_{i+1/2,j}^R, \quad x \geq 0.
\end{aligned}
\right.
\label{eqn:f0 reconstruct 1st}
\end{equation}
with first-order accuracy and
\begin{equation}
\hat{\boldsymbol f}(0,\mathbf x,\mathbf u_j)=\left\{
\begin{aligned}
&\hat{\boldsymbol f}_{i+1/2,j}^L + \partial_\mathbf x \hat{\boldsymbol f}_{i,j} \mathbf x, \quad x < 0, \\
&\hat{\boldsymbol f}_{i+1/2,j}^R + \partial_\mathbf x \hat{\boldsymbol f}_{i+1,j} \mathbf x, \quad x \geq 0.
\end{aligned}
\right.
\label{eqn:f0 reconstruct 2nd}
\end{equation}
up to second order, where $\hat{\boldsymbol f}_{i+1/2,j}^{L,R}$ are the reconstructed particle distribution around the interface, and $\{\partial_\mathbf x \hat{\boldsymbol f}_{i,j},\partial_\mathbf x \hat{\boldsymbol f}_{i+1,j}\}$ are their slopes in the neighboring cells.

The macroscopic conservative variables in the gPC expansions at the interface can be evaluated by taking moments over velocity space,
\begin{equation}
\hat{\boldsymbol w}=\int_{u_j>0} \hat{\boldsymbol f}_{i+1/2,j}^L\varpi d\mathbf u_j +\int_{u_j<0} \hat{\boldsymbol f}_{i+1/2,j}^R\varpi d\mathbf u_j,
\end{equation}
from which the equilibrium distribution function can be defined.
The equilibrium distribution around a cell interface can be constructed with respect to desired order of accuracy, e.g. for second-order accuracy, 
\begin{equation}
    \hat{\boldsymbol m}(t, \mathbf x, \mathbf u)=\hat{\boldsymbol m}^{0}(1+\mathbf a \cdot \mathbf x+A t).
\end{equation}
The space and time derivatives of Maxwellian are related with macroscopic slopes, and can be determined with the help of Euler equations,
\begin{equation}
\begin{aligned}
    \frac{\partial \hat{\boldsymbol w}}{\partial t}=\int A \hat{\boldsymbol m}^{0} \varpi d \mathbf u, \\
    \nabla_\mathbf x \hat{\boldsymbol w} =\int \mathbf a \hat{\boldsymbol m}^{0} \varpi d \mathbf u.
\end{aligned}
\end{equation}

After all the coefficients are obtained, the time-dependent interface distribution function can be written as,
\begin{equation}
\begin{aligned}
    f_{N}(t,0,\mathbf u_j) =& \sum_{i}^N \hat f_{i}(t,0,\mathbf u_j) = \hat{\boldsymbol f}^T \hat{\boldsymbol \Phi}, \\
    \hat{\boldsymbol f}\left(0, t, u_{j}\right)=&\left(1-e^{-\nu t}\right) \hat{\boldsymbol m}_{j}^{0} \\
    &+\left[\left(-1+e^{-\nu t}\right) / \nu +t e^{-\nu t}\right] \mathbf u \cdot \mathbf a \hat{\boldsymbol m}_{j}^{0} \\
    &+\left[\left(\nu t-1+e^{-\nu t}\right) / \nu \right] A \hat{\boldsymbol m}_{j}^{0} \\
    &+e^{-\nu t}\left[\left(\hat{\boldsymbol f}_{i+1 / 2, j}^{L}-\mathbf u_{j} t \partial_\mathbf x \hat{\boldsymbol f}_{i, j}\right) H\left[\mathbf u_{j}\right]\right.\\
    &+\left(\hat{\boldsymbol f}_{i+1 / 2, j}^{R}-\mathbf u_{j} t \partial_\mathbf x \hat{\boldsymbol f}_{i+1, j}\right)\left(1-H\left[\mathbf u_{j}\right]\right),
\end{aligned}
\label{eqn:interface distribution}
\end{equation}
where $H$ is the heaviside step function.
The corresponding fluxes of particle distribution function and conservative flow variables can be obtained via
\begin{equation}
\begin{aligned}
    &{{F}}_{N}^f(t,0,\mathbf u_j,\mathbf z)= \mathbf u_j f_{N}(t,0,\mathbf u_j,\mathbf z) , \\
    &{\mathbf{F}}_{N}^W(t,0,\mathbf z)=\int \mathbf u_j f_{N}(t,0,\mathbf u_j,\mathbf z) \varpi d\mathbf u_j ,
\end{aligned}
\end{equation}
and the time-integrated fluxes in Eq.(\ref{eqn:galerkin micro update}) and (\ref{eqn:galerkin macro update}) can be evaluated with respect to time in Eq.(\ref{eqn:interface distribution}).

Besides the construction of the interface flux, the collision term needs to be evaluated inside control volume for the update of particle distribution function within each time step.
In the solution algorithm, Eq.(\ref{eqn:galerkin macro update}) will be updated first, and the obtained macroscopic variables will be used to construct the Maxwellian at $t^{n+1}$.
As a result, an implicit update of collision term can be achieved based on the explicit solver framework.
Let us rewrite the update algorithm for the $k$-th gPC coefficient of particle distribution function in control volume $(\Omega_i, \Omega_j)$,
\begin{equation}
\begin{aligned}
&\hat f_{i,j,k}^{n+1} + \frac{\Delta t \sum_p^N \sum_q^N \hat \nu_p^{n+1} \hat f_q^{n+1} \langle \Phi_p \Phi_q \Phi_k \rangle}{\gamma_k} \\
=&\hat f_{i,j,k}^n + \frac{1}{\Omega_{i}}\int_{t^n}^{t^{n+1}} \sum_{S_{r} \in \partial \Omega_{i}} S_r \hat {F}_{r,j,k}^f dt \\
&+\frac{\Delta t \sum_p^N \sum_q^N \hat \nu_p^{n+1} \hat m_q^{n+1} \langle \Phi_p \Phi_q \Phi_k \rangle}{\gamma_k},
\end{aligned}
\label{eqn:galerkin linear system}
\end{equation}
which forms a linear system in $A\hat{\boldsymbol f}=B$ manner.
The system can be directly solved, but brings considerable computational cost as the gPC expansion order $N$ increases.
It can be solved in a more elegant way with the hybridization of Galerkin and collocation methods proposed \cite{xiao2020stochasticflow}.
The main idea of this method can be summarized as to solve an intrusive SG system with gPC expansions by using collocation points.
To make use of it, in the solution algorithm, we first update the gPC macroscopic variables to $t^{n+1}$ step, and the distribution function to the intermediate step $t^*$,
\begin{equation}
\hat {\mathbf{W}}_{i,k}^{n+1}=\hat{\mathbf{W}}_{i,k}^n+\frac{1}{\Omega_{i}}\int_{t^n}^{t^{n+1}}\sum_{\mathbf S_{r} \in \partial \Omega_{i}} \mathbf{S}_{r} \cdot \hat{\mathbf{F}}_{r,k}^{W} dt,
\end{equation}
\begin{equation}
\hat f_{i,j,k}^{*}=\hat f_{i,j,k}^n+\sum_{S_{r} \in \partial \Omega_{i}} S_r \hat {F}_{r,j,k}^f dt,
\end{equation}
which is then evaluated on the quadrature points $z_q$,
\begin{equation}
f_{i,j,q}^*=f_{Ni,j}^*(z_q) = \sum_m^N \hat f_{i,j,k}^{*} (z_q) \Phi_k (z_q).
\end{equation}
Afterwards, the collision term is solved via
\begin{equation}
\begin{aligned}
f_{i,j,q}^{n+1}=&f_{i,j,q}^* + \Delta t \nu_{i,j,q}^{n+1} (\mathcal M_{i,j,q}^{n+1} - f_{i,j,q}^{n+1}) \\
=&( f_{i,j,q}^* + \Delta t \nu_{i,j,q}^{n+1} \mathcal M_{i,j,q}^{n+1} )/(1+\Delta t \nu_{i,j,q}^{n+1}).
\end{aligned}
\label{eqn:micro update 1d}
\end{equation}
The updated distribution function can be reabsorbed into the gPC expansion,
\begin{equation}
\hat f^{n+1}_{i,j,k} = \frac{\langle f^{n+1}_{i,j},\Phi_k \rangle}{\langle \Phi_{k}^2\rangle},
\end{equation}
and the final solution in gPC expansion at $t^{n+1}$ is,
\begin{equation}
f_{Ni,j}^{n+1}=\sum_{k}^N \hat f^{n+1}_{i,j,k} \Phi_k.
\end{equation}
With the latter hybrid Galerkin-collocation method, the computational efficiency can be improved with orders of magnitude.

\section{\label{sec:experiments}Numerical Experiments}

In this section, we are going to conduct the numerical experiments covering different flow regimes.
Different kinds of uncertainties will be coupled with the flow evolving processes throughout the simulations.
The motivation of this section, on one hand, is to investigate multi-scale gas dynamic system and analyze typical flow phenomena in conjunction with propagation of uncertainties.
On the other hand, it serves to provide the first-hand benchmark solutions of uncertainty quantification in non-equilibrium flows.

For convenience, dimensionless variables will be introduced in the simulations,
\begin{equation*}
\begin{aligned}
    & \tilde{\mathbf x}=\frac{\mathbf x}{L_0}, \ \tilde{\rho}=\frac{\rho}{\rho_0}, \  \tilde{T}=\frac{T}{T_0}, \ \tilde{\mathbf u}=\frac{\mathbf u}{(2RT_0)^{1/2}}, \  \tilde{\mathbf U}=\frac{\mathbf U}{(2RT_0)^{1/2}}, \\
    & \tilde{f}=\frac{f}{\rho_0 (2RT_0)^{3/2}}, \ \tilde{\mathbf T}=\frac{\mathbf T}{\rho_0 (2RT_0)}, \ \tilde{\mathbf q}=\frac{\mathbf q}{\rho_0 (2RT_0)^{3/2}},
\end{aligned}
\end{equation*}
where $R$ is the gas constant, $\mathbf T$ is stress tensor, and $\mathbf q$ is heat flux. 
The denominators with subscript zero are characteristic variables in the reference state. 
For brevity, the tilde notation for dimensionless variables will be removed henceforth.

\subsection{Homogeneous relaxation of non-equilibrium distribution}

First let us consider the homogeneous relaxation of particles from an initial non-equilibrium distribution.
The evolution system writes
\begin{equation*}
    f_t=\nu(\mathcal M - f),\quad f(t=0,u)=u^2 e^{-u^2}.
\end{equation*}
The uncertainty originates from collision kernel, and results in a stochastic collision frequency $\nu \sim \mathcal N(1,0.2^2)$.
It can be written into the gPC expansion,
\begin{equation*}
    \nu=1+0.2z,
\end{equation*}
where $1$ and $z$ are the first two polynomials in the Hermite system.
The theoretical solution can be constructed following the integral solution of homogeneous kinetic model equation.
Therefore, the particle distribution function obeys a log-normal distribution in the random space, and the expected value and standard deviation can be constructed as,
\begin{equation*}
\begin{aligned}
    &\mathbb E(f) = f_0 \exp(-t + 0.04 t^2/2) + \mathcal M (1 - \exp(-t + 0.04 t^2/2)), \\
    &\mathbb S(f) = \left[ (f_0 - \mathcal M)^2 (\exp(0.04 t^2) - 1) \exp(-2t + 0.04 t^2) \right]^{1/2}.
\end{aligned}
\end{equation*}
This case serves as a benchmark validation of the current numerical scheme.
The detailed computational setup can be found in the following table.
\begin{table}[htbp]
	\caption{Computational setup of homogeneous relaxation.} 
	\centering
	\begin{tabular}{lllllll} 
		\toprule 
		t & $\Delta t$ & $u$ & $N_u$ & Integral & $N$ &  \\ 
		\hline
		$[0,10]$ & $0.01$ & $[-6,6]$ & $201$ & Newton-Cotes & $[0,9]$ &  \\ 
		\hline
		$N_q$ & Polynomial & $\nu$ \\ 
		\hline
		$[1,17]$ & Hermite & $\mathcal N(1,0.2^2)$ \\ 
		\hline
	\end{tabular} 
	\label{tab:relax}
\end{table}

\begin{figure}
\subfloat[Expectation\label{sfig:testa}]{%
    \includegraphics[width=.7\linewidth]{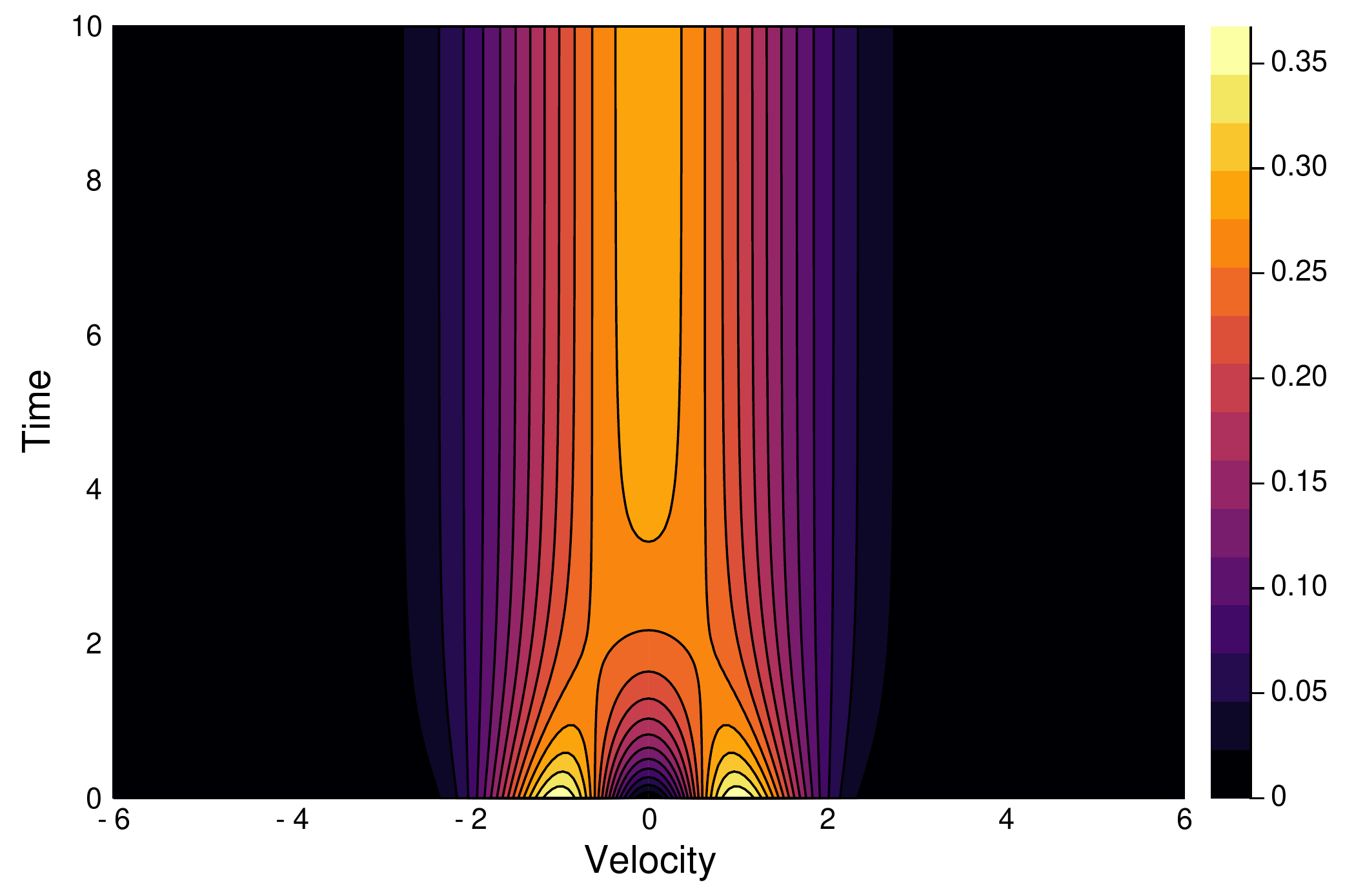}%
}\hfill
\subfloat[Standard deviation\label{sfig:testa}]{%
    \includegraphics[width=.7\linewidth]{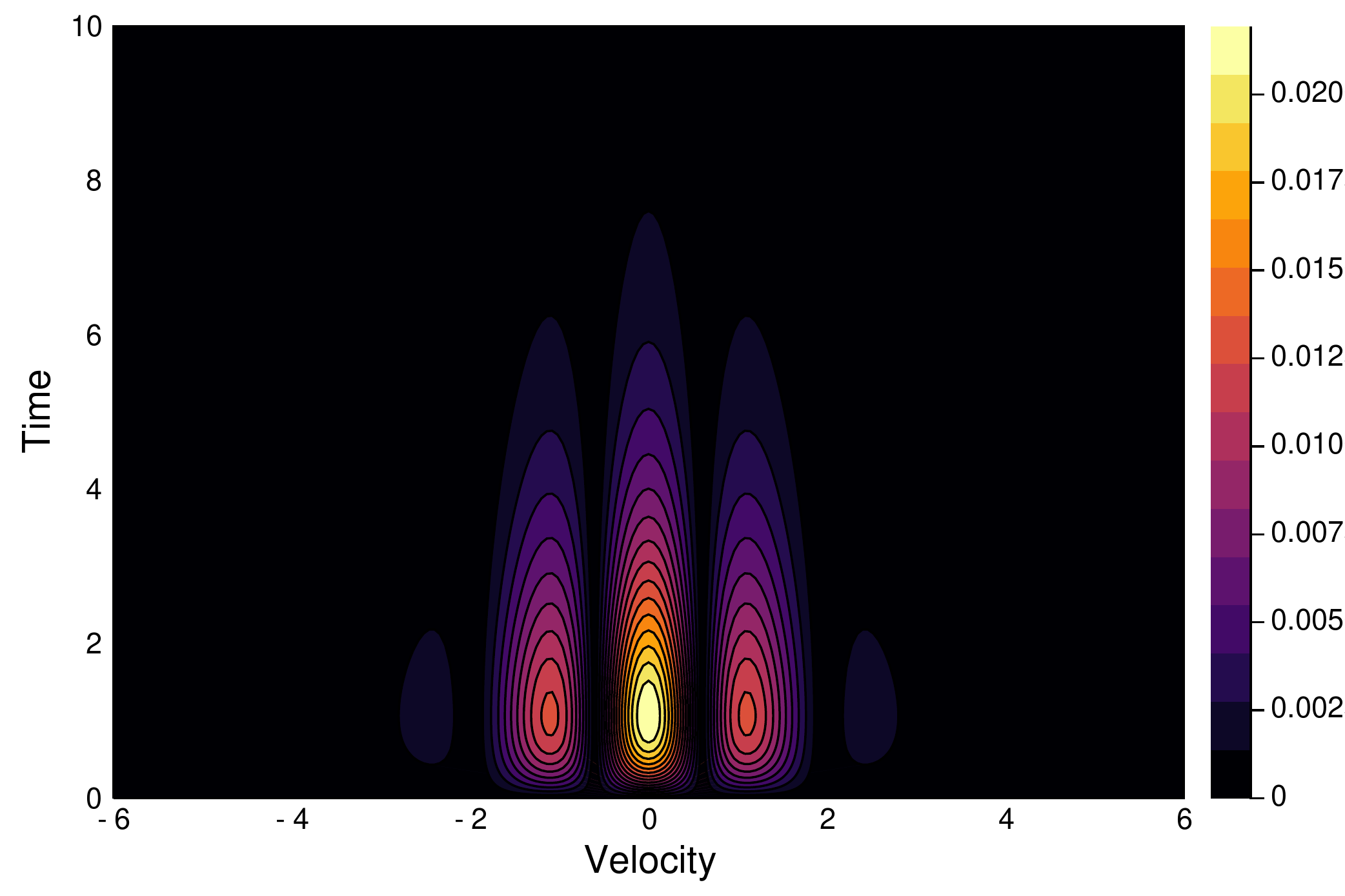}%
}
\caption{Expectation value and standard deviation of particle distribution within $\{t,u\}\in [0,10]\times[-6,6]$ in the homogeneous relaxation problem.}
\label{pic:relaxation micro evolution}
\end{figure}

The stochastic evolution of particle distribution function in its expectation and standard deviation in the phase space $\{ t\times u \}$ are presented in Fig.\ref{pic:relaxation micro evolution}.
With the occurrence of intermolecular interactions, the particle distribution function approaches the Maxwellian gradually from initial bimodal non-equilibrium.
As analyzed in Sec.\ref{sec:asymptotic homogeneous}, a maximum of standard deviation emerges close to the time axis.
It can be understood either with the exact solution Eq.(\ref{eqn:sgbgk homogeneous solution 1}), or from a physical point of view.
The stochastic collision frequency results in prominent uncertainties where the particle collisions are happening significantly.
As time goes with $t>8$, the distribution function is close to equilibrium state and thus kept in a dynamical balance with the Maxwellian which is set to be deterministic in this case.
Therefore, the collision term plays no more incentive effects on the uncertainty propagation.

\begin{figure}
    \includegraphics[width=.7\linewidth]{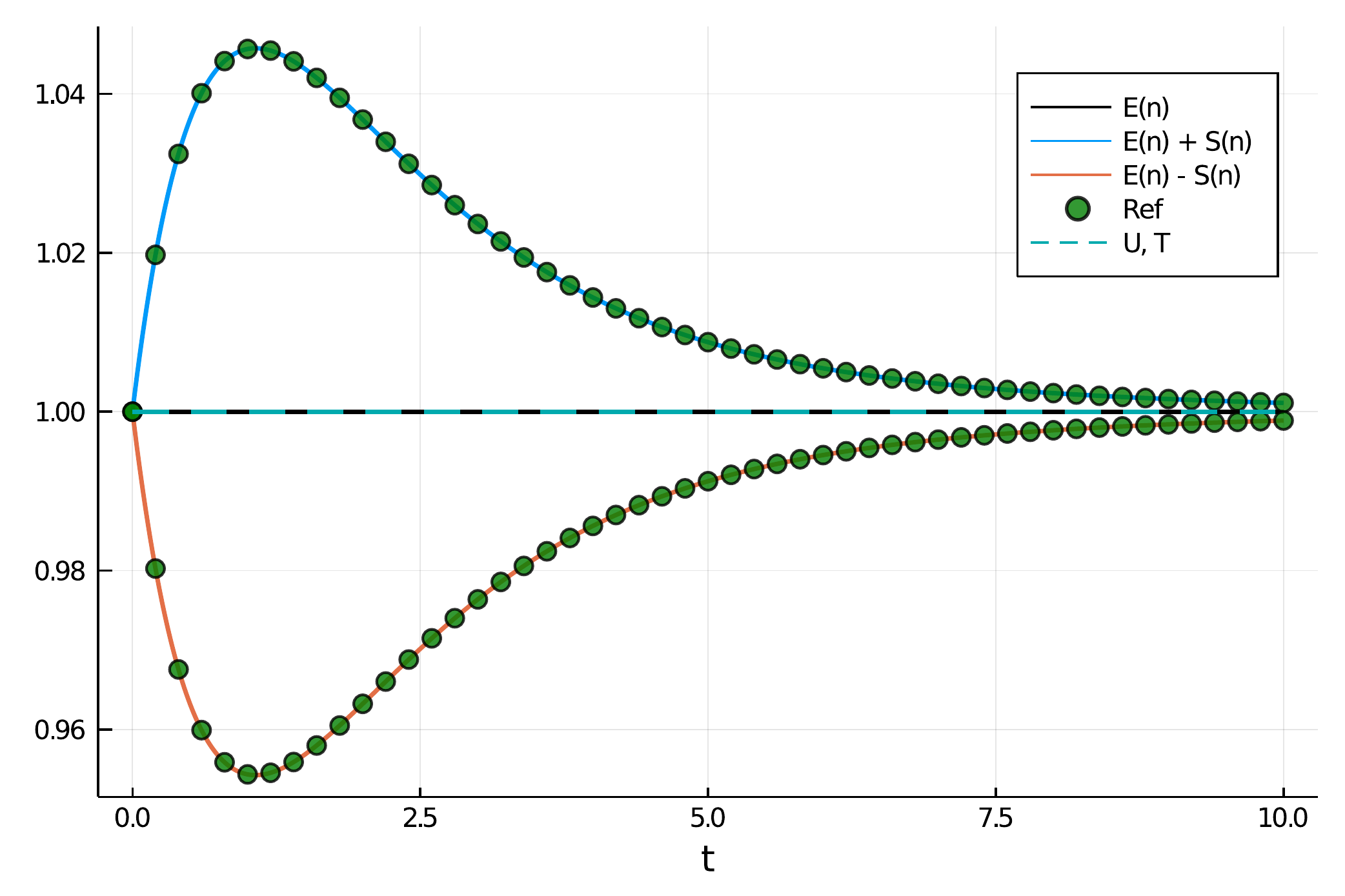}%
\caption{Evolution of macroscopic density, velocity and temperature within $t\in [0,10]$ in the homogeneous relaxation problem. The results are normalized by the initial values.}
\label{pic:relaxation macro evolution}
\end{figure}

As the microscopic particle distribution has a one-to-one correspondence with its macroscopic system, we can easily get the macroscopic evolution by taking moments in the velocity space.
Fig.\ref{pic:relaxation macro evolution} presents the time evolution of number density, velocity and temperature.
As is shown, the stochastic collision term here plays as a scalar multiplier and only affects gas density.

For the validation of the current scheme, we plot the $L_1$ and $L_2$ errors of the numerical solutions with respect to varying order $N$ for gPC expansions.
As is shown, the spectral convergence of the scheme in the probabilistic space is clearly identified. 

\begin{figure}
    \subfloat[$L_1$ error]{%
        \includegraphics[width=.7\linewidth]{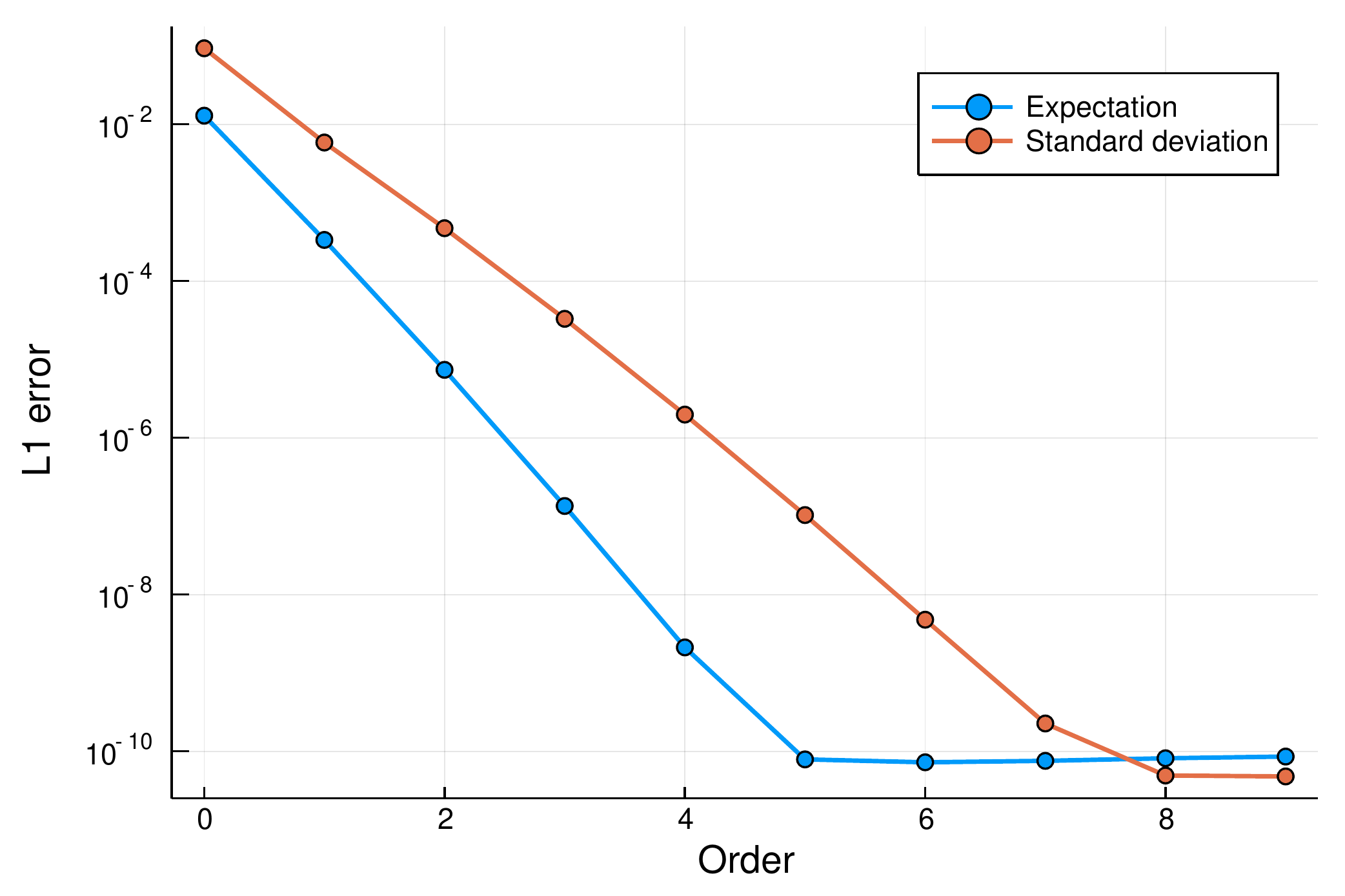}%
    }\hfill
    \subfloat[$L_2$ error]{%
        \includegraphics[width=.7\linewidth]{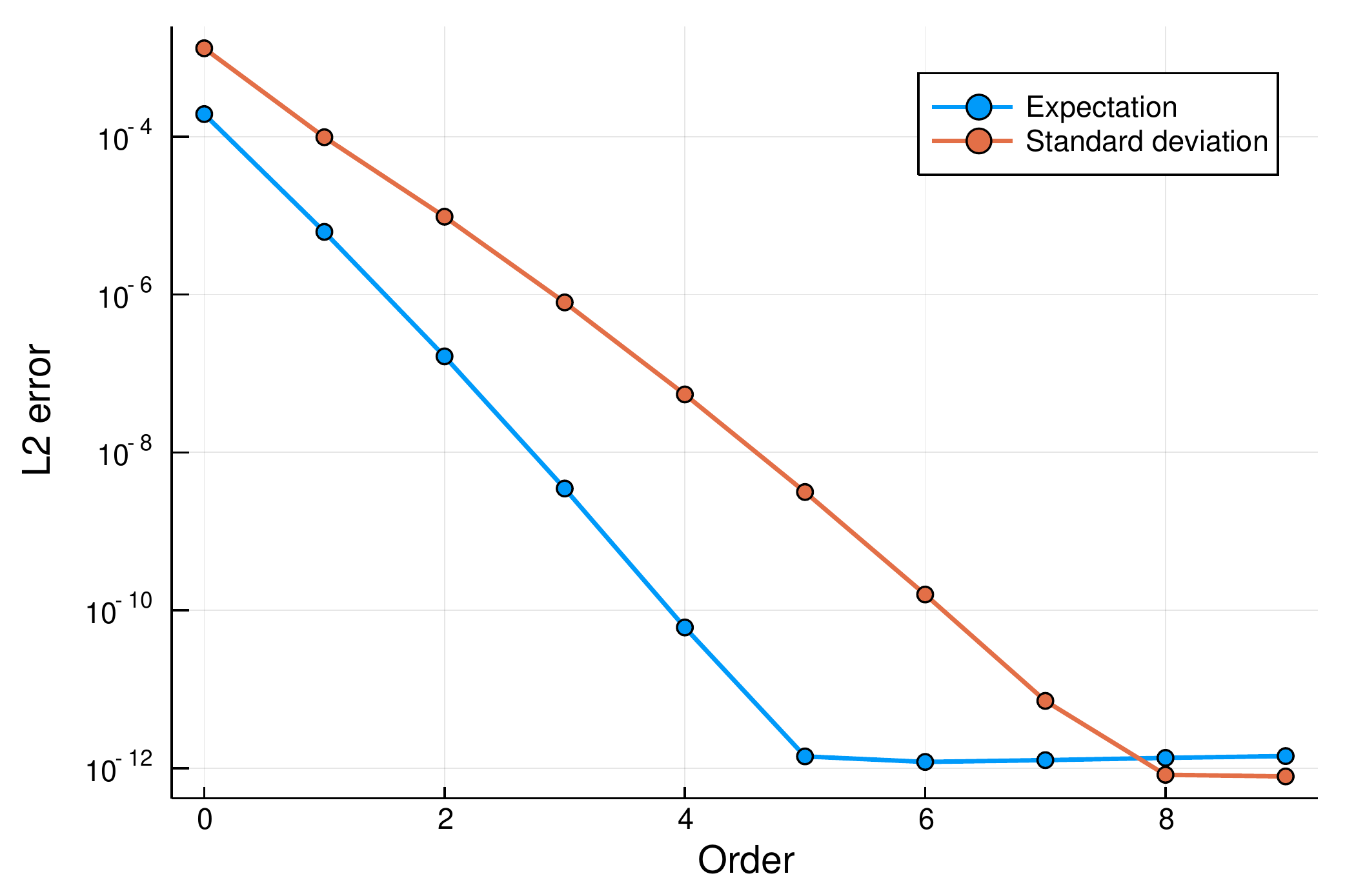}%
    }
\caption{Errors of expectation value and standard deviation of particle distribution function within $\{t,u\}\in [0,10]\times[-6,6]$ in the homogeneous relaxation problem.}
\label{pic:relaxation error}
\end{figure}

\subsection{\label{sec:shock structure}Normal shock structure}

In the following we turn to cases with nonuniform distribution in space.
The first example is normal shock structure, which is highly dissipative and related to strong non-equilibrium effects.
Based on the reference frame of shock wave, the stationary upstream and downstream status can be described via the well-known Rankine-Hugoniot relation,
\begin{equation*}
\begin{aligned}
&\frac{\rho_+}{\rho_-}=\frac{(\gamma+1)\rm{Ma}^2}{(\gamma-1)\rm{Ma}^2+2},\\
&\frac{U_+}{U_-}=\frac{(\gamma-1)\rm{Ma}^2+2}{(\gamma+1)\rm{Ma}^2},\\
&\frac{T_+}{T_-}=\frac{ ((\gamma-1)\rm{Ma}^2+2) (2\gamma\rm{Ma}^2-\gamma+1) }{(\gamma+1)^2 \rm{Ma}^2},
\end{aligned}
\end{equation*}
where $\gamma$ is the ratio of specific heat.
The upstream and downstream conditions are denoted with $\{ \rho_-, U_-, T_- \}$ and $\{ \rho_+, U_+, T_+ \}$.

The collision frequency in the kinetic equation can be derived from transport phenomena,
\begin{equation*}
\nu = \frac{p}{\mu},
\end{equation*}
where $p$ is the pressure and $\mu$ is the viscosity coefficient.
In this case, we continue dealing with random collision term.
A stochastic variable $\xi$ is introduced in the variable hard-sphere (VHS) model for the evaluation of viscosity, which reads
\begin{equation*}
\mu=\xi\mu_{0} \left( \frac{T}{T_{0}} \right)^\eta,
\end{equation*}
and the viscosity coefficient in the reference state is connected with the Knudsen number,
\begin{equation*}
\mu_{0} =  \frac{5 (\alpha + 1) (\alpha + 2) \sqrt\pi}  {4 \alpha (5 - 2 \omega) (7 - 2 \omega)}   \mathrm{Kn}_{0},
\end{equation*}
where $\{ \alpha, \omega, \eta \}$ are parameters for the VHS model.
The computational setup for this case is presented in Table.\ref{tab:shock}.

\begin{table}[htbp]
	\caption{Computational setup of normal shock structure.} 
	\centering
	\begin{tabular}{lllllll} 
		\toprule 
		$x$ & $N_x$ & $u$ & $N_u$ & Integral & $N$ &  \\ 
		\hline
		$[-35,35]$ & 100 & $[-12,12]$ & 101 & Newton-Cotes & 5 \\ 
		\hline
		$N_q$ & Polynomial & $\xi$ & Ma & Kn & CFL  \\ 
		\hline
		9 & Legendre & $\mathcal U(0.6,1.4)$ & $[2,3]$ & 1 & 0.5   \\ 
		\hline
		$\gamma$ & $\alpha$ & $\omega$ & $\eta$ &    \\ 
		\hline
		3 & 1 & 0.5 & 0.81 &    \\ 
		\hline
	\end{tabular} 
	\label{tab:shock}
\end{table}

\begin{figure}
    \subfloat[$\mathbb E(\rho)$]{%
        \includegraphics[width=.7\linewidth]{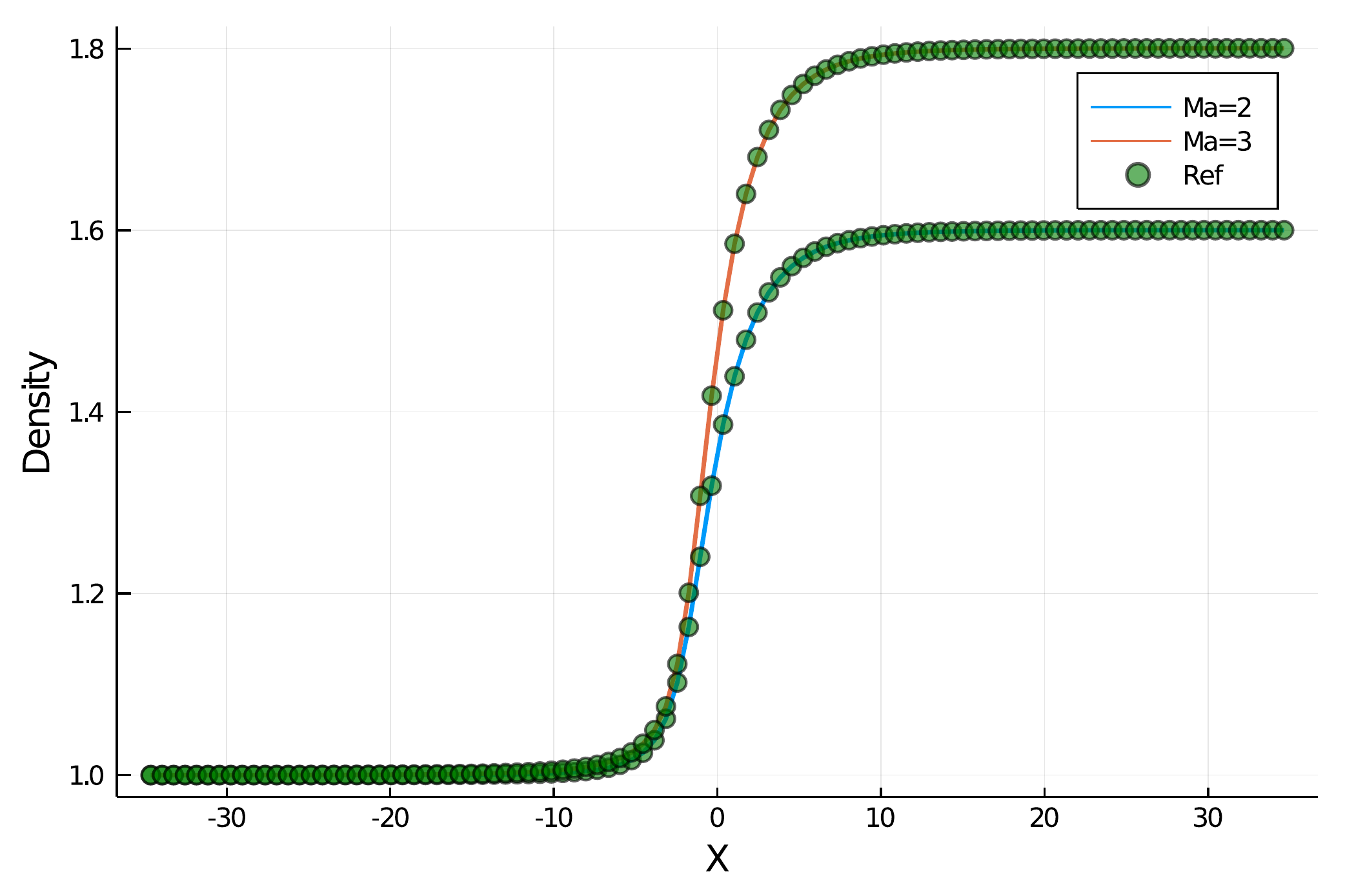}%
    }\hfill
    \subfloat[$\mathbb E(U)$]{%
        \includegraphics[width=.7\linewidth]{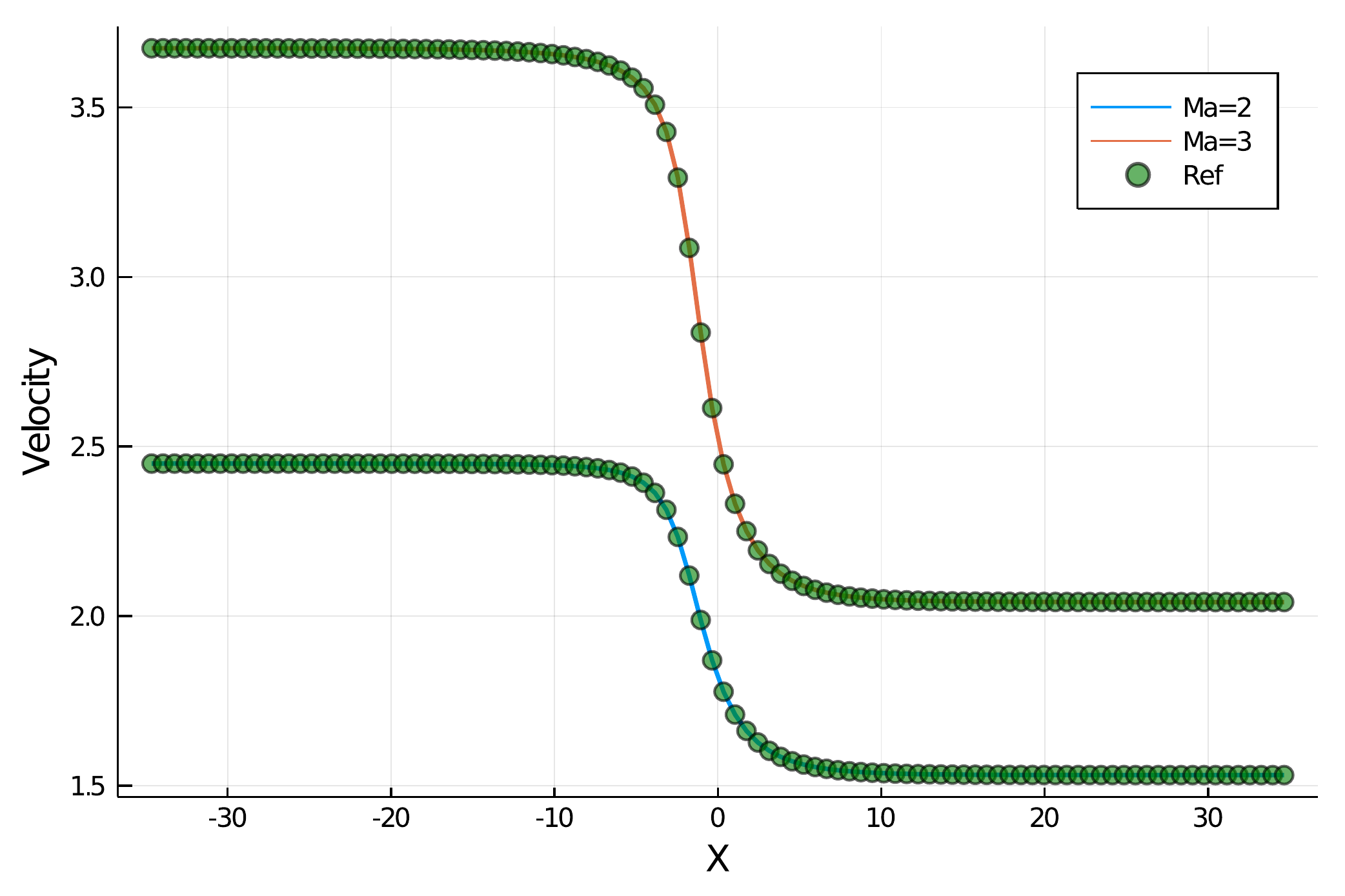}%
    }\hfill
    \subfloat[$\mathbb E(T)$]{%
        \includegraphics[width=.7\linewidth]{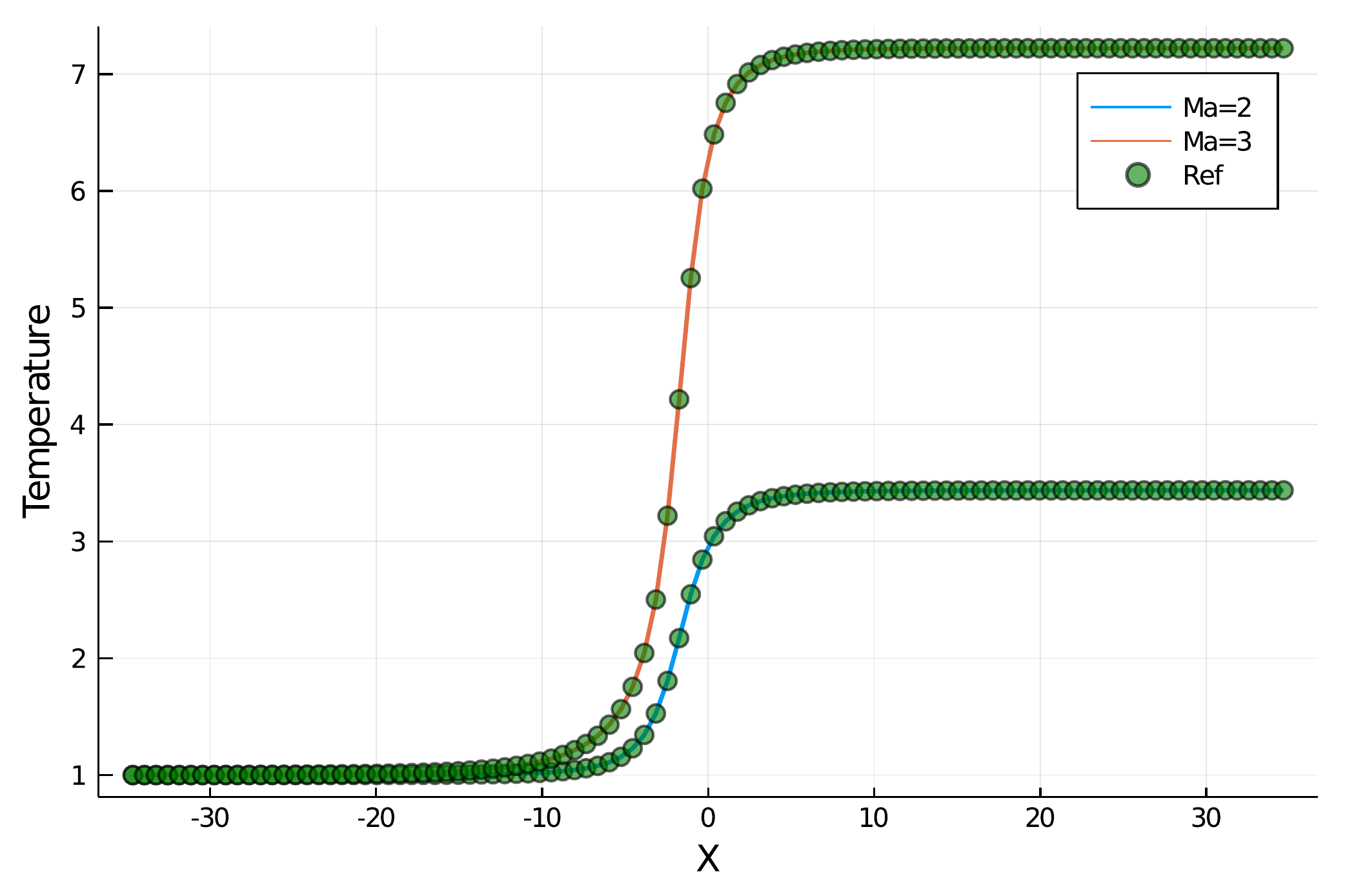}%
    }
\caption{Expectation values of macroscopic density, velocity and temperature in the normal shock structure.}
\label{pic:shock mean}
\end{figure}

\begin{figure}
    \subfloat[$\mathbb S(\rho)$]{%
        \includegraphics[width=.7\linewidth]{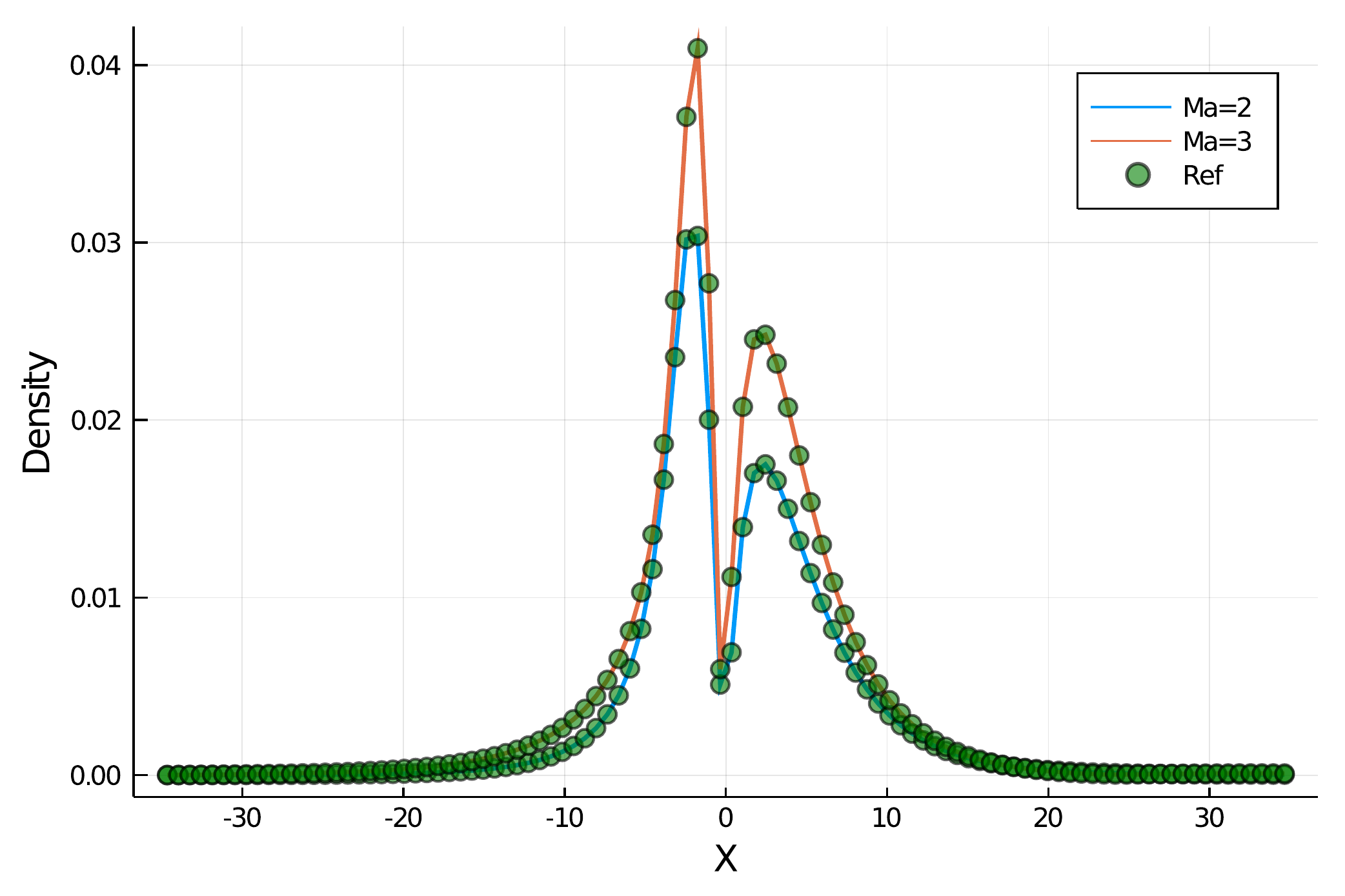}%
    }\hfill
    \subfloat[$\mathbb S(U)$]{%
        \includegraphics[width=.7\linewidth]{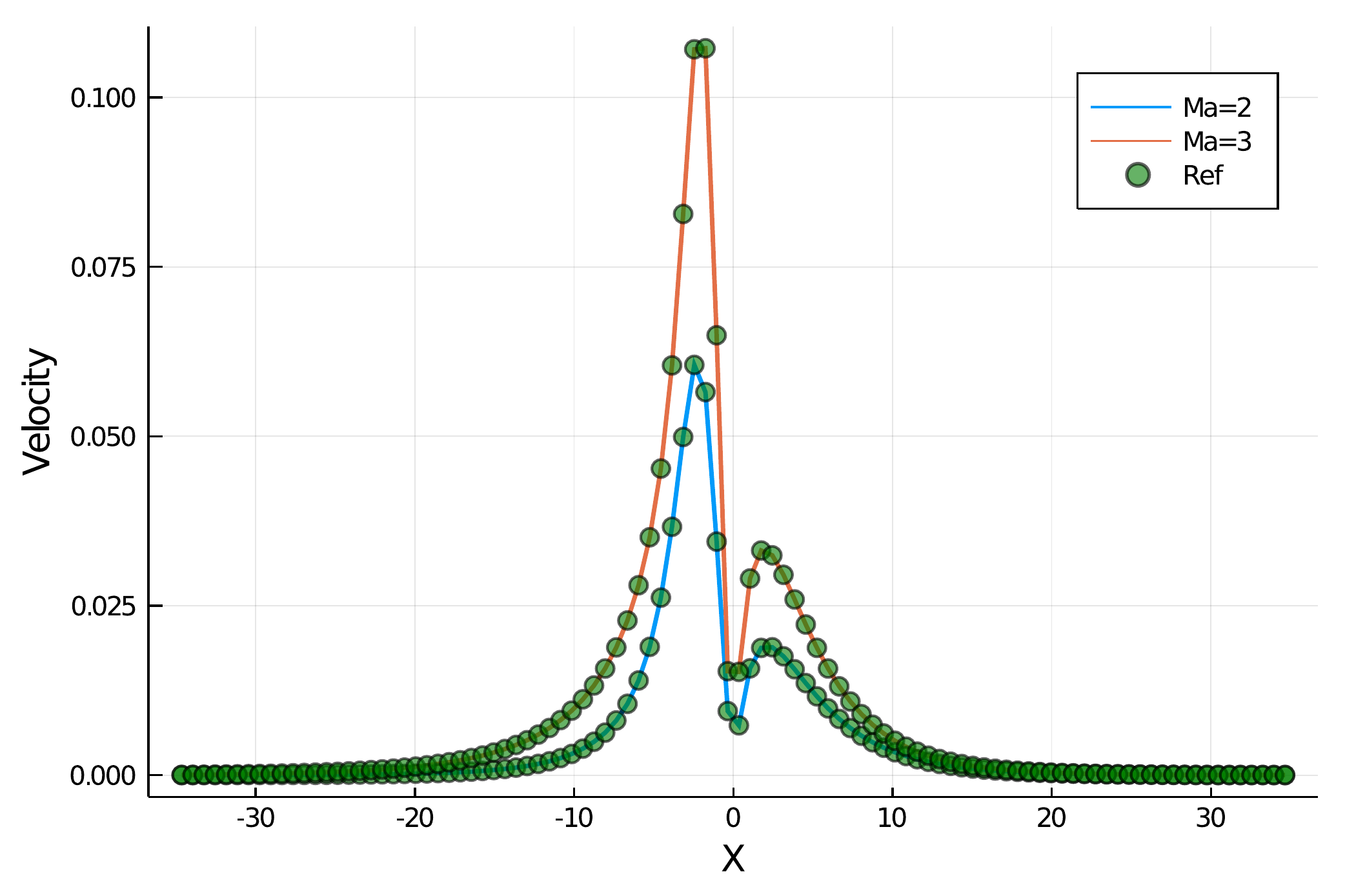}%
    }\hfill
    \subfloat[$\mathbb S(T)$]{%
        \includegraphics[width=.7\linewidth]{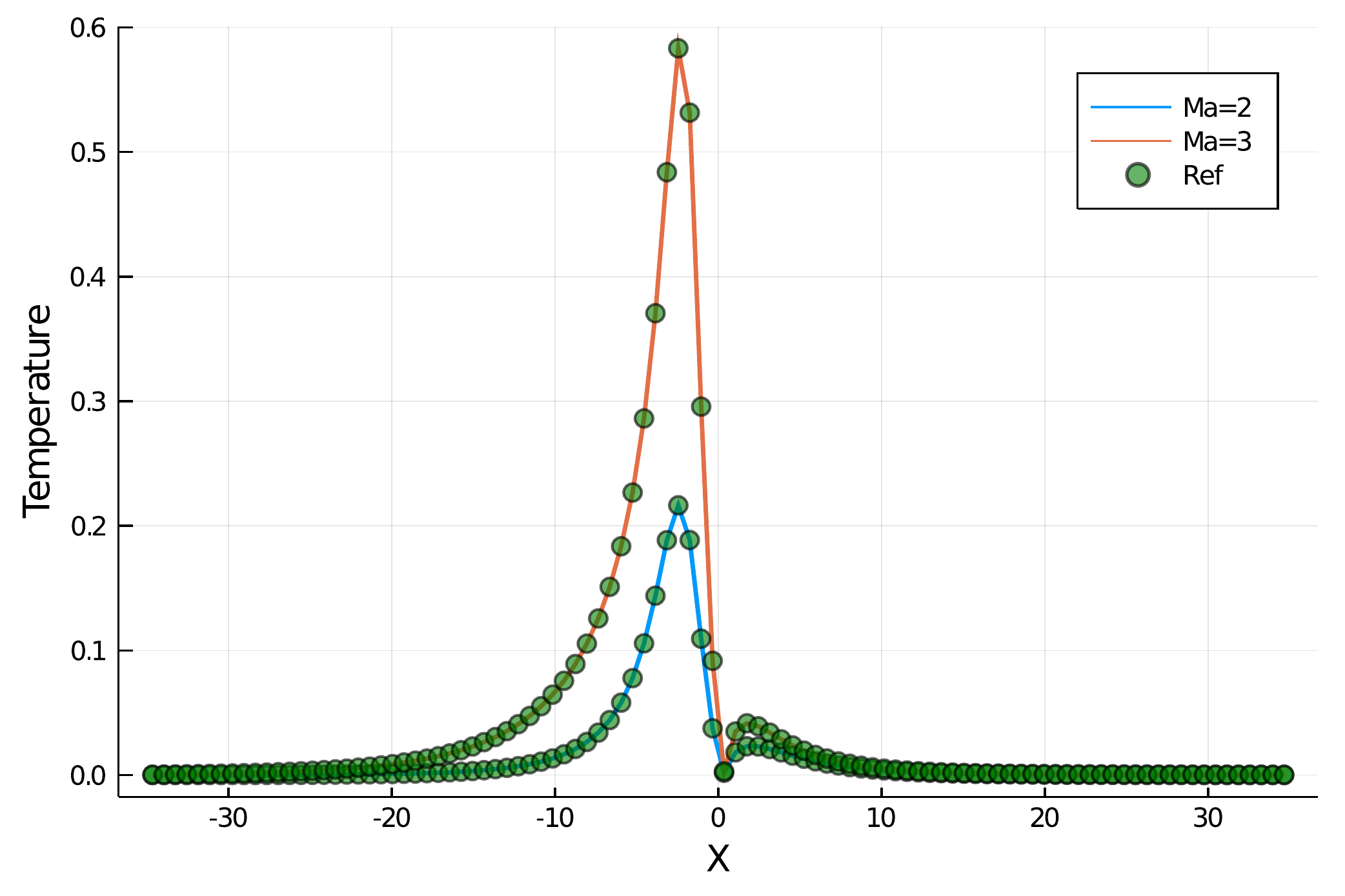}%
    }
\caption{Standard deviations of macroscopic density, velocity and temperature in the normal shock structure.}
\label{pic:shock std}
\end{figure}

The numerical profiles of macroscopic quantities at different upstream Mach numbers $\rm Ma=2$ and $3$ are presented in Fig.\ref{pic:shock mean} and \ref{pic:shock std}.
For the validation of current scheme, the reference solutions produced by the unified gas-kinetic scheme \cite{xiao2017well} with 10000 Monte-Carlo samplings are plotted.
As shown in Fig.\ref{pic:shock mean}, with the increasing Mach number, the expected shock profile becomes wider due to the increasing demand of particle momentum and energy exchanges.

Now let us turn to the stochasticity.
From Fig.\ref{pic:shock std}, it is clear that the shock wave serves as a main source for uncertainties with significant intermolecular interactions inside.
Given the fixed Rankine-Hugoniot relationship, the status at the central point of
shock $x = 0$ are basically determined by the upstream and downstream variables.
Therefore, the uncertainties of flow field present raise a bimodal pattern inside the shock profile.
In other words, the stochastic collision kernel affects the width and shape of the shock wave structure.
A positive correlation can be identified between the steepness of variances and upstream Mach number.
Looking into the figures, we can find it seems that the upstream half of shock center is more sensitive than the downstream part, resulting in a sharper distribution of flow variables.
Among all the quantities, the deviation between upstream and downstream of temperature variance is most striking, indicating a higher sensitivity of higher order moments of particle distribution function.

\begin{figure}
    \subfloat[$\rm Ma=2$]{%
        \includegraphics[width=.7\linewidth]{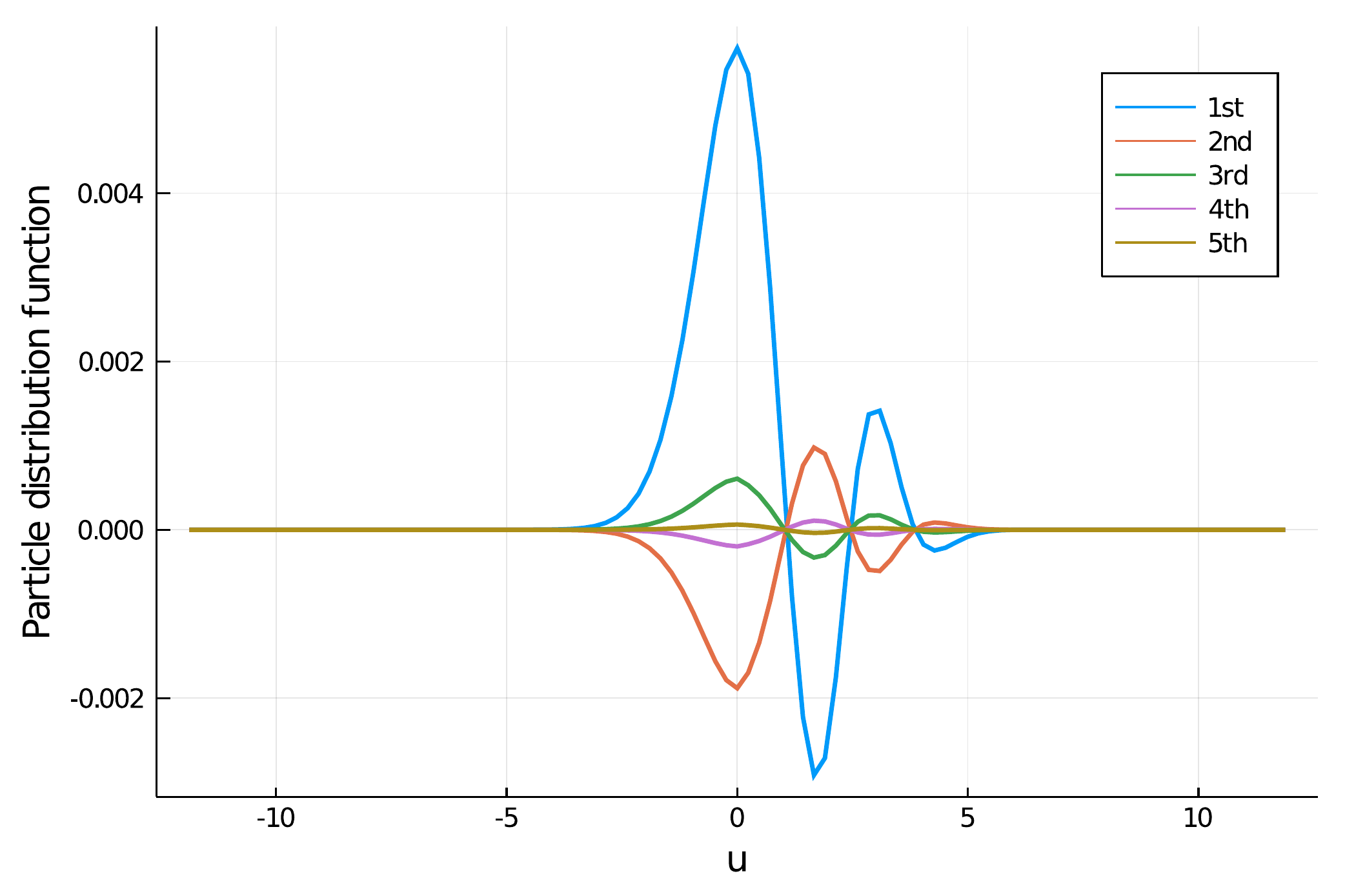}%
    }\hfill
    \subfloat[$\rm Ma=3$]{%
        \includegraphics[width=.7\linewidth]{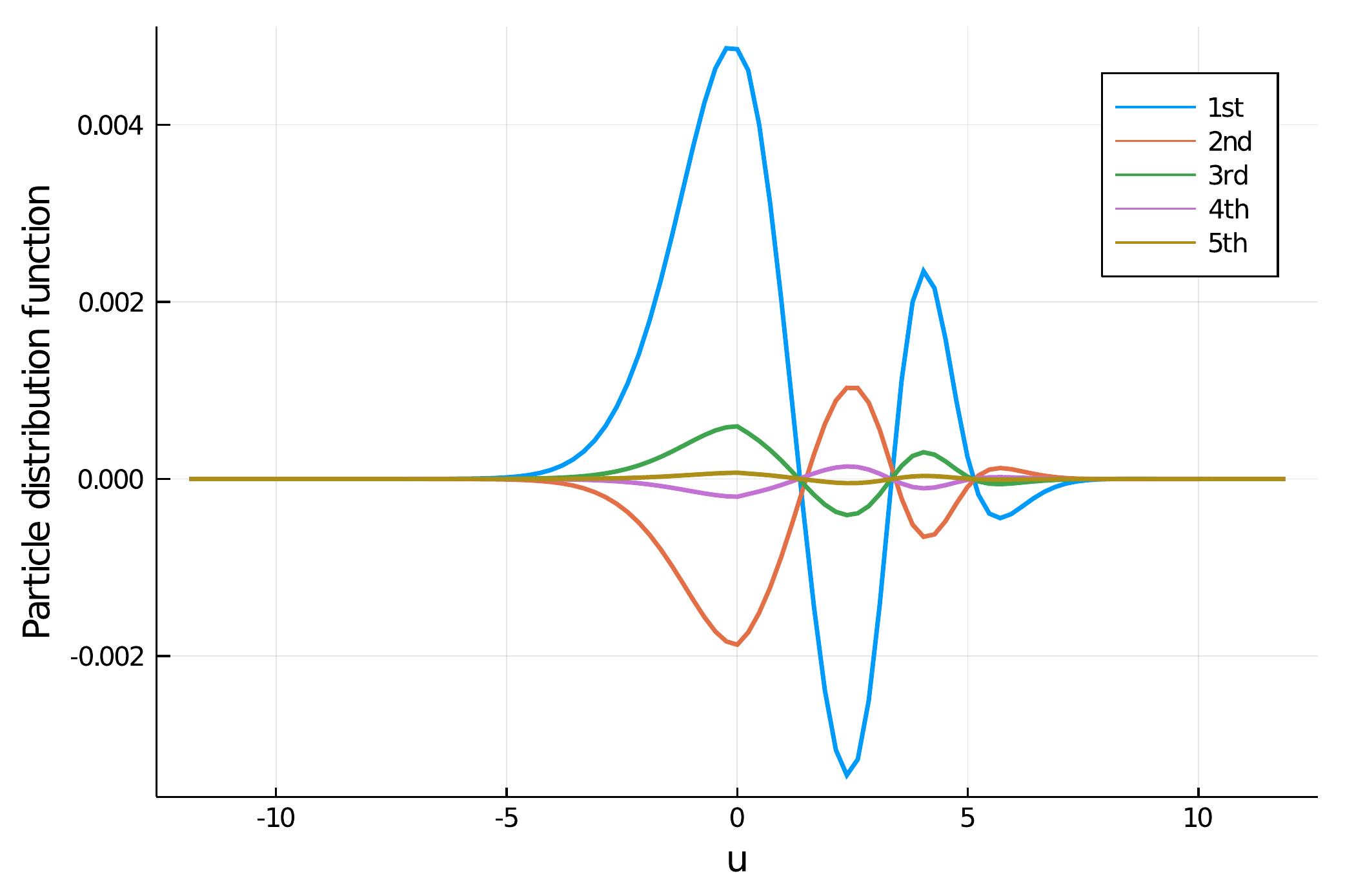}%
    }
\caption{Polynomial chaos expansion coefficients of particle distribution function at the center of normal shock structure.}
\label{pic:shock gpc}
\end{figure}

The gas kinetic modeling and simulation provide us a chance to study the distribution of particles at mesoscopic level.
As analyzed in Sec.\ref{sec:asymptotic}, even under the simple viscosity with linear distribution in the random space, Eq.(\ref{eqn:sgbgk homogeneous solution}) ensures a cascade evolution mechanism to correlate all the gPC expansion orders nonlinearly.
Fig. \ref{pic:shock gpc} presents the gPC coefficients of particle distribution function at the central point of shock from first order.
From the first order gPC , the uncertainties have been delivered to higher order moments with descending magnitudes.
Counting the contributions from different orders via Eq.(\ref{eqn:gpc mean and variance}), we get the corresponding expectation and variance of particle distribution function.
Given the higher temperature, there is a wider distribution of particles along velocity space at $\rm Ma=3$. 
Similar as the profile in physical space, we see the contributions from either side of particle distribution function in velocity space.
Three local maximums emerge on the standard deviation profile, which correspond to most probable velocity, and its upstream and downstream.

\begin{figure}
    \subfloat[$\mathbb E(f)$]{%
        \includegraphics[width=.7\linewidth]{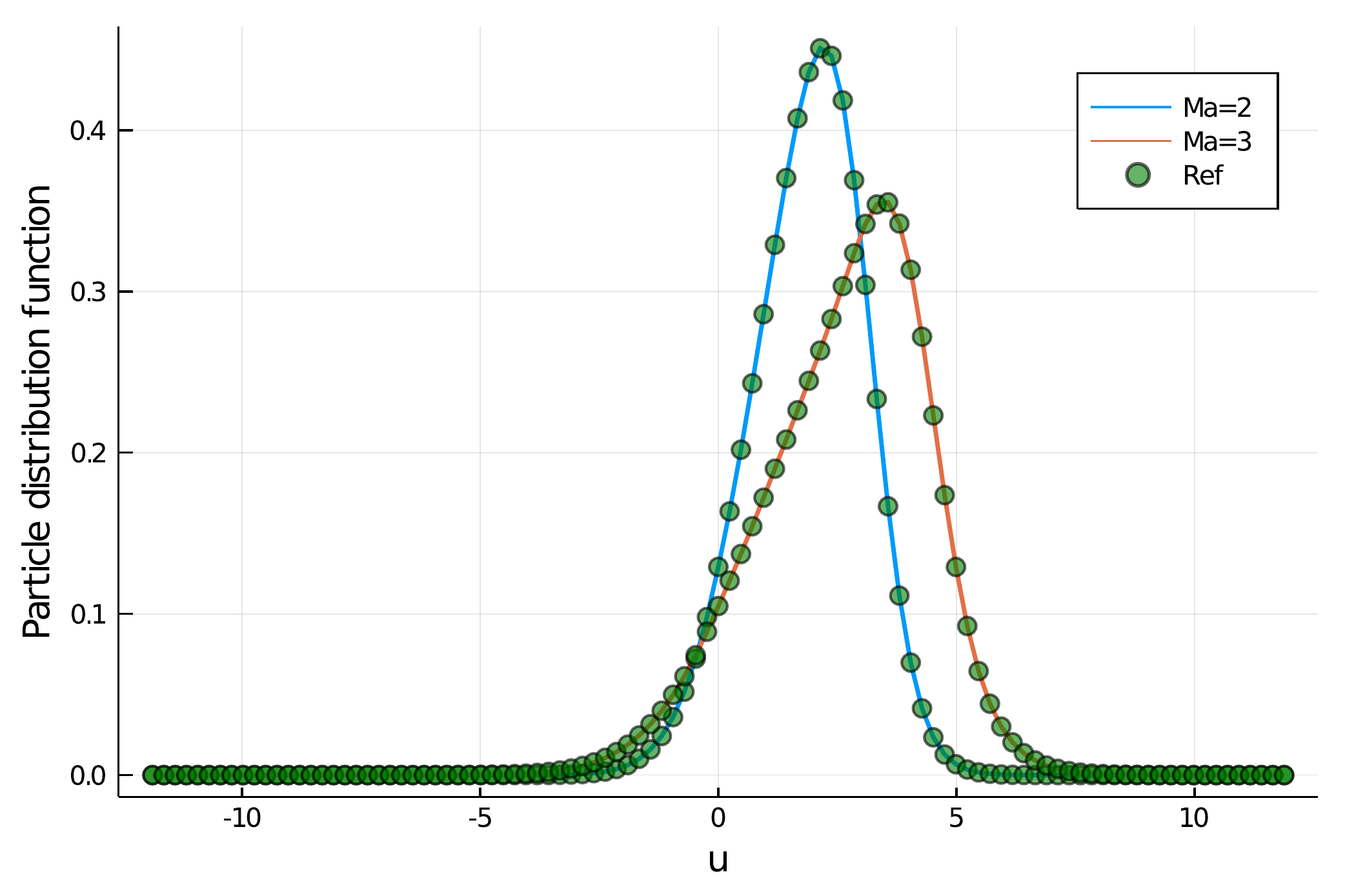}%
    }\hfill
    \subfloat[$\mathbb S(f)$]{%
        \includegraphics[width=.7\linewidth]{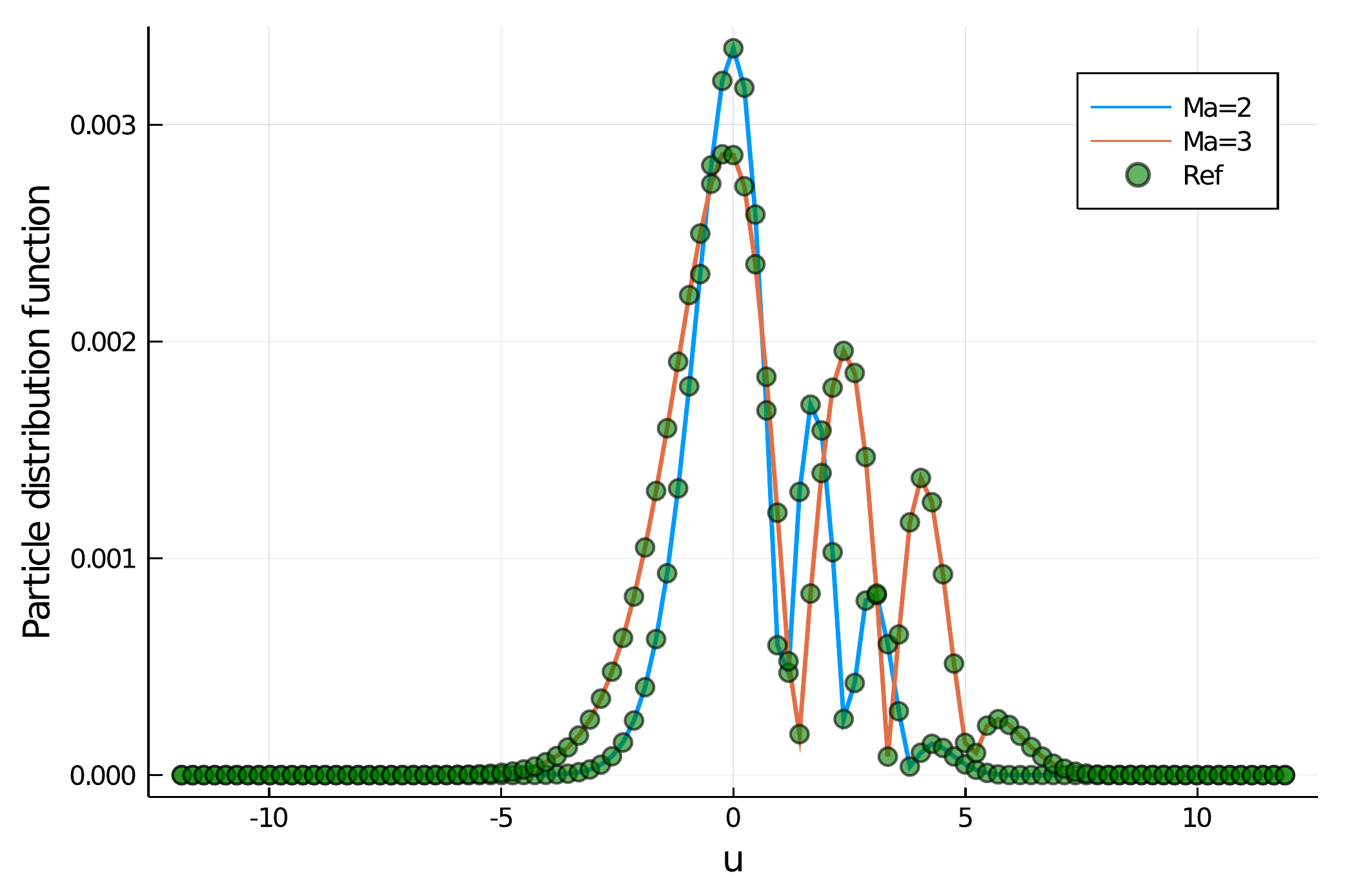}%
    }
\caption{Expectation value and standard deviation of particle distribution function at the center of normal shock structure.}
\label{pic:shock pdf}
\end{figure}

\subsection{Shear layer}

Now let us consider the flow problems in which the transverse processes dominates.
This case comes from the literature \cite{xiao2020velocity}, and we hereby rewrite it into dimensionless form.
A two-dimensional shear layer exists in the flow domain, with the initial condition
\begin{equation*}
\left[\begin{array}{c}
\rho \\
U \\
V \\
T \\
\end{array}\right]_{{L}}=\left[\begin{array}{c}
1 \\
0  \\
\xi \\
1 \\
\end{array}\right],
\end{equation*}
for the left half, and
\begin{equation*}
\left[\begin{array}{c}
\rho \\
U \\
V \\
T \\
\end{array}\right]_{{R}}=\left[\begin{array}{c}
1 \\
0  \\
-1 \\
0.5 \\
\end{array}\right],
\end{equation*}
for the right half.
The reference viscosity coefficient is evaluated in the same way as described in Sec.\ref{sec:shock structure}, and then used to determine the collision time with 
\begin{equation*}
    \tau_0=\frac{\mu_0}{p_0}.
\end{equation*}
The simulation is conducted within the time period $t\in[0,100 \tau_0]$.
The computational setup is detailed in Table.\ref{tab:layer}.

\begin{table}[htbp]
	\caption{Computational setup of shear layer.} 
	\centering
	\begin{tabular}{lllllll} 
		\toprule 
		$t$ & $x$ & $N_x$ & $u$ & $N_u$ & v \\ 
		\hline
		$[0,100\tau_0]$ & $[-1,1]$ & $1000$ & $[-4.5,4.5]$ & 32 & $[-4.5,4.5]$ \\ 
		\hline
		$N_v$ & Integral & $N$ & $N_q$ & Polynomial & $\xi$  \\ 
		\hline
		64 & rectangle & 5 & 9 & Legendre & $\mathcal U(0.9,1.1)$   \\ 
		\hline
		Kn & CFL & $\gamma$ & $\alpha$ & $\omega$ & $\eta$ &    \\ 
		\hline
		$0.005$ & $0.5$ & $1.67$ & $1$ & $0.5$ & $0.81$   \\ 
		\hline
	\end{tabular} 
	\label{tab:layer}
\end{table}

\begin{figure}
    \subfloat[Expected density and temperature]{%
        \includegraphics[width=.7\linewidth]{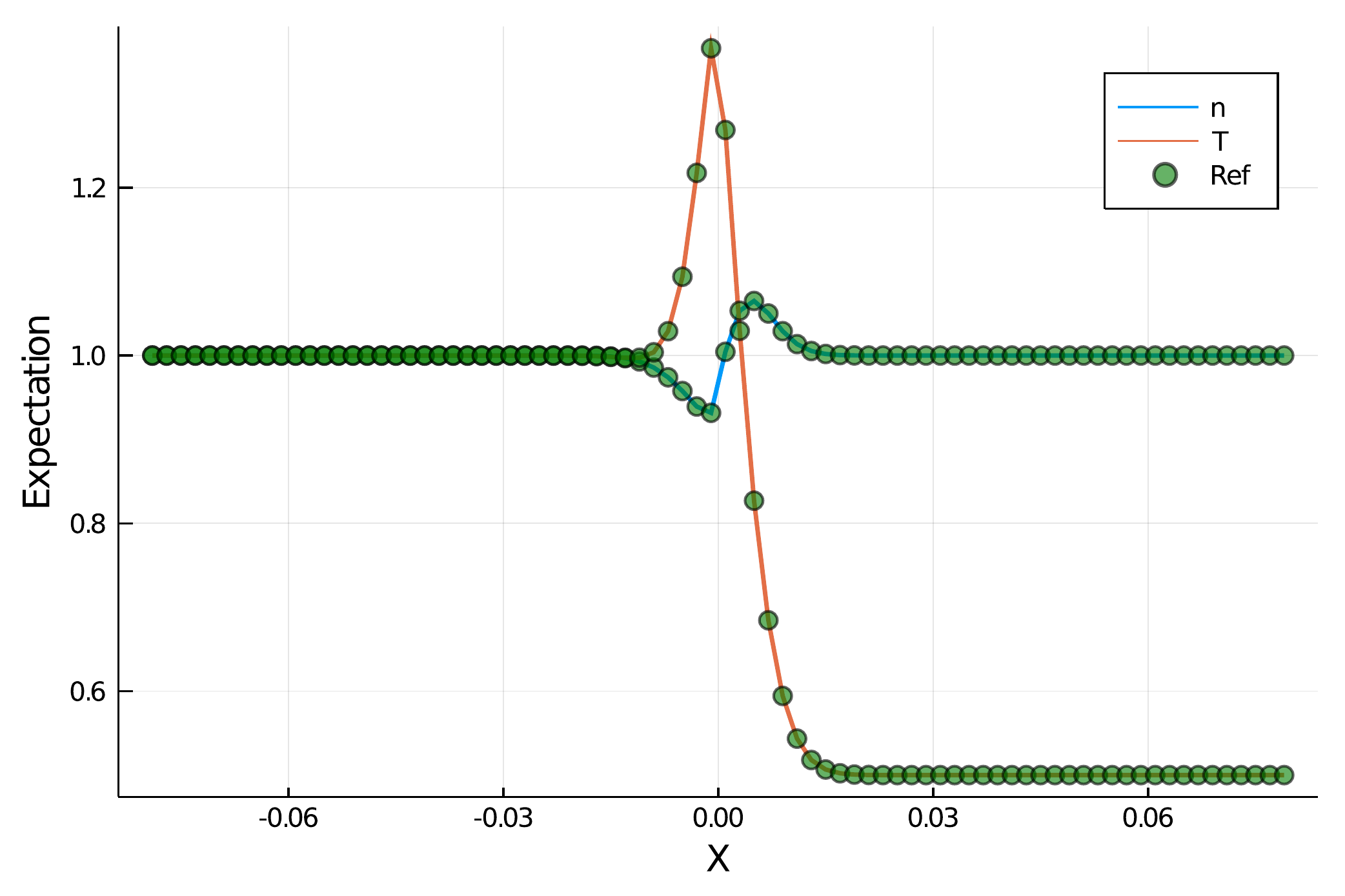}%
    }\hfill
    \subfloat[Expected velocity]{%
        \includegraphics[width=.7\linewidth]{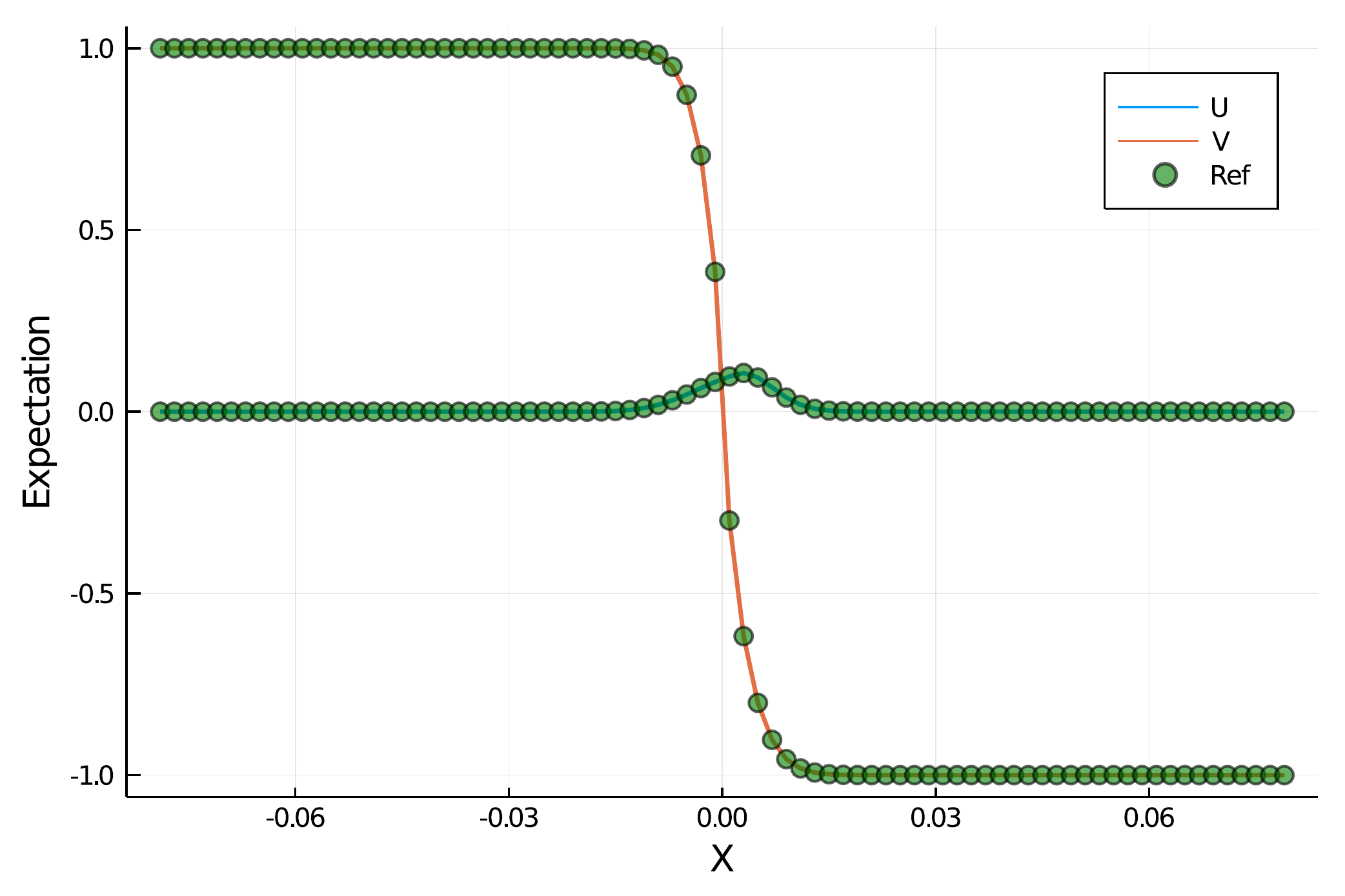}%
    }\hfill
    \subfloat[Standard deviation]{%
        \includegraphics[width=.7\linewidth]{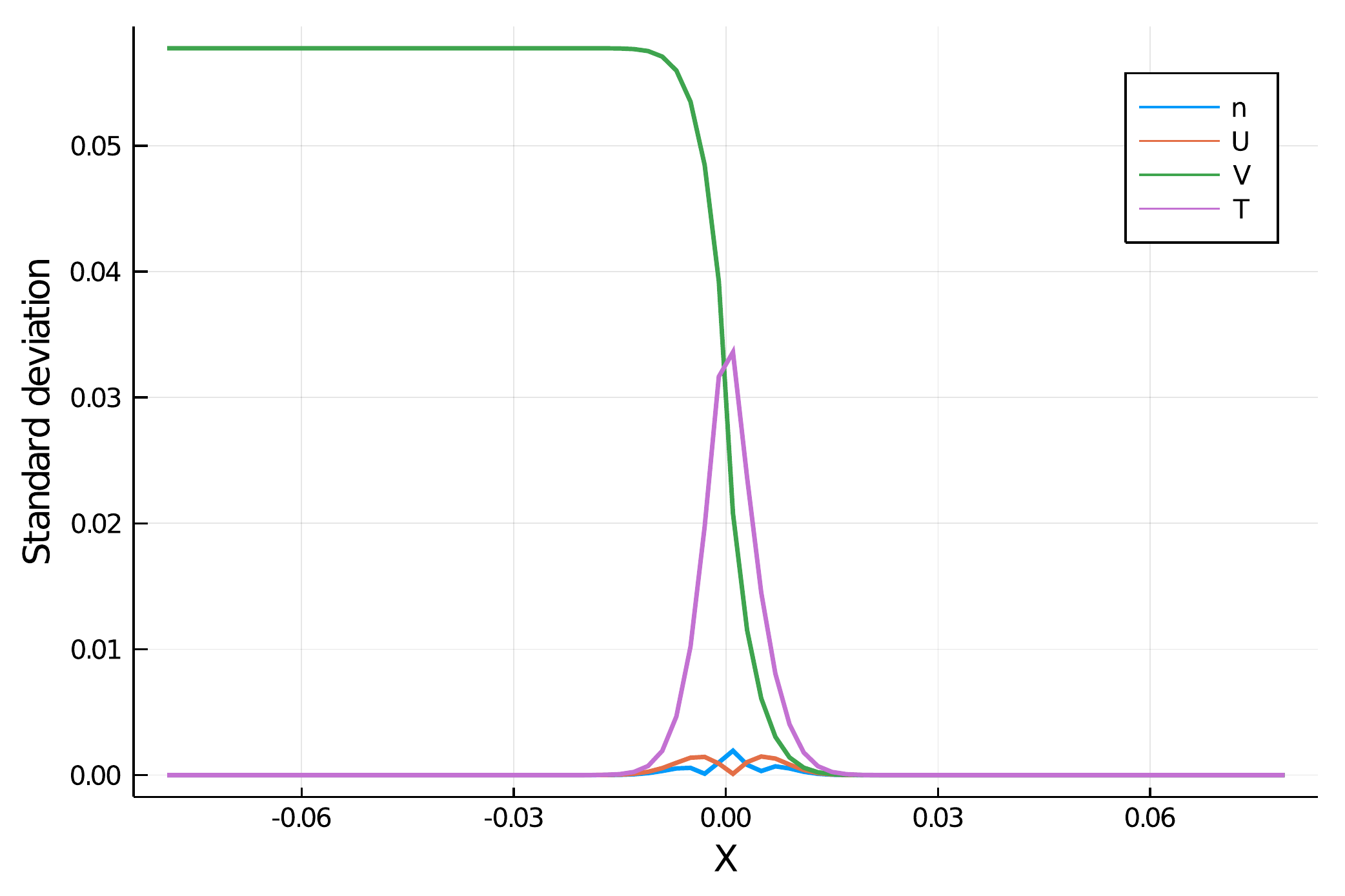}%
    }
\caption{Expectation value and standard deviation of macroscopic density, velocity and temperature at $t=\tau_0$ in the shear layer.}
\label{pic:layer t1}
\end{figure}

\begin{figure}
    \subfloat[Expected density and temperature]{%
        \includegraphics[width=.7\linewidth]{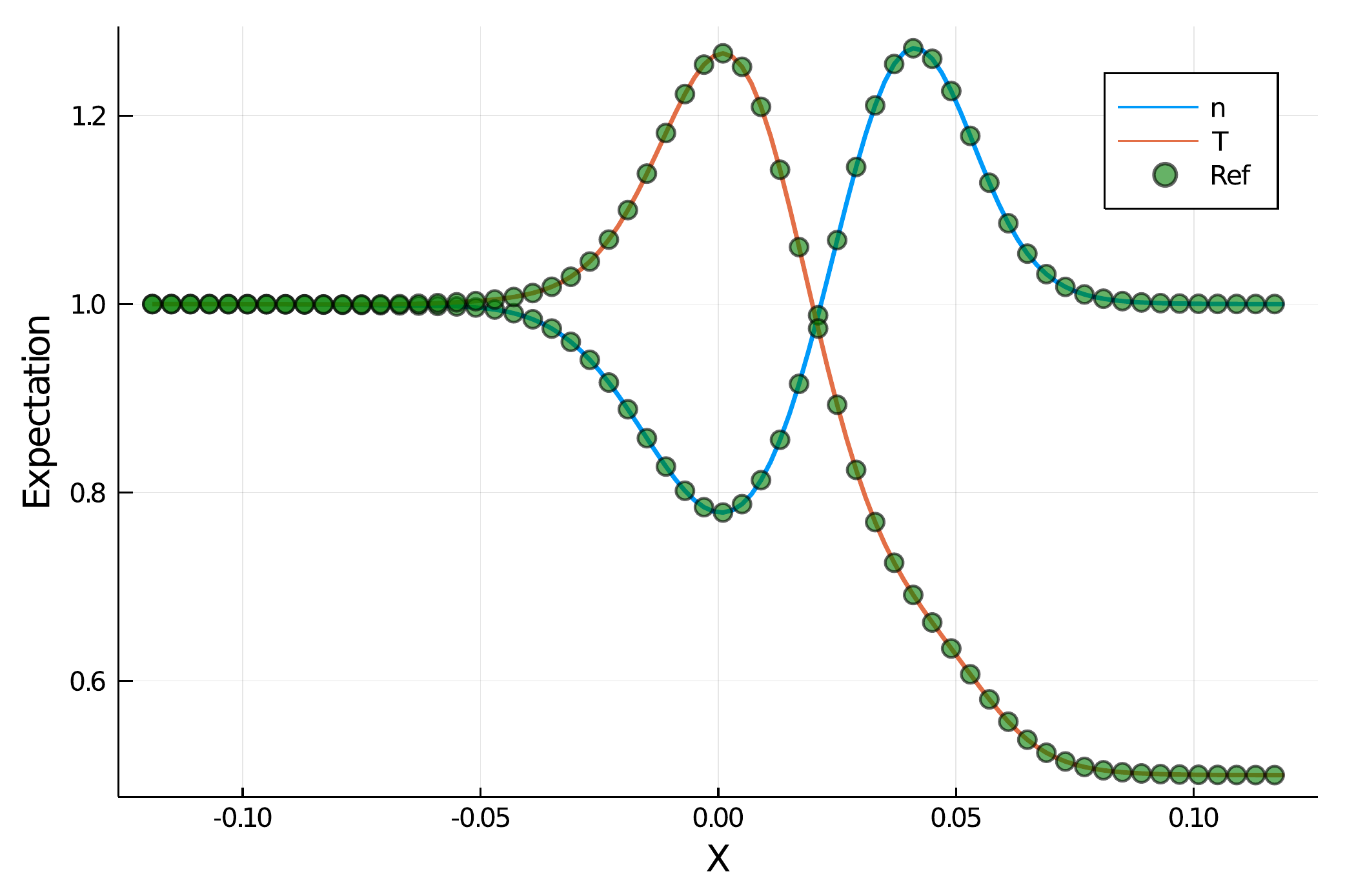}%
    }\hfill
    \subfloat[Expected velocity]{%
        \includegraphics[width=.7\linewidth]{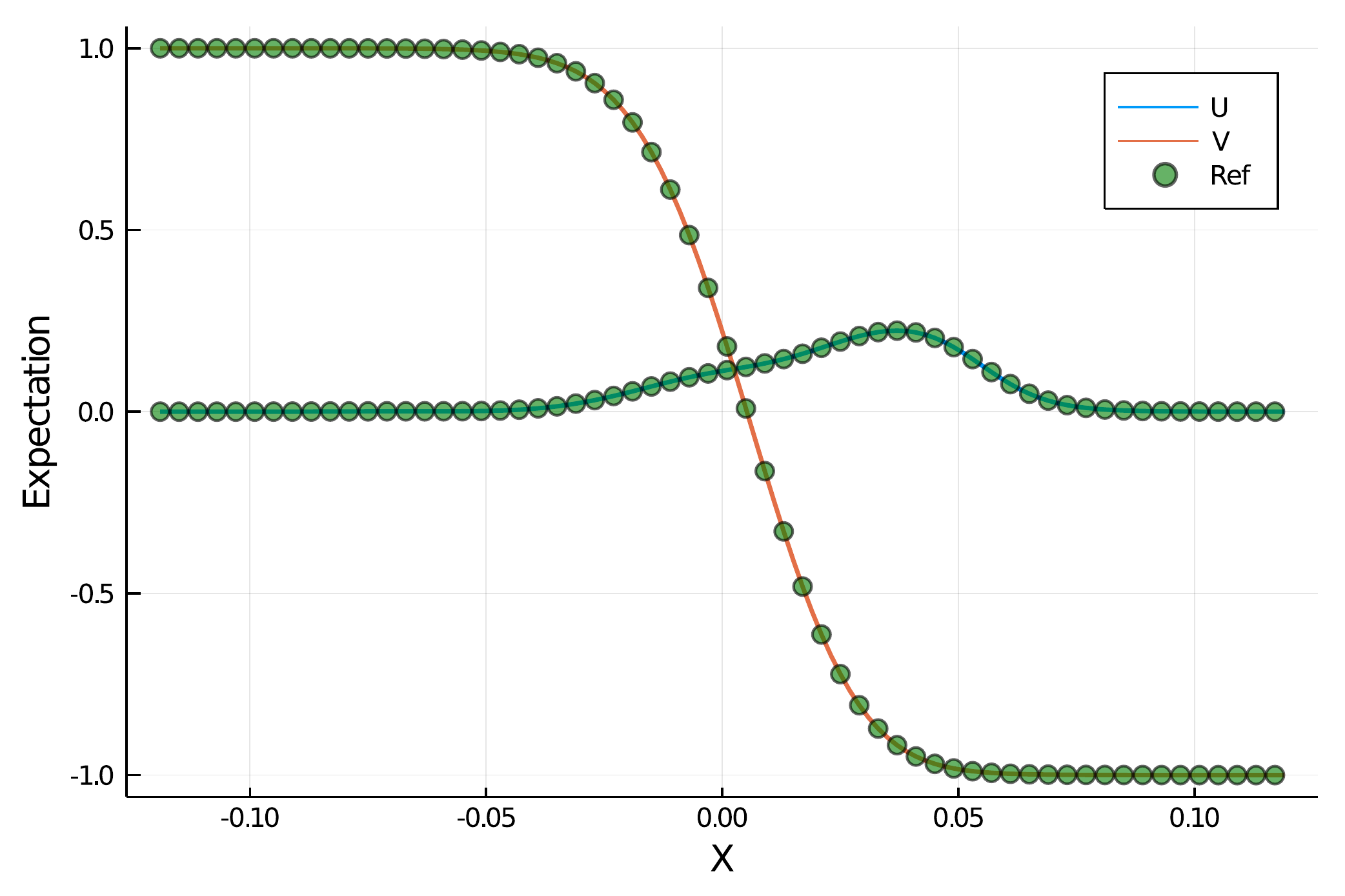}%
    }\hfill
    \subfloat[Standard deviation]{%
        \includegraphics[width=.7\linewidth]{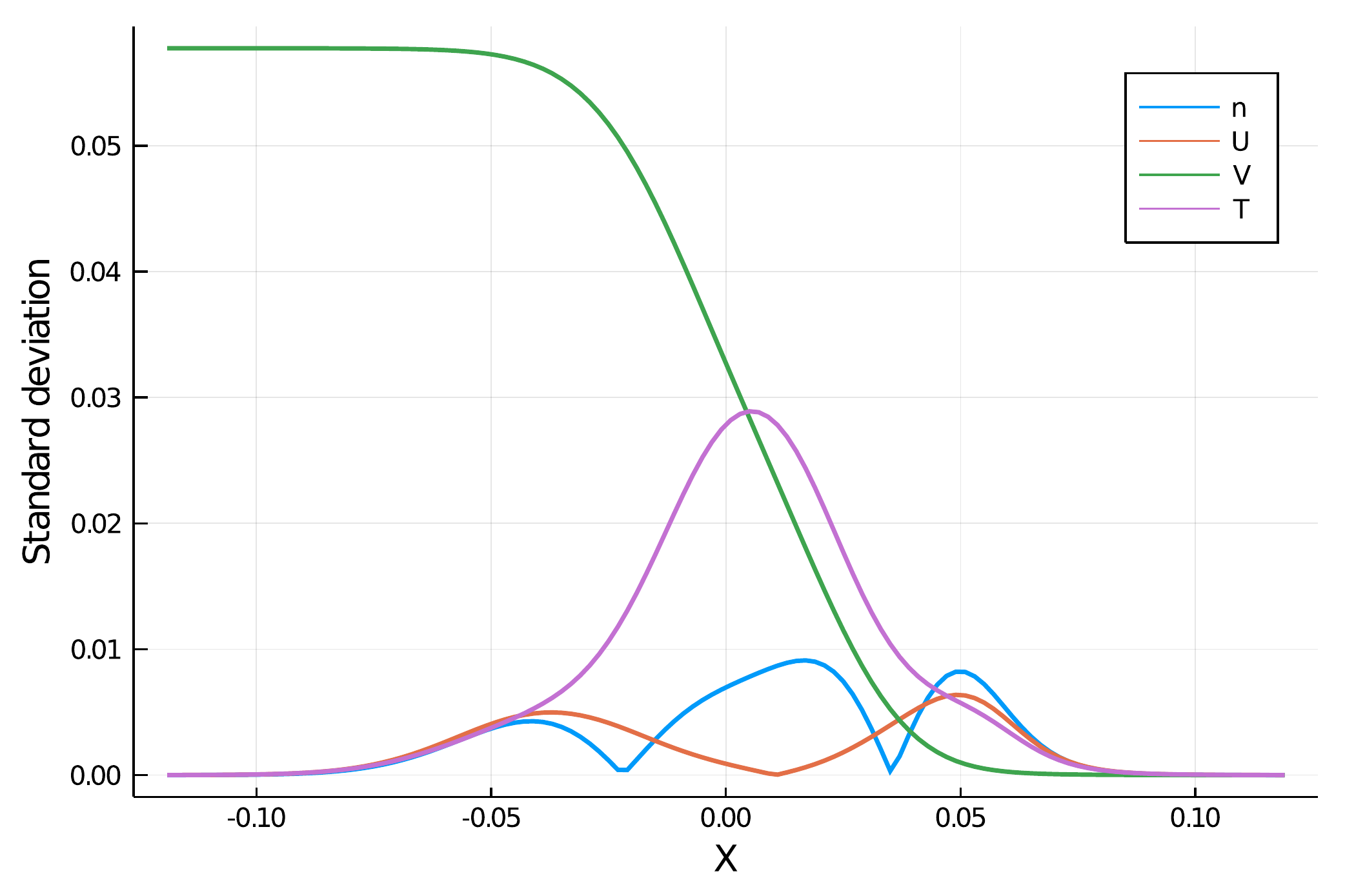}%
    }
\caption{Expectation value and standard deviation of macroscopic density, velocity and temperature at $t=10\tau_0$ in the shear layer.}
\label{pic:layer t2}
\end{figure}

\begin{figure}
    \subfloat[Expected density and temperature]{%
        \includegraphics[width=.7\linewidth]{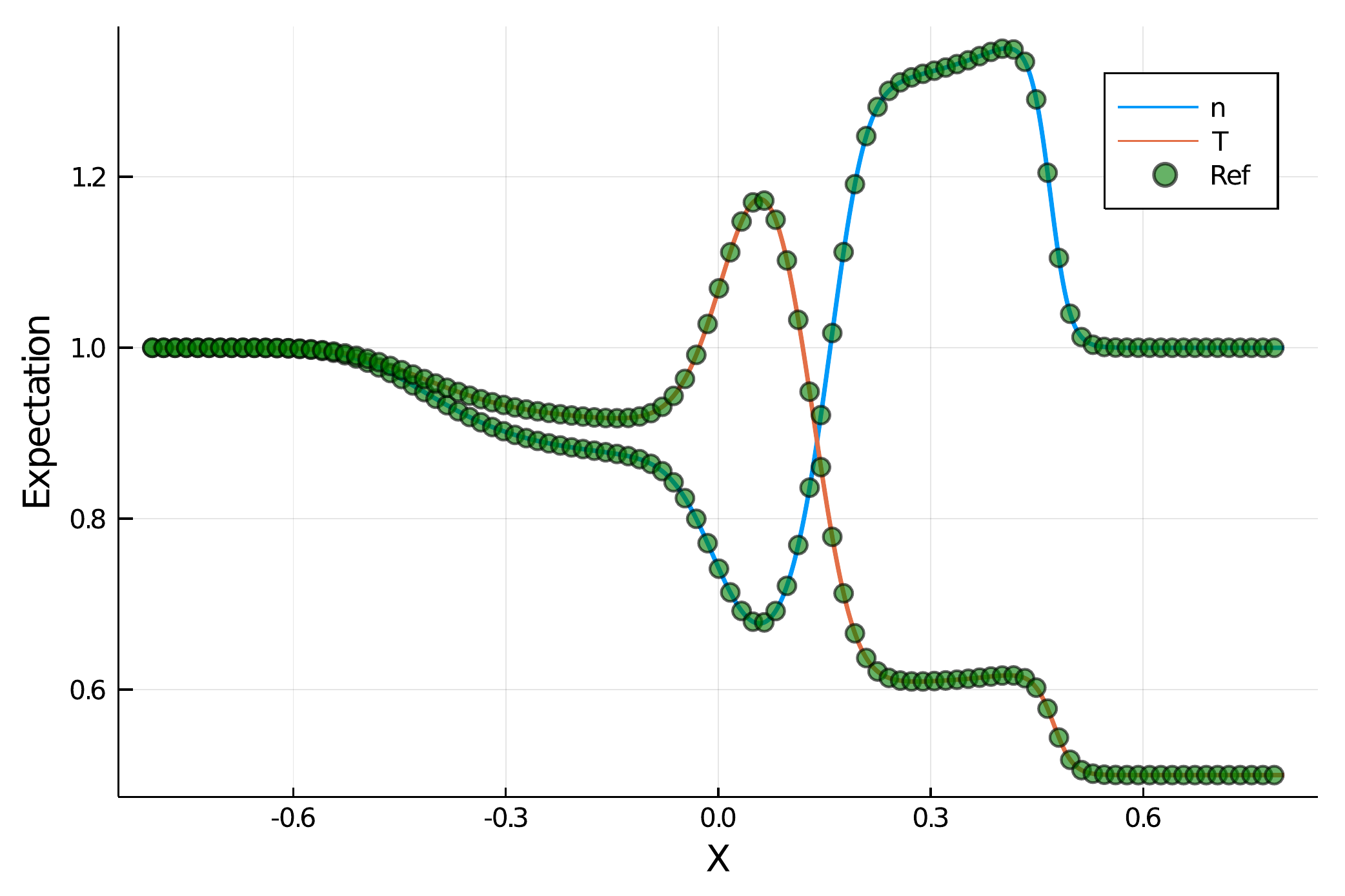}%
    }\hfill
    \subfloat[Expected velocity]{%
        \includegraphics[width=.7\linewidth]{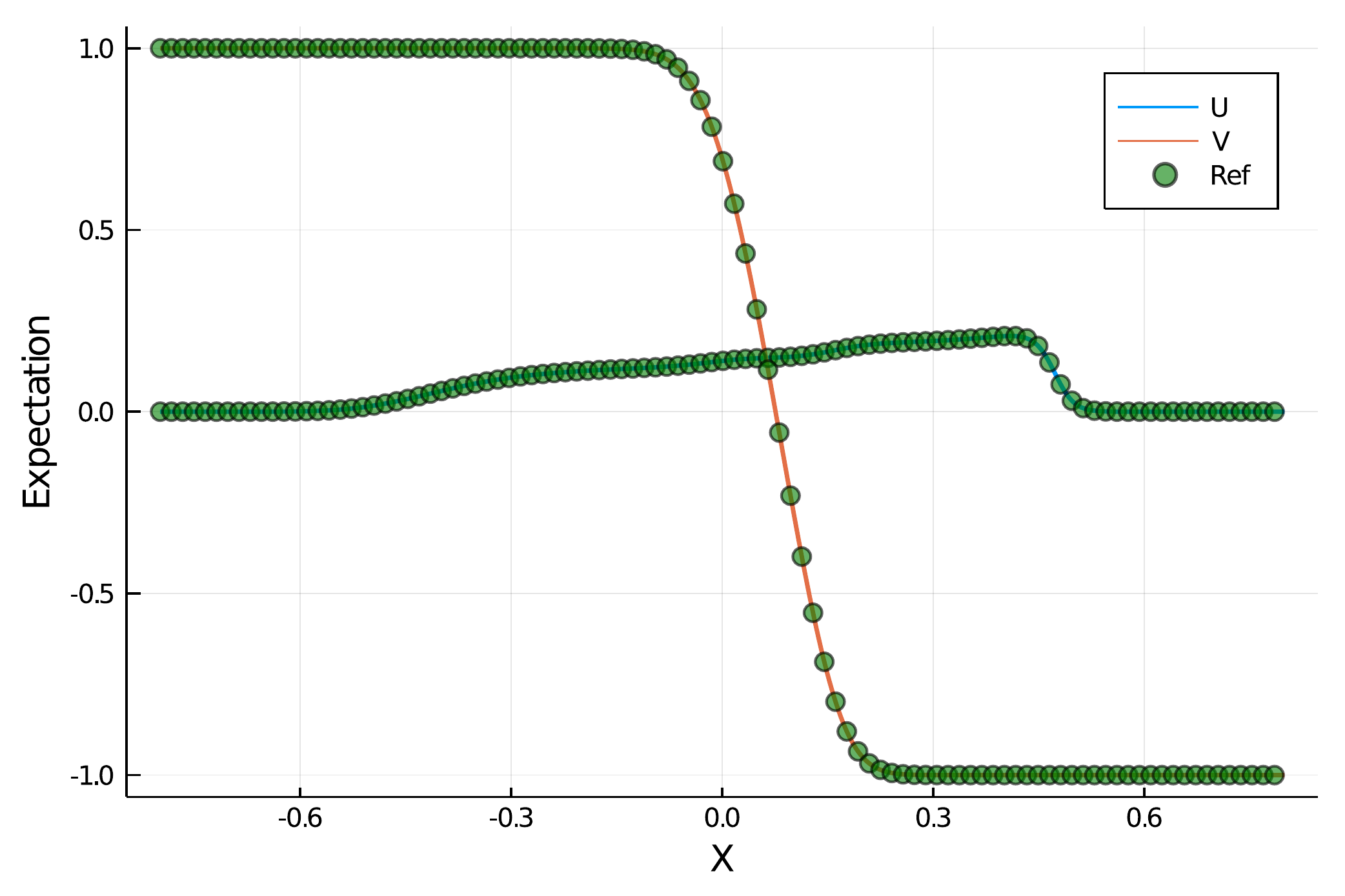}%
    }\hfill
    \subfloat[Standard deviation]{%
        \includegraphics[width=.7\linewidth]{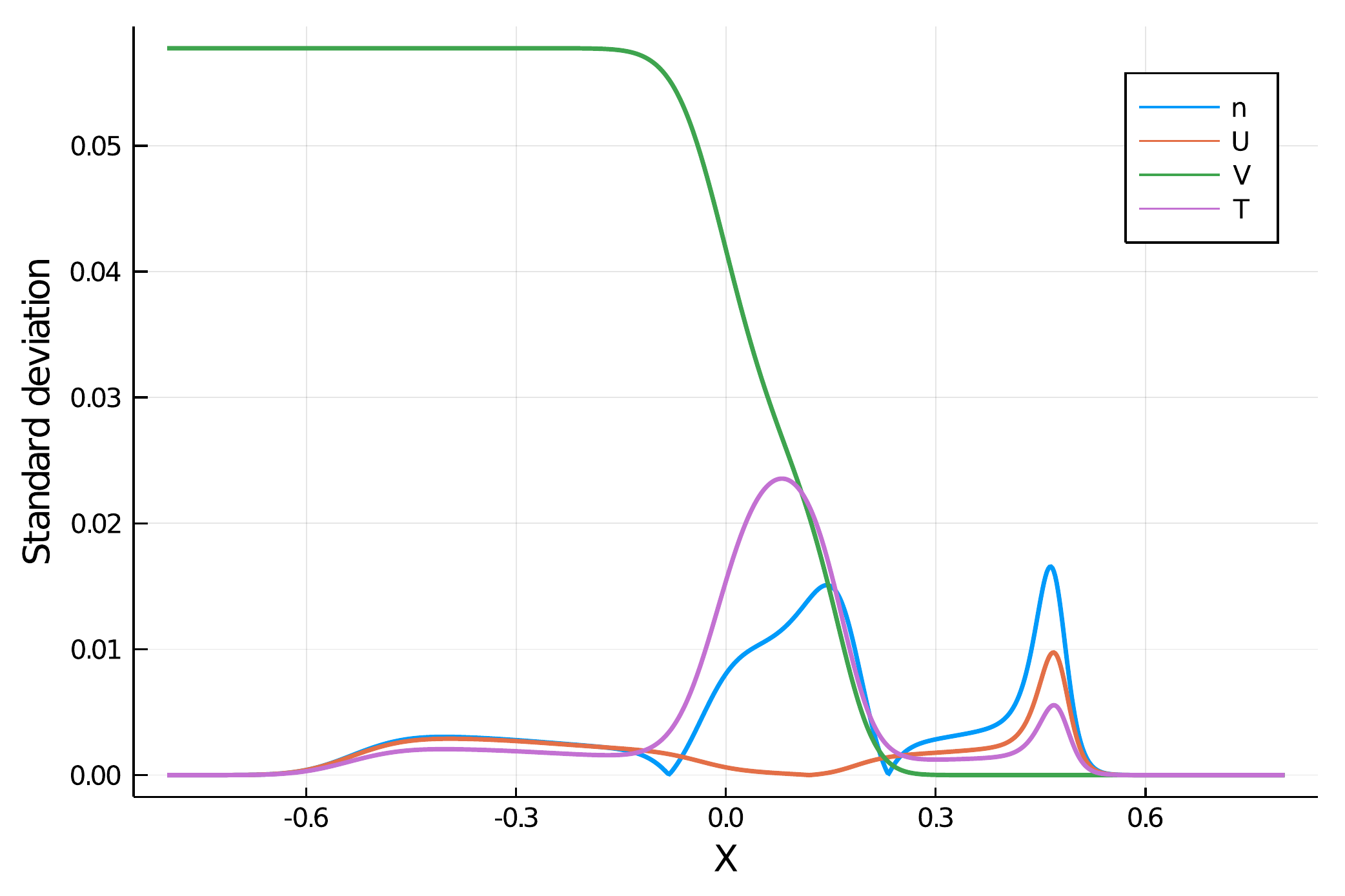}%
    }
\caption{Expectation value and standard deviation of macroscopic density, velocity and temperature at $t=100\tau_0$ in the shear layer.}
\label{pic:layer t3}
\end{figure}

Fig.\ref{pic:layer t1}, \ref{pic:layer t2} and \ref{pic:layer t3} show the macroscopic flow variables at $t=\tau_0$, $10\tau_0$ and $100\tau_0$.
Given the pressure difference, a positive $U$-velocity emerges around the initial interface, and the heat is transferred from left to right.
Each of the density and temperature presents a non-monotonic profile at the center of flow domain.
As a result, a transition layer is formed that links the left and right status of the flow field.
As time evolves, the shear layer expands gradually, with the $V$-velocity diffused inside it.

The stochastic simulation provides us the chance to study the propagation of uncertainties along with the bulk flow.
Generally speaking from Fig.\ref{pic:layer t1}(c), \ref{pic:layer t2}(c) and \ref{pic:layer t3}(C), the uncertainties travel along with the main flow structure of expectation values and present similar propagating patterns.
At different time instants, one can find the one-to-one correspondence between the mean flow organizations with their local maximums of variance.
Compared with density and velocity, the temperature distribution is related to the second-order moments of particle distribution function and indicates a higher sensitivity with respect to randomness.

Besides, it seems the magnitude of variances is positively associated with the gradients of flow variables inside the flow domain.
The underlying principle can be identified with the help of stochastic Galerkin equations.
For brevity, we take the first-order truncation of the Boltzmann moments system, i.e. the Navier-Stokes, as an example to illustrate the contribution of spatial distribution of flow variables onto stochastic evolution.
Let us write down the second and the energy equation in the Navier-Stokes system,
\begin{equation*}
    \frac{\partial (\rho E)}{\partial t} + \nabla_\mathbf x \cdot 
    \left({\mathbf U} {\rho} E\right)
    = -\nabla_\mathbf x \cdot
    \left(\mathbf P \cdot \mathbf U\right) - \nabla_\mathbf x \cdot \mathbf q,
\end{equation*}
and project it into one-dimensional stochastic Galerkin equation, which yields, 
\begin{equation*}
\begin{aligned}
    &\frac{\partial (\hat{\rho E})_i}{\partial t} + \frac{\partial}{\partial x} \frac{\sum_j \sum_k (\hat{\rho E})_j \hat{U}_k \langle \Phi_j \Phi_k, \Phi_i\rangle}{\gamma_i} \\
    &=- \frac{\partial}{\partial x} \frac{\sum_j \sum_k \left((\hat P_{xx})_j \hat{U}_k + (\hat P_{xy})_j \hat{V}_k \right)
    \langle \Phi_j \Phi_k, \Phi_i\rangle}{\gamma_i}
    -\frac{\partial (\hat{q}_x)_i}{\partial x}. 
\end{aligned}
\end{equation*}
We consider the initial status of the shear layer, i.e. the gas is still in $x$ direction and only $V$-velocity possesses non-zero first order gPC coefficient $\hat V_1$.
Therefore, the above equation reduces to
\begin{equation*}
    \frac{\partial (\hat{\rho E})_i}{\partial t} =- \frac{\partial}{\partial x} \frac{\sum_j \sum_k \left((\hat P_{xy})_j \hat{V}_k \right)
    \langle \Phi_j \Phi_k, \Phi_i\rangle}{\gamma_i}
    -\frac{(\partial \hat{q}_x)_i}{\partial x}. 
\end{equation*}
where the $j$-th gPC component of stress $P_{xy}$ is 
\begin{equation*}
    (\hat P_{xy})_j=\frac{\int \sum_p \sum_q u(v - \hat V_p) \hat f_q \langle \Phi_p \Phi_q, \Phi_j \rangle dudv}{\gamma_j}.
\end{equation*}
Therefore, the discontinuous distribution of first order gPC coefficients for $V$-velocity results in a nonlinear increase at the same order of energy, which turns the initial deterministic temperature into stochastic one.
It also explains the sensitivity of temperature with respect to randomness since this nonlinear correlation is absent in mass and momentum equations.

\begin{figure}
    \subfloat[Expectation value]{%
        \includegraphics[width=.7\linewidth]{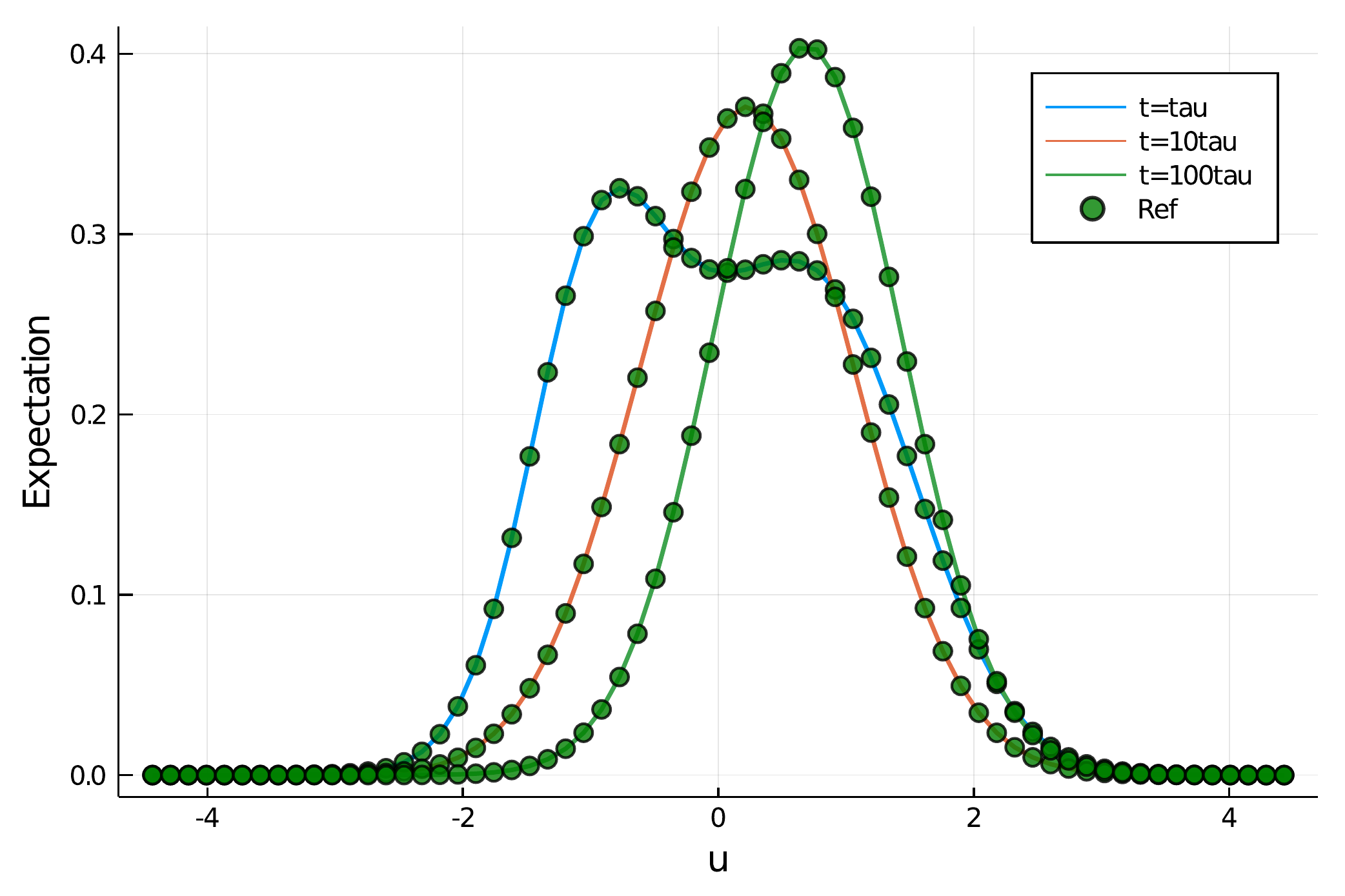}%
    }\hfill
    \subfloat[Standard deviation]{%
        \includegraphics[width=.7\linewidth]{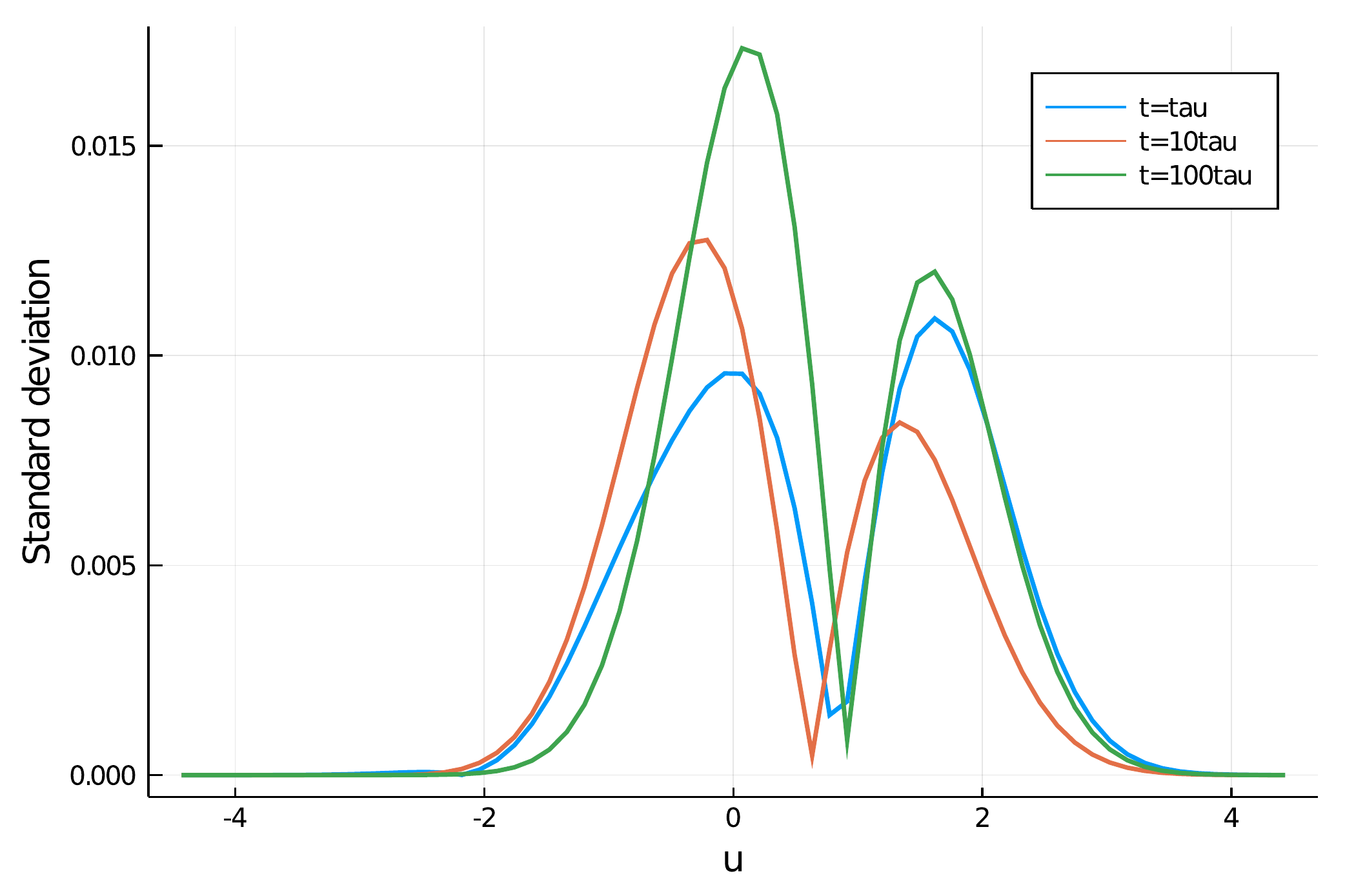}%
    }
\caption{Expectation value and standard deviation of particle distribution function at the domain center $x=0$.}
\label{pic:layer pdf}
\end{figure}

The particle distribution function along $v$ velocity at the domain center $x=0$ is shown in Fig.\ref{pic:layer pdf}.
Over time, the particle distribution function evolves from initial non-equilibrium bimodal distribution towards Maxwellian, resulting in moderate profiles of conservative variables.
Similar as macroscopic variables, the clear upstream and downstream effects can be observed in the standard deviations of particle distribution function, where each contributes a major source for randomness.
During the gas evolutionary process, the magnitudes of variances are persistently amplified.

\subsection{Lid-driven cavity}

The lid-driven cavity is a complex system under the synergy of boundary effect, shearing process, swirling flow and heat transfer.
In the simulation, the gas is enclosed by four solid walls, while the upper wall is moving in the transverse direction with $\{U_w=\xi$, $V_w=0\}$.
The Maxwell boundary condition is adopted for all the walls, and thus the boundary distribution function for flux evaluation is constructed as follows,
\begin{equation}
    f_w = \mathcal M_w H(\mathbf u \cdot \mathbf n) + f_{in} \left(1-H(\mathbf u \cdot \mathbf n)\right),
    \label{eqn:maxwell boundary}
\end{equation}
where $\mathcal M_w$ is the Maxwellian at solid wall, $f_{in}$ is the distribution function extrapolated from inner flow field, $\mathbf n$ is the unit direction vector of boundary, and $H(x)$ is the heaviside step function.
The detailed computational setup is listed in Table.\ref{tab:cavity}.

\begin{table}[htbp]
	\caption{Computational setup of lid-driven cavity.} 
	\centering
	\begin{tabular}{lllllll} 
		\toprule 
		$x$ & $y$ & $N_x$ & $N_y$ & $u$ & $N_u$ \\ 
		\hline
		$[0,1]$ & $[0,1]$ & $45$ & $45$ & $[-5,5]$ & 60 \\ 
		\hline
		$v$ & $N_v$ & Integral & $N$ & $N_q$ & Polynomial &  \\ 
		\hline
		$[-5,5]$ & 60 & rectangle & 4 & 7 & Legendre &   \\ 
		\hline
		$\xi$ & Kn & CFL & $\gamma$ & $\alpha$ & $\omega$ &    \\ 
		\hline
		$\mathcal U(0.9,1.1)$ & $[0.001,0.1]$ & $0.8$ & $1.67$ & $1$ & $0.5$ &    \\ 
		\hline
		$\eta$ & Boundary &     \\ 
		\hline
		$0.72$ & Maxwell &    \\ 
		\hline
	\end{tabular} 
	\label{tab:cavity}
\end{table}

\begin{figure}
    \subfloat[$\rm Kn=0.001$]{%
        \includegraphics[width=.33\linewidth]{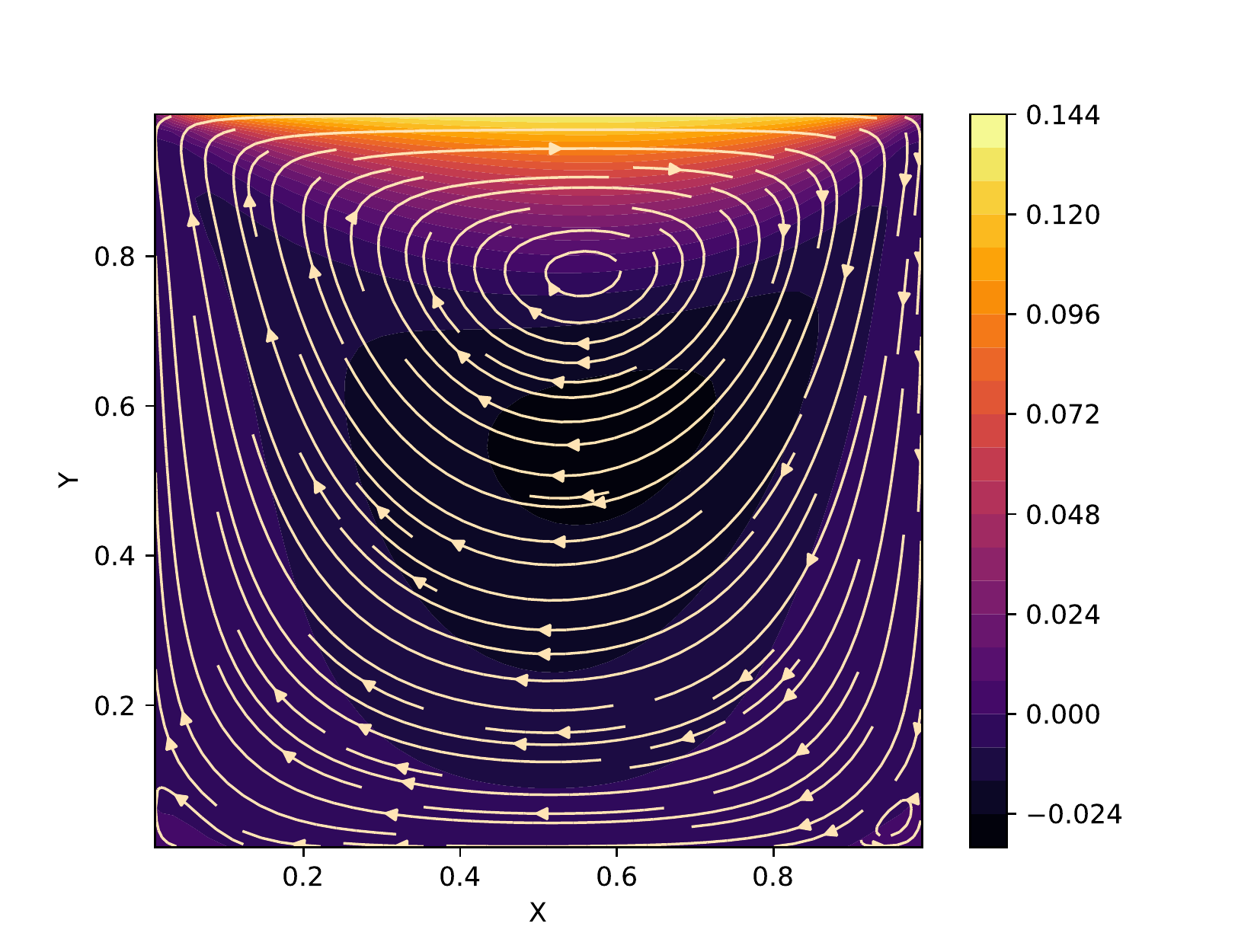}%
    }\hfill
    \subfloat[$\rm Kn=0.075$]{%
        \includegraphics[width=.33\linewidth]{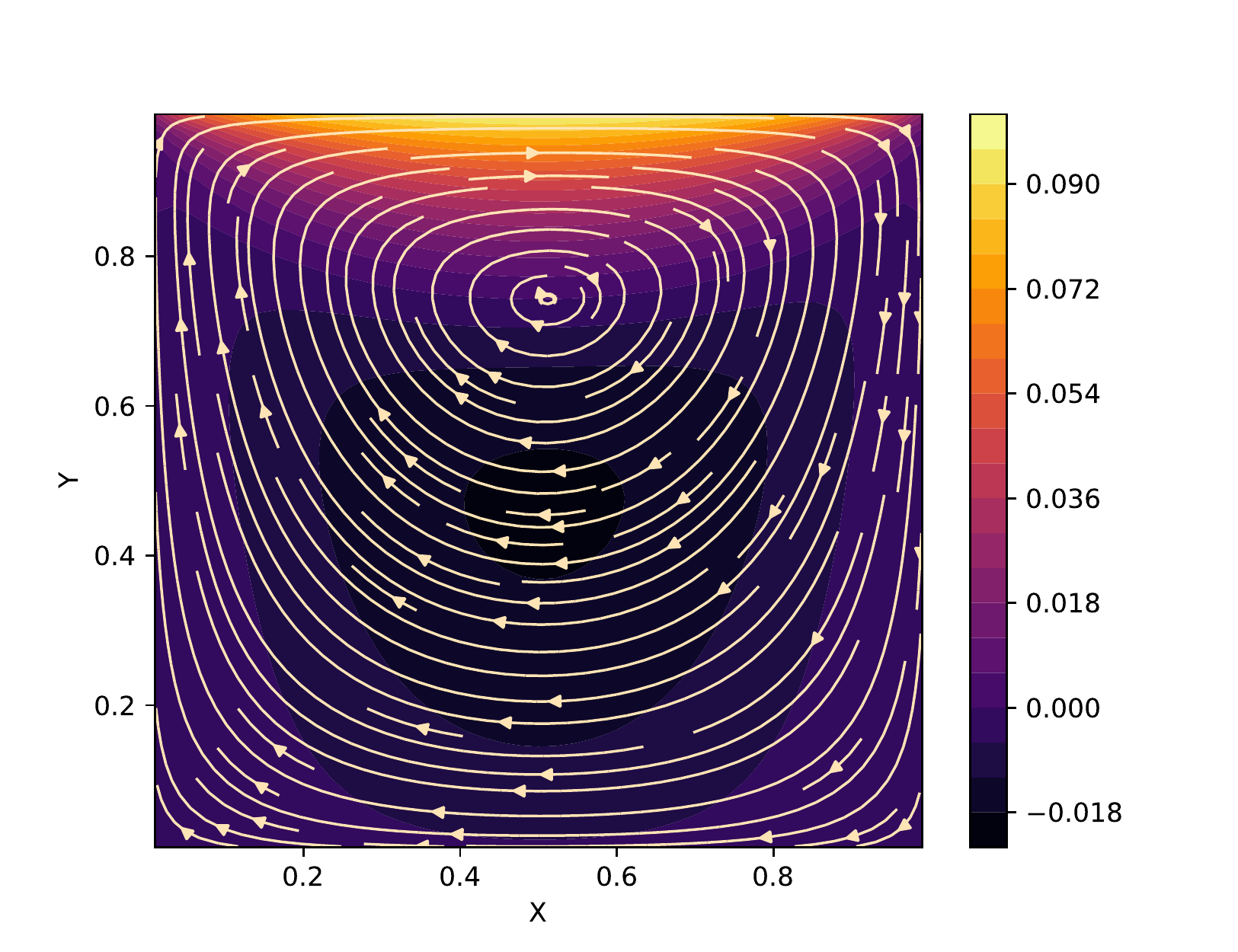}%
    }\hfill
    \subfloat[$\rm Kn=0.5$]{%
        \includegraphics[width=.33\linewidth]{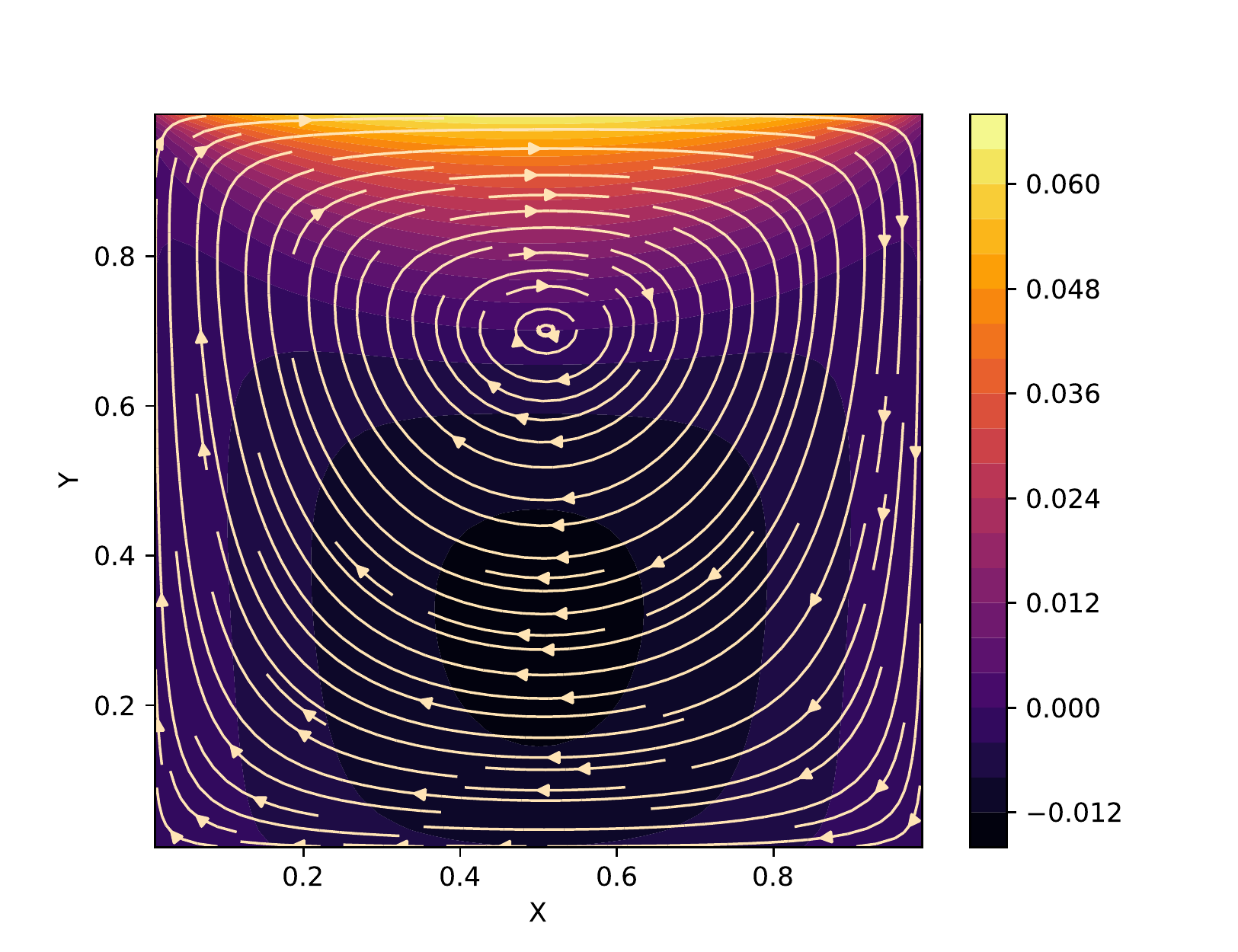}%
    }\hfill
    \subfloat[$\rm Kn=0.001$]{%
        \includegraphics[width=.33\linewidth]{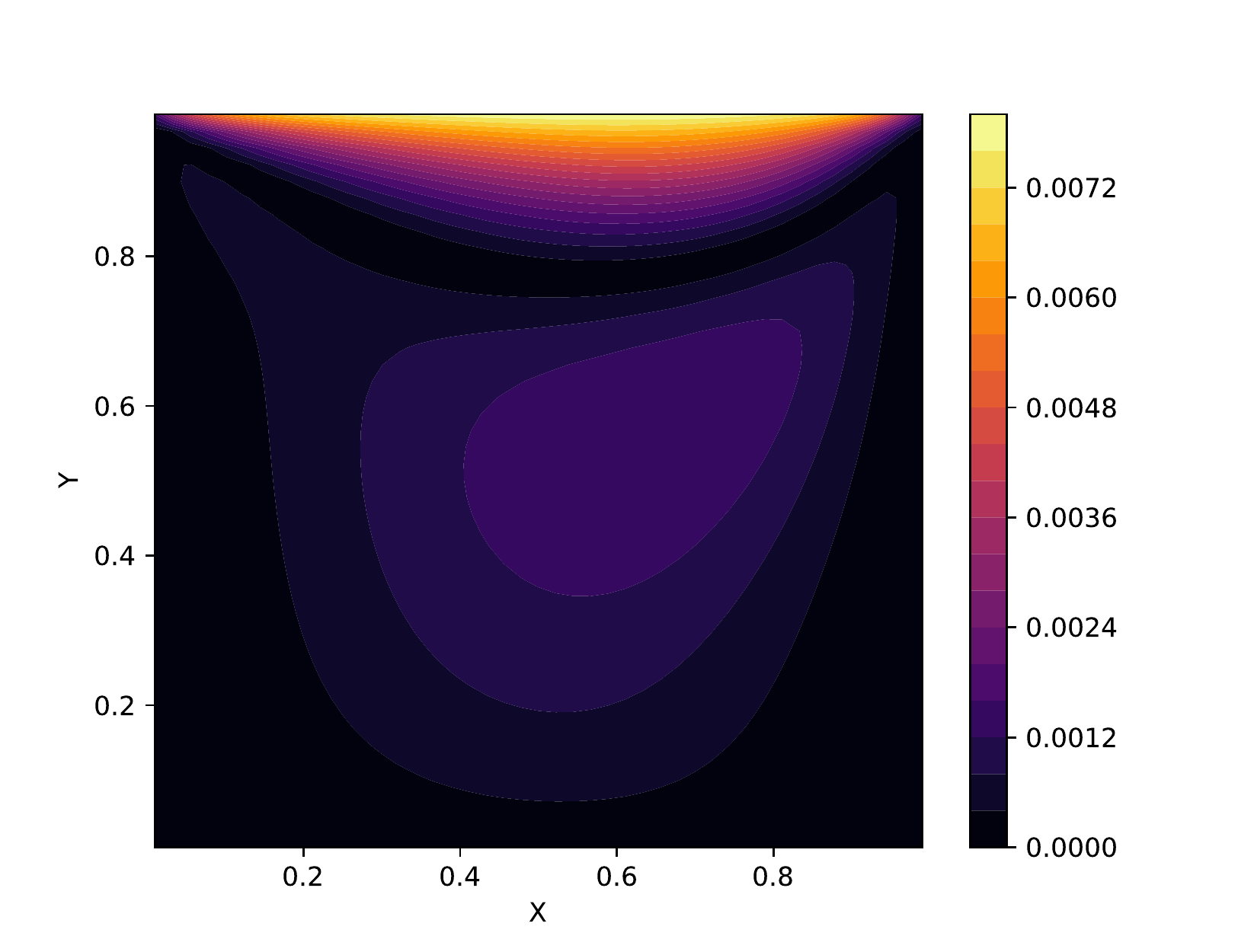}%
    }\hfill
    \subfloat[$\rm Kn=0.075$]{%
        \includegraphics[width=.33\linewidth]{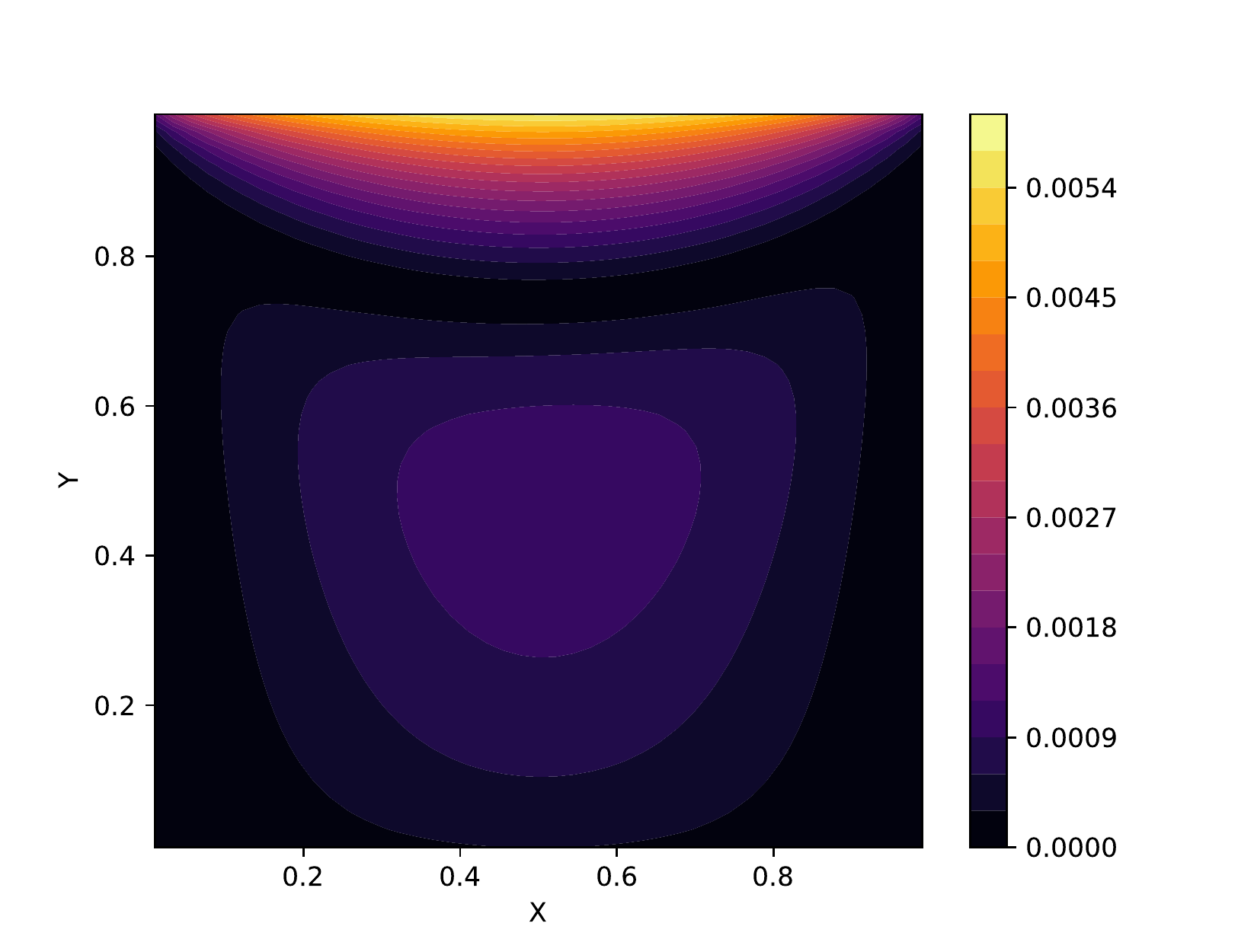}%
    }\hfill
    \subfloat[$\rm Kn=0.5$]{%
        \includegraphics[width=.33\linewidth]{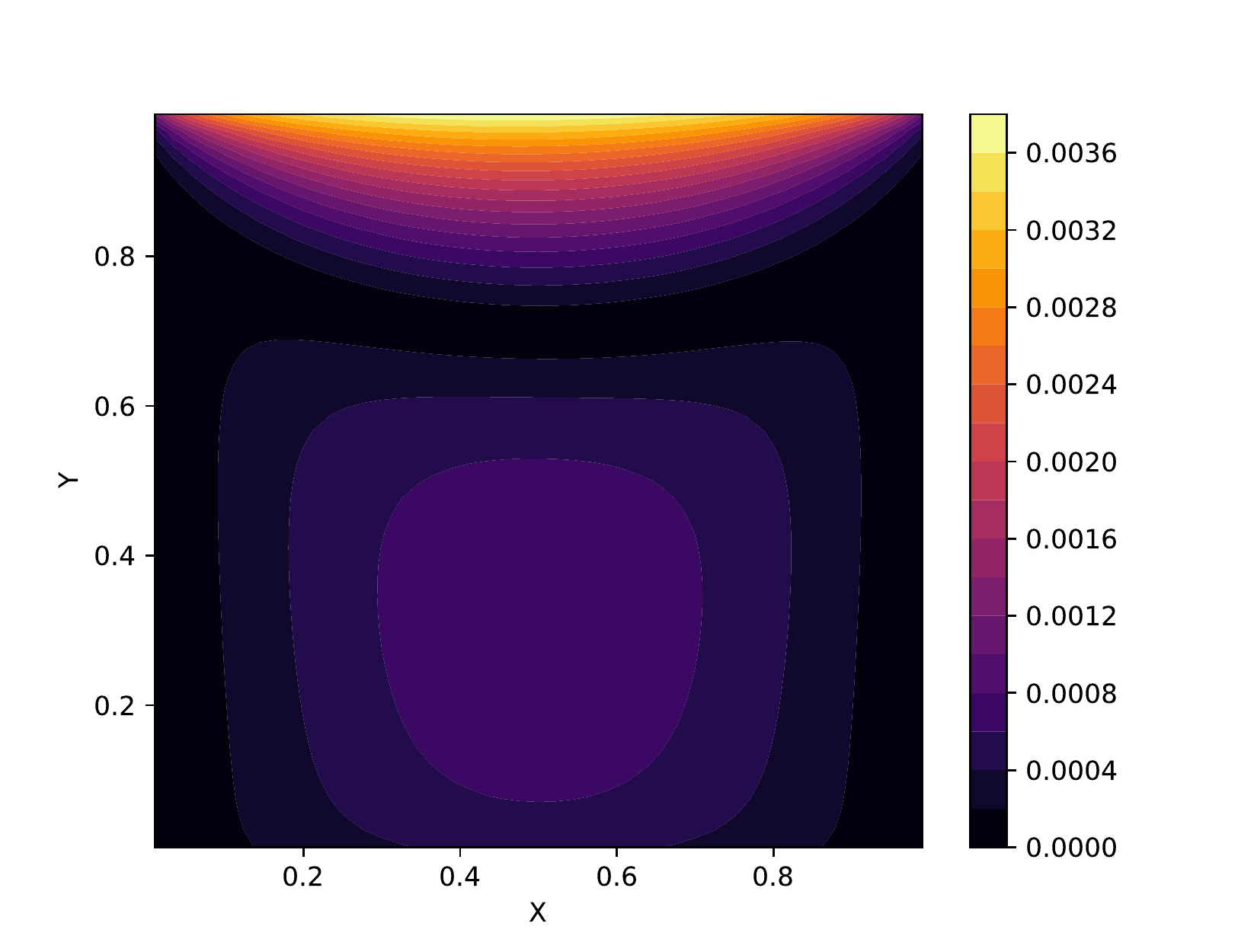}%
    }
\caption{Expectation values (first row) and standard deviations (second row) of $U$-velocity (contour) and streamline (vector) at different reference Knudsen numbers in the lid-driven cavity.}
\label{pic:cavity velocity}
\end{figure}

In this case, the movement of upper surface is the source for the inner fluid motion.
The non-equilibrium shearing transfers vorticity downwards and helps forming the eddies inside the cavity.
Fig.\ref{pic:cavity velocity} presents expectation values and standard deviations of $U$-velocity contours along with velocity vectors.
As is shown, a steady main vortex is formed in all cases with different Knudsen numbers in the reference state.
At $\rm Kn=0.001$ two small corner vortices exist along with the main eddy, while then disappears as $\rm Kn$ increases.
Th results here are consistent with the solutions shown in the literature \cite{xiao2018investigation}.

\begin{figure}
    \subfloat[$\rm Kn=0.001$]{%
        \includegraphics[width=.33\linewidth]{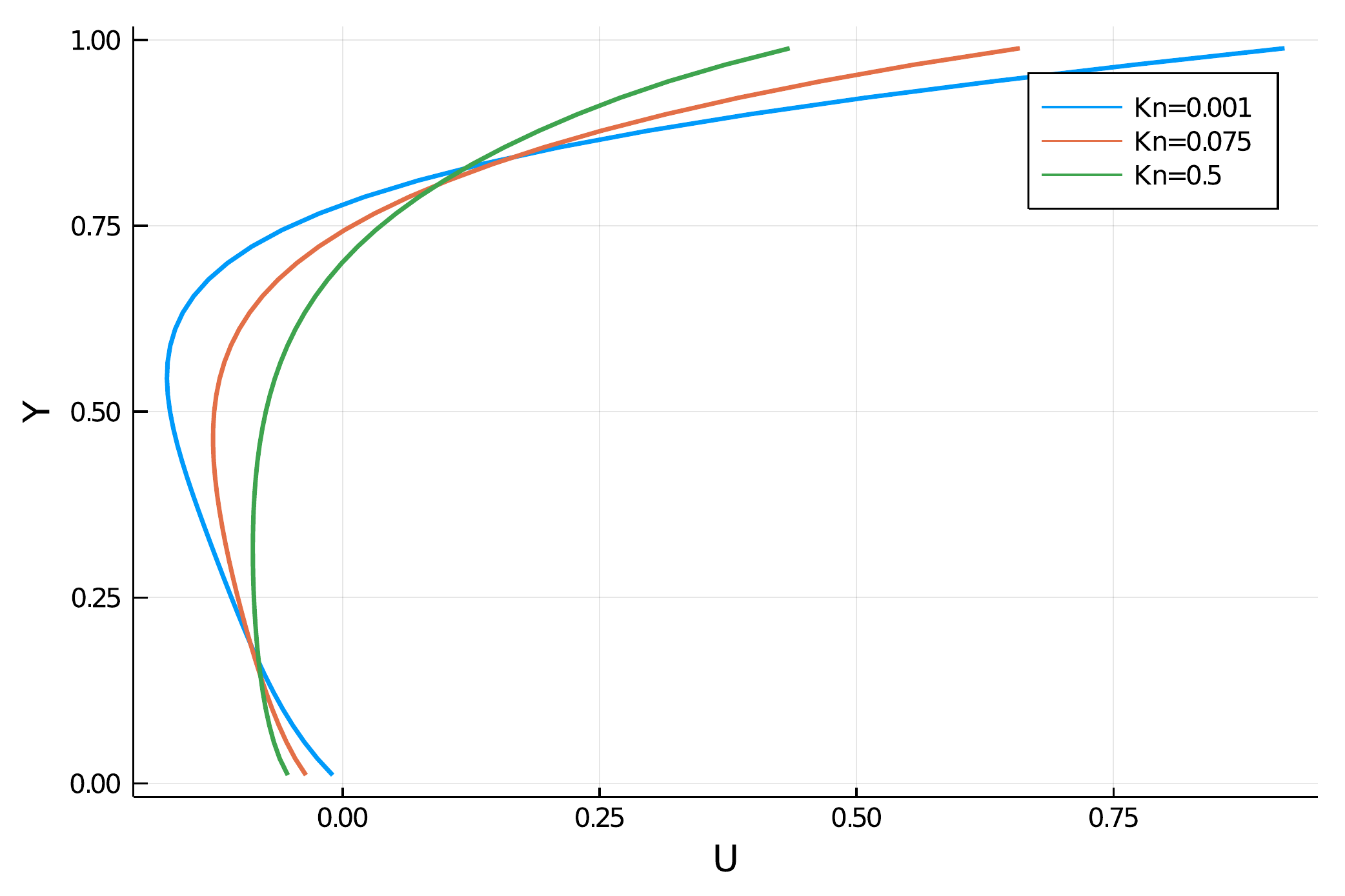}%
    }
    \subfloat[$\rm Kn=0.075$]{%
        \includegraphics[width=.33\linewidth]{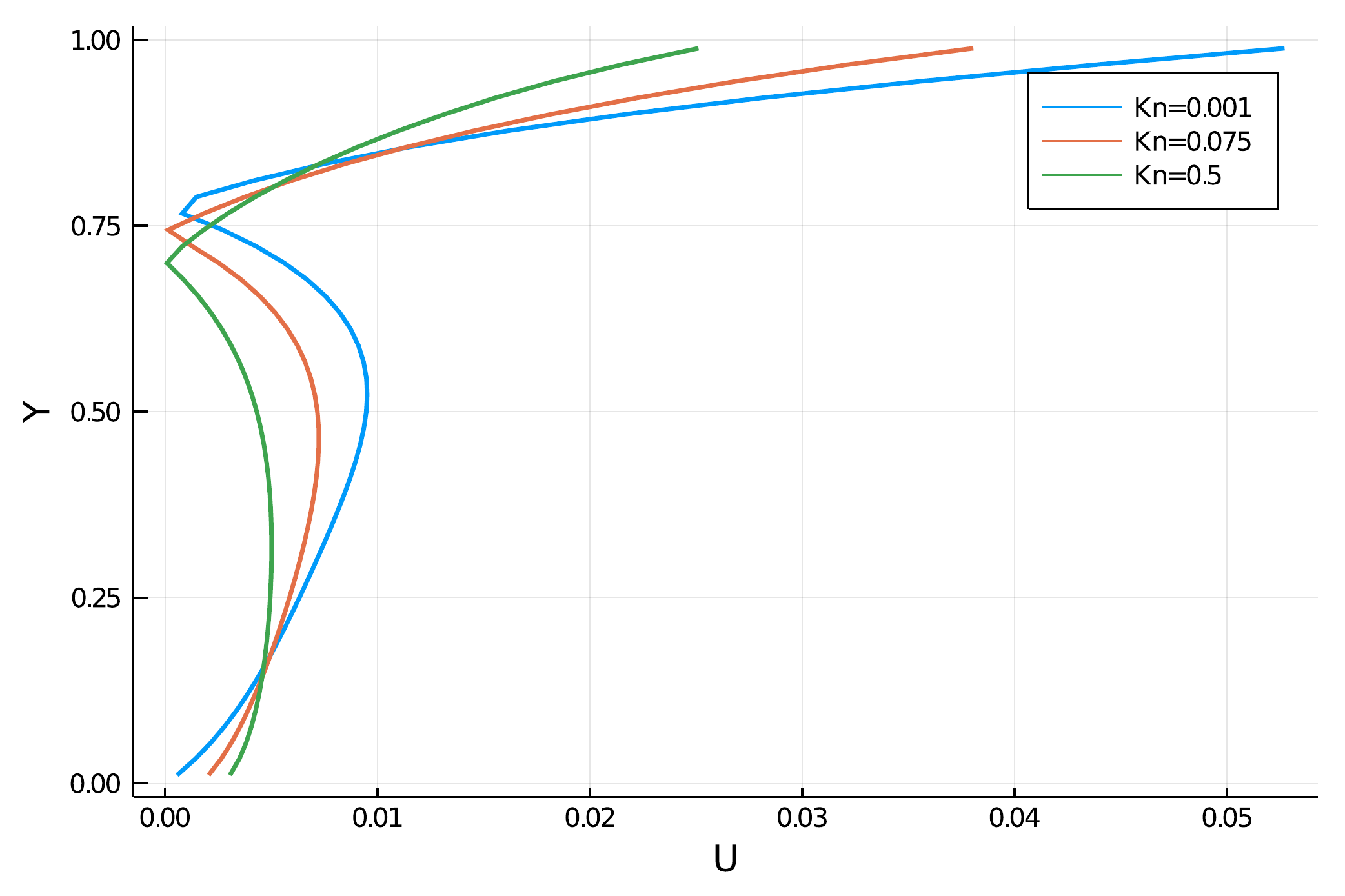}%
    }
    \caption{Expectation values and standard deviations of $U$-velocity along the vertical center line $x=0.5$ at different reference Knudsen numbers in the lid-driven cavity. The velocities have been normalized by $U_w$.}
    \label{pic:cavity vcenter line}
\end{figure}

\begin{figure}
    \subfloat[$\rm Kn=0.001$]{%
        \includegraphics[width=.33\linewidth]{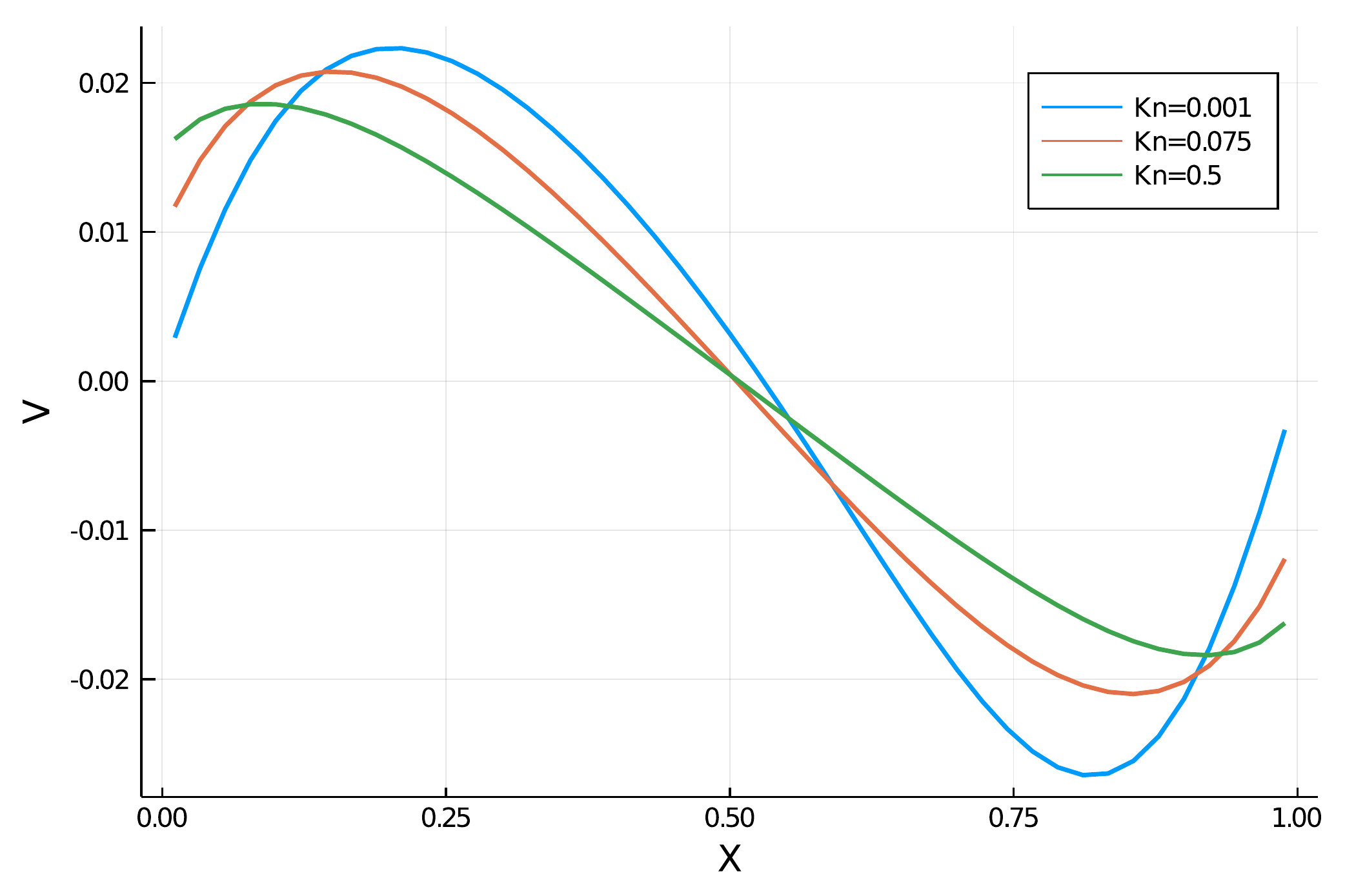}%
    }
    \subfloat[$\rm Kn=0.075$]{%
        \includegraphics[width=.33\linewidth]{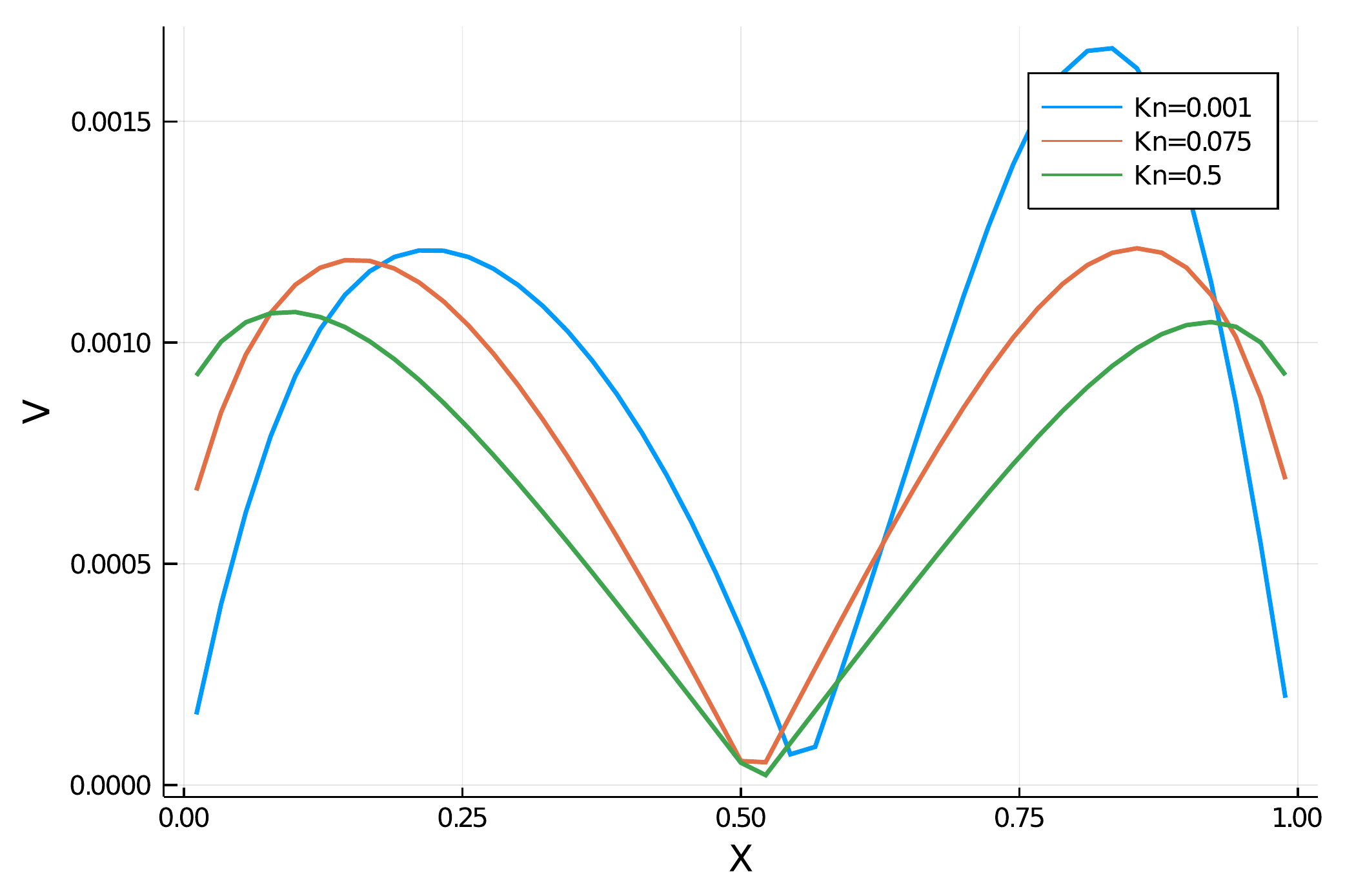}%
    }
    \caption{Expectation values and standard deviations of $V$-velocity along the horizontal center line $y=0.5$ at different reference Knudsen numbers in the lid-driven cavity. The velocities have been normalized by $U_w$.}
    \label{pic:cavity hcenter line}
\end{figure}

Fig.\ref{pic:cavity velocity}(d)-(e) shows the standard deviations of $U$-velocity inside the cavity.
It is clear that the main flow field presents a similar pattern as its variance.
From the results, we see that the upper boundary and main vortex are two driving forces of uncertainty.
As shown in Eq.(\ref{eqn:maxwell boundary}), the stochastic wall speed defines the Maxwellian distribution at the boundary, and then participates in the flux evolution.
Due to slip effect for rarefied gas-surface interaction, the magnitude of velocity variance decreases persistently with the increasing Knudsen number.
On the other hand, the magnitude of variance caused by the eddy, however, varies quite moderately throughout all the cases.
Fig.\ref{pic:cavity vcenter line} and \ref{pic:cavity hcenter line} provide the velocity distributions along the vertical $(x=0.5)$ and horizontal $(y=0.5)$ center lines.
With the increasing Kn, the intensity of vortex is damped by the enhanced viscosity, resulting in milder distribution along the center lines.
From Fig.\ref{pic:cavity vcenter line}, we see clearly the slip effect at boundary in the transition regime of flow dynamics, and its influence on the expected velocity profile and its variance.
In Fig.\ref{pic:cavity vcenter line}, the positive correlation between vortex intensity and its variance, although in a mild way, can be also identified.

\begin{figure}
    \subfloat[$\rm Kn=0.001$]{%
        \includegraphics[width=.33\linewidth]{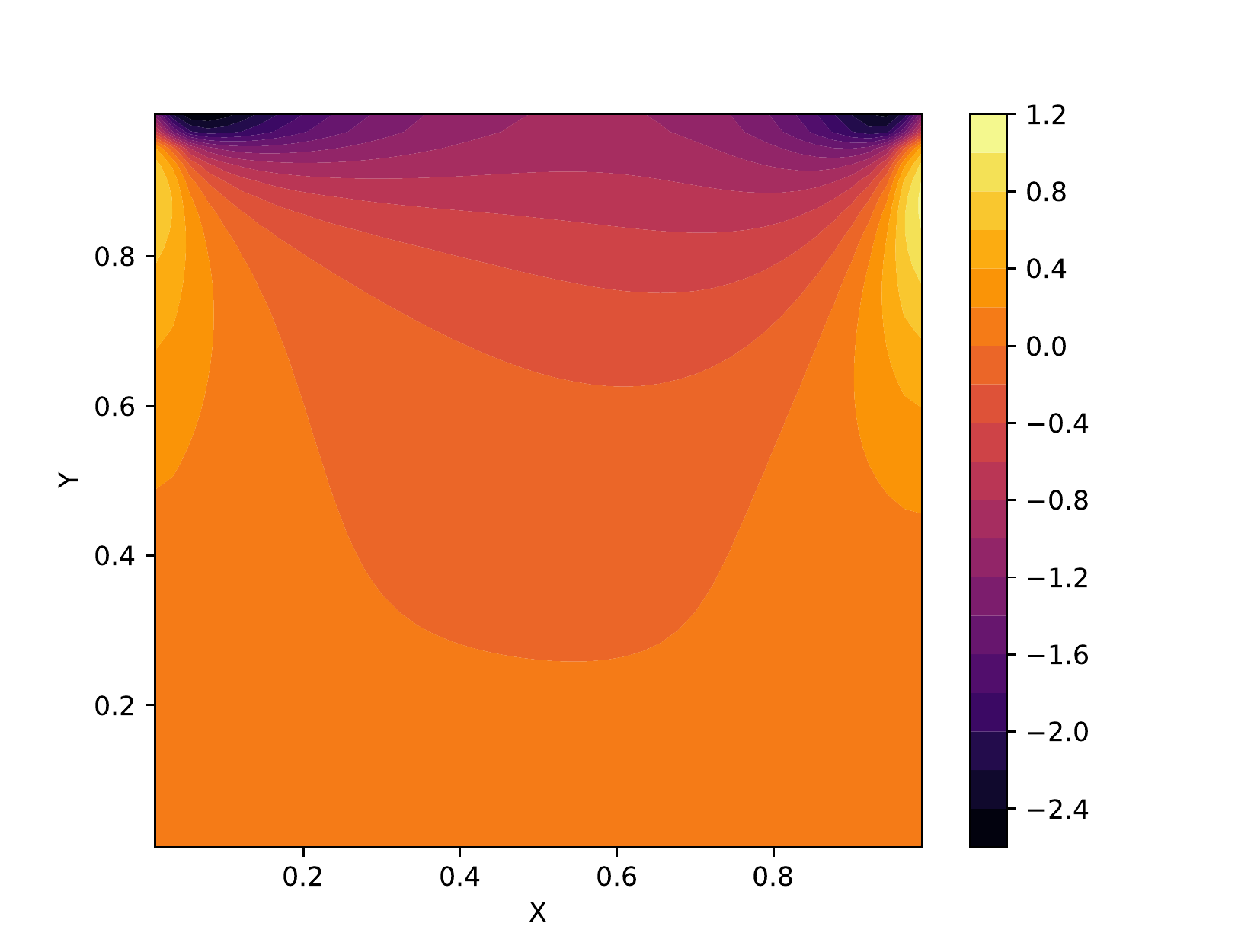}%
    }\hfill
    \subfloat[$\rm Kn=0.075$]{%
        \includegraphics[width=.33\linewidth]{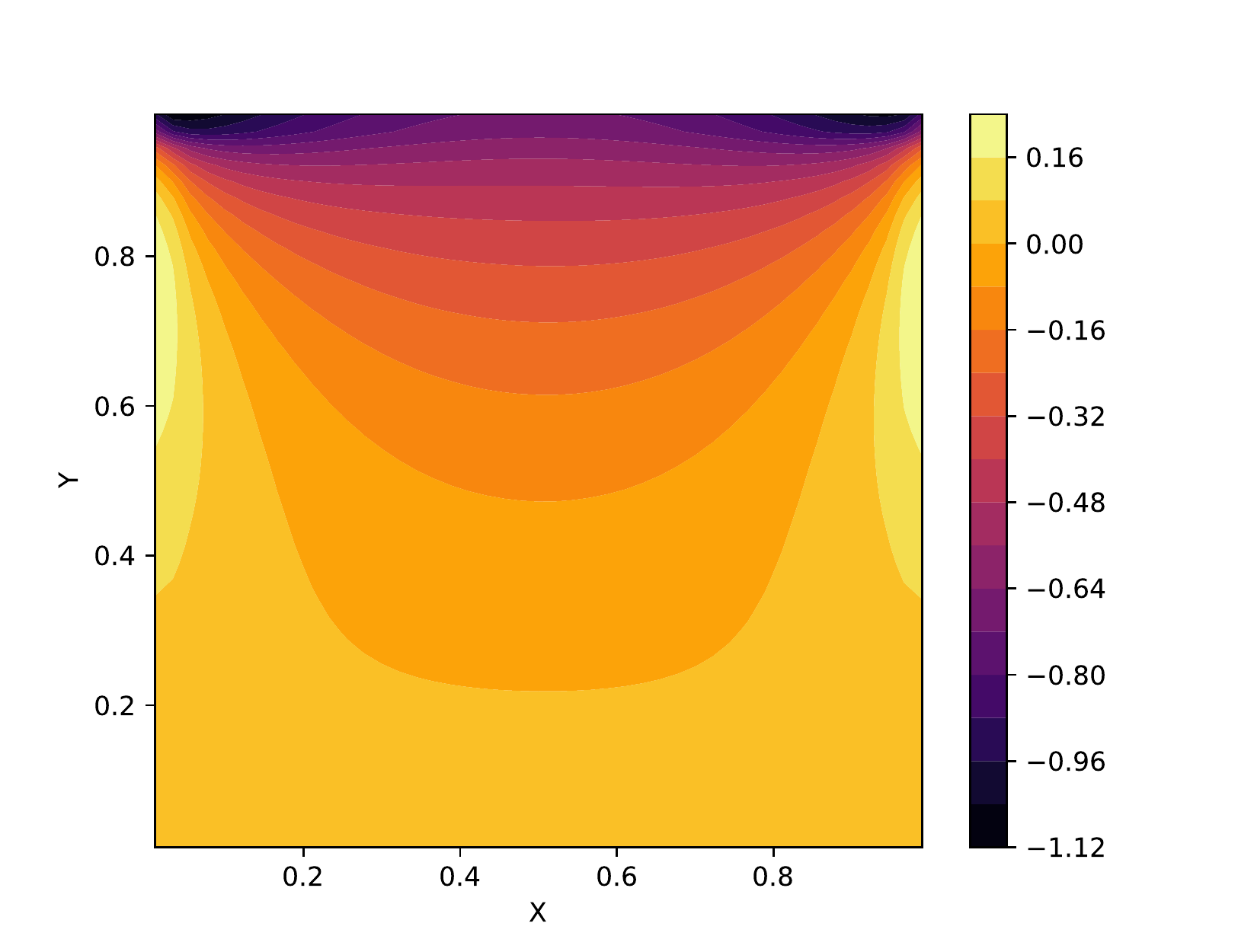}%
    }\hfill
    \subfloat[$\rm Kn=0.5$]{%
        \includegraphics[width=.33\linewidth]{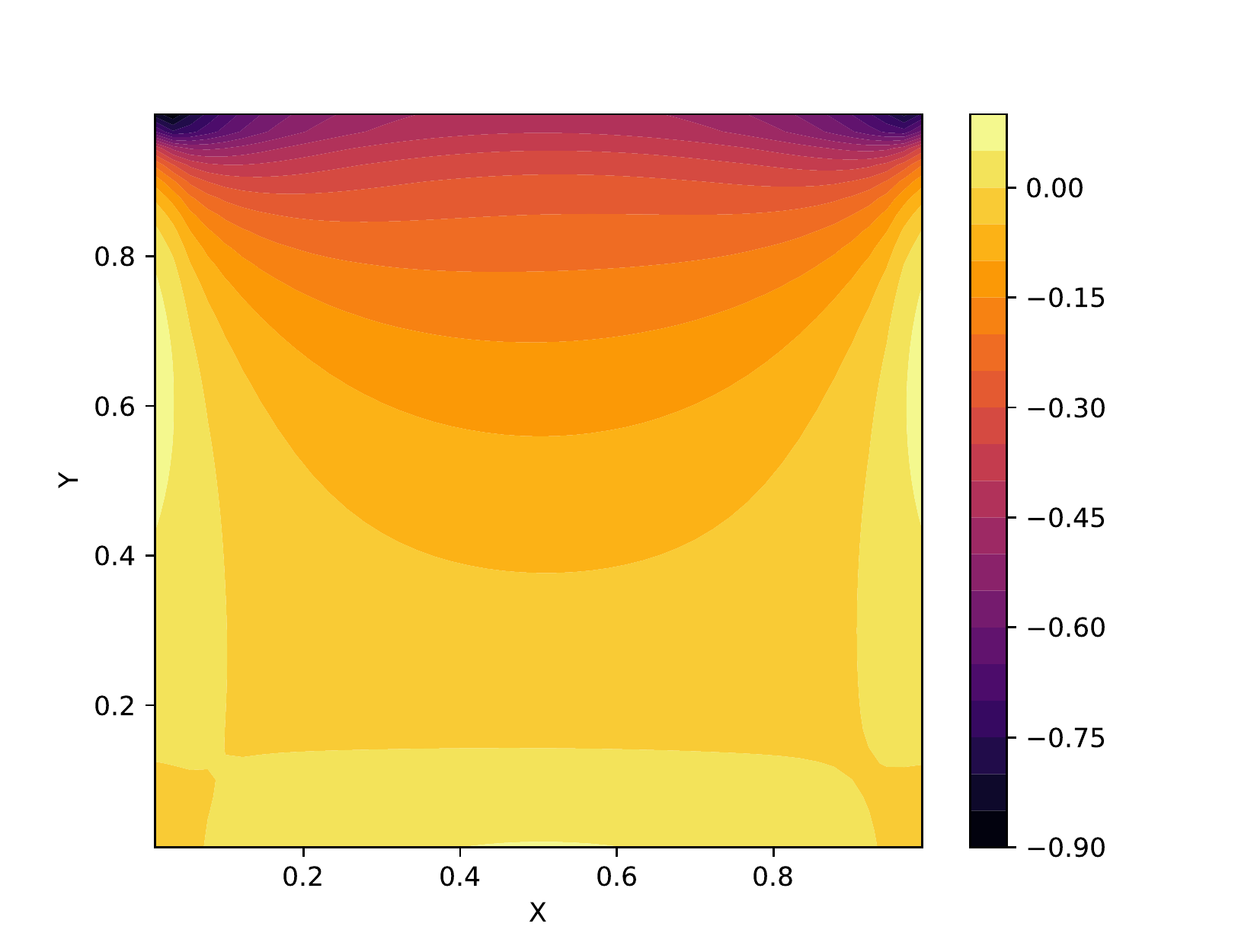}%
    }\hfill
    \subfloat[$\rm Kn=0.001$]{%
        \includegraphics[width=.33\linewidth]{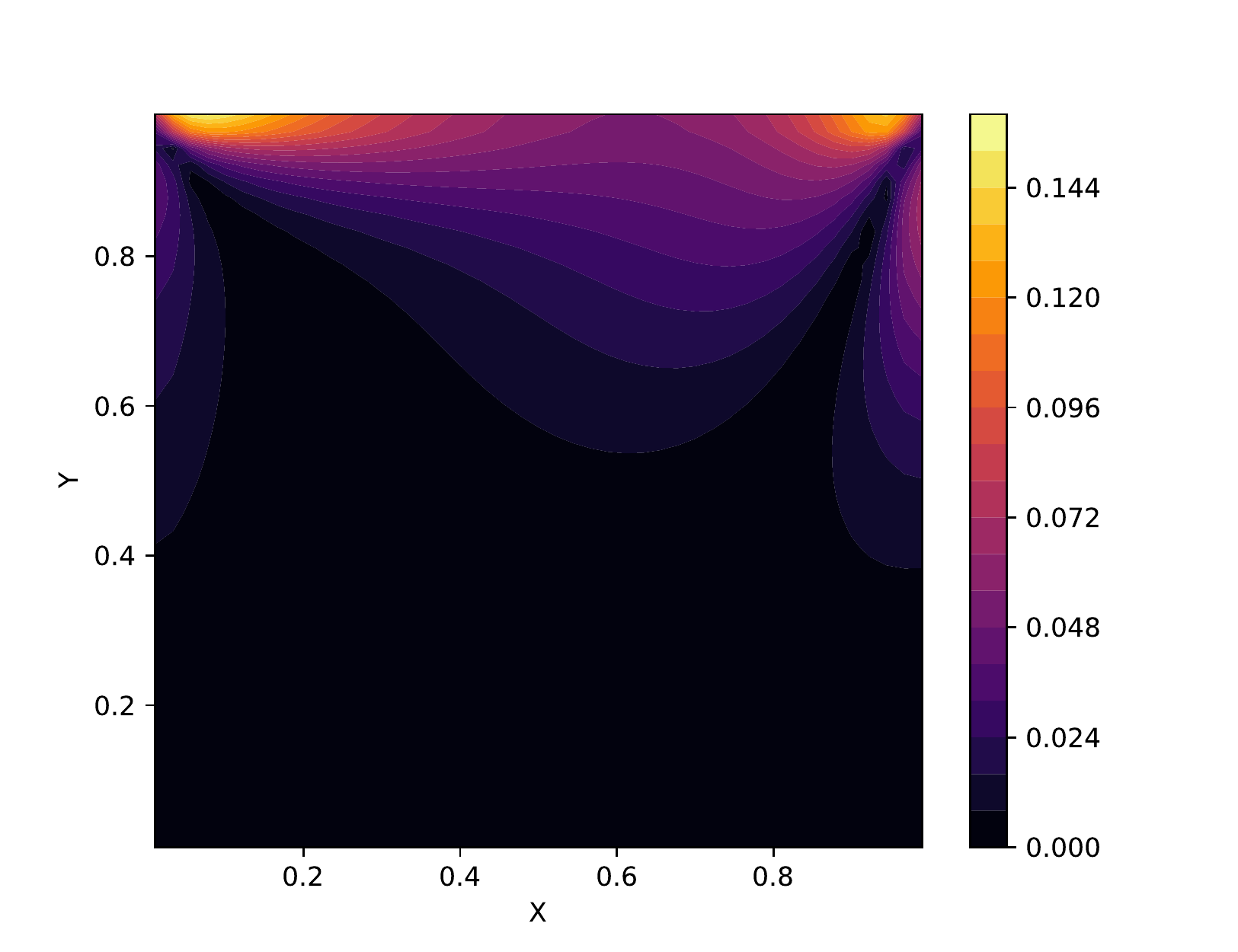}%
    }\hfill
    \subfloat[$\rm Kn=0.075$]{%
        \includegraphics[width=.33\linewidth]{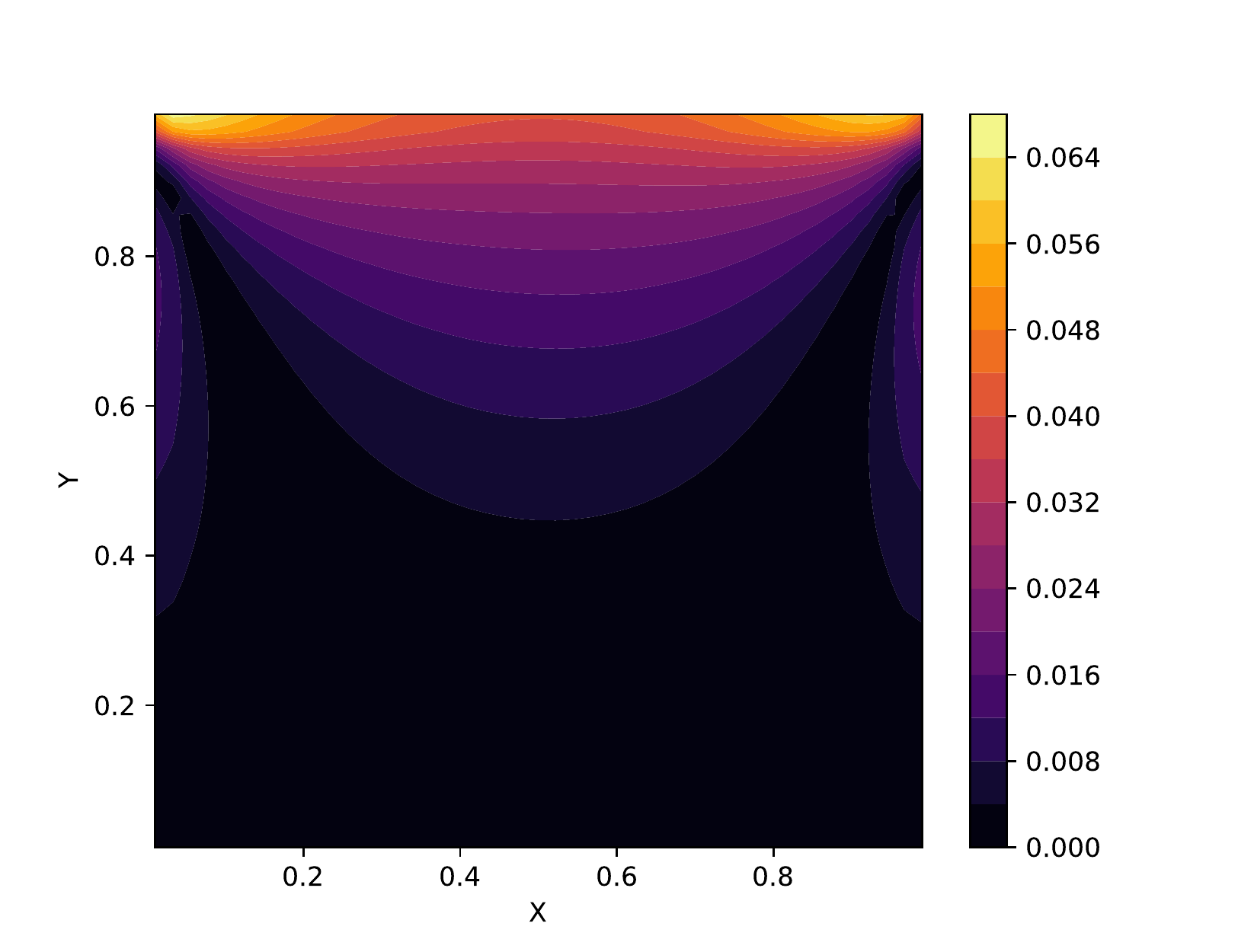}%
    }\hfill
    \subfloat[$\rm Kn=0.5$]{%
        \includegraphics[width=.33\linewidth]{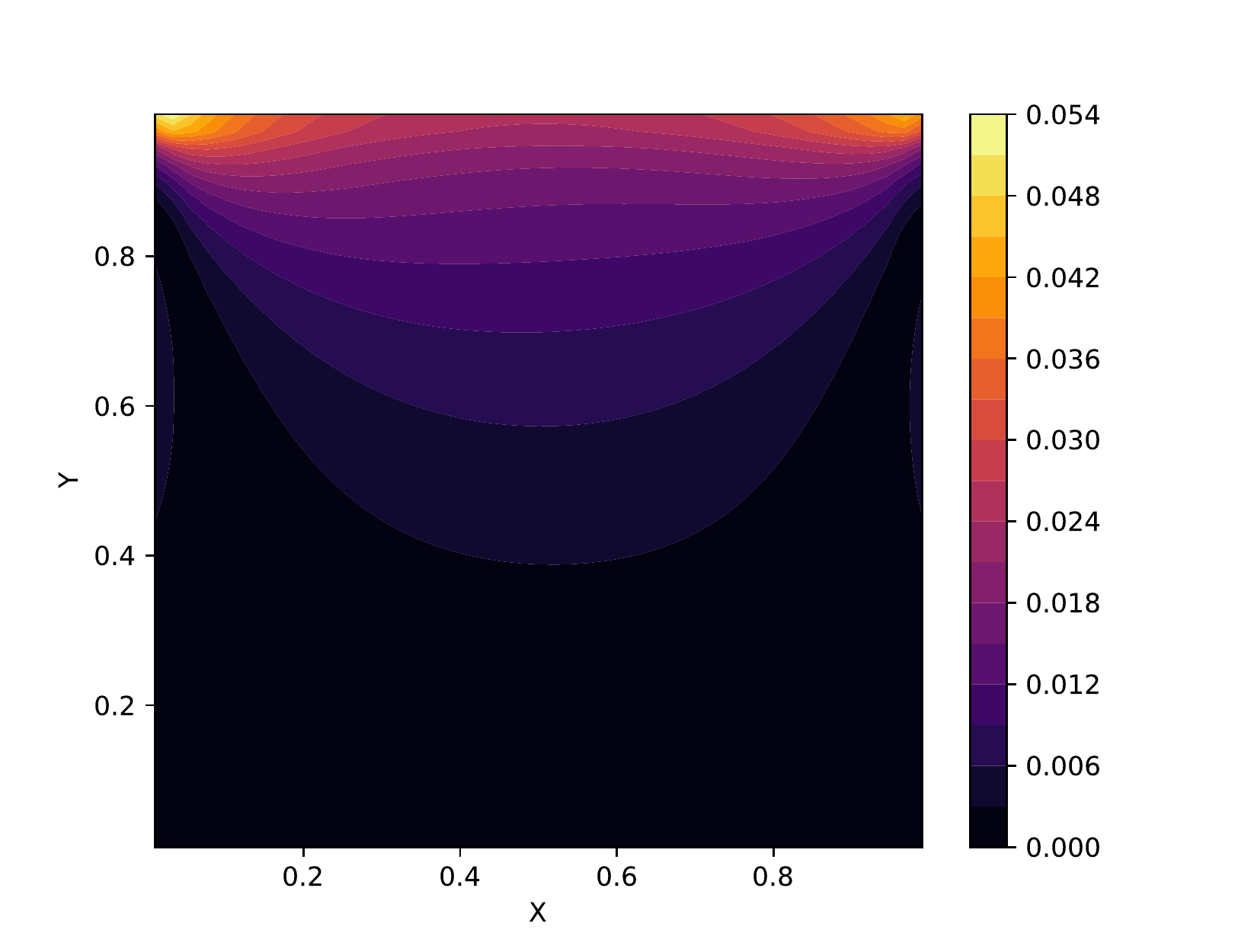}%
    }
\caption{Expectation values (first row) and standard deviations (second row) of vorticity at different reference Knudsen numbers in the lid-driven cavity.}
\label{pic:cavity vorticity}
\end{figure}

For a better understanding of the transverse process of flow dynamics inside the cavity, the distributions of vorticity are presented in Fig.\ref{pic:cavity vorticity}.
Since the eddy structure away from walls is similar as point vortex, the vorticity around vortex center is negligible.
The shearing process near boundaries plays a major role in vorticity transport.
Besides the upper moving wall, the rest three static walls contribute to vorticity evolution at the same time, given the shearing between boundary and inner flow.
Obviously, the slip effect also has significant influence on the magnitude of boundary vorticity flux.
This effect can be observed even more clearly in the variance distribution, thanks to its higher sensitivity compared with mean field.

\begin{figure}
    \subfloat[$\rm Kn=0.001$]{%
        \includegraphics[width=.33\linewidth]{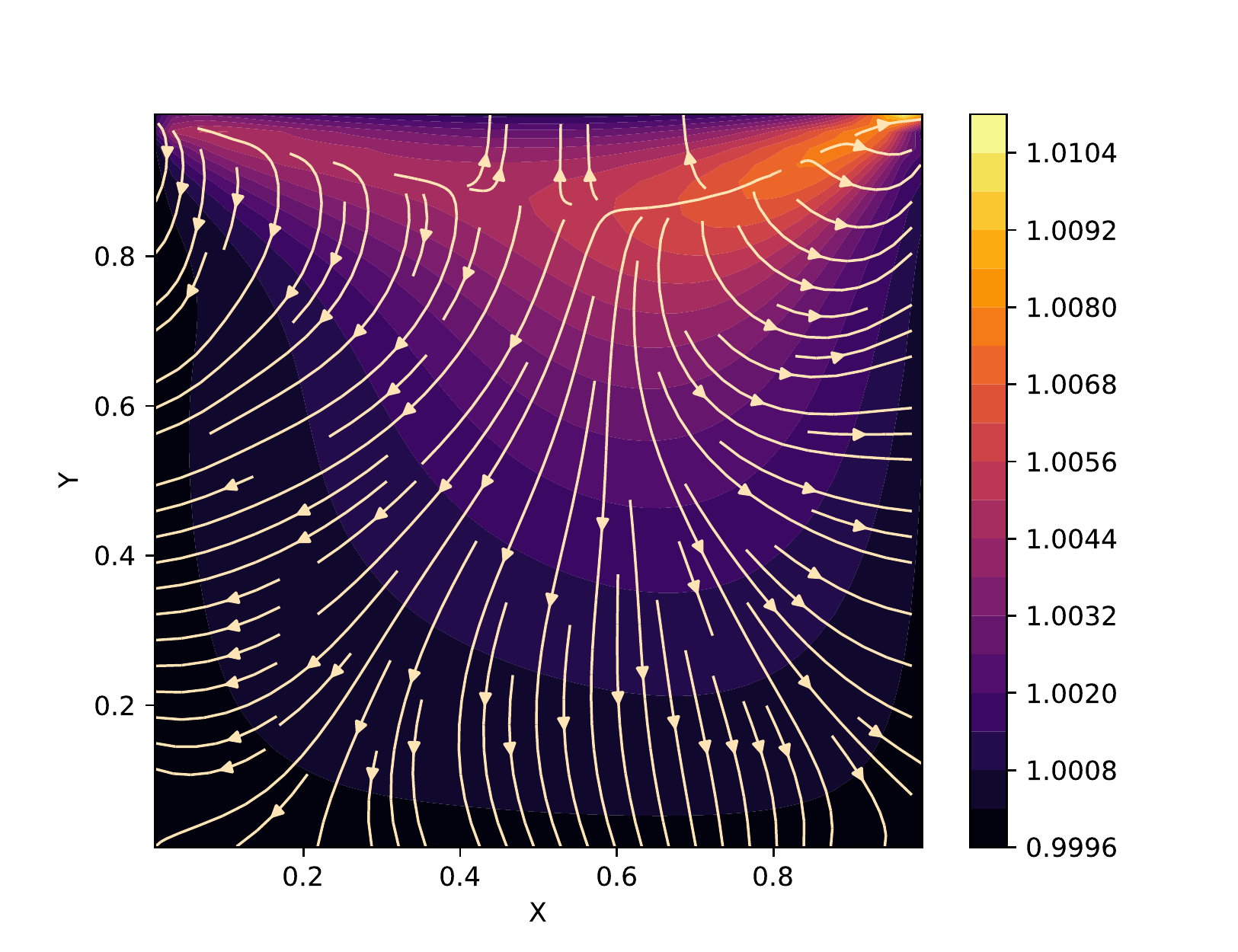}%
    }\hfill
    \subfloat[$\rm Kn=0.075$]{%
        \includegraphics[width=.33\linewidth]{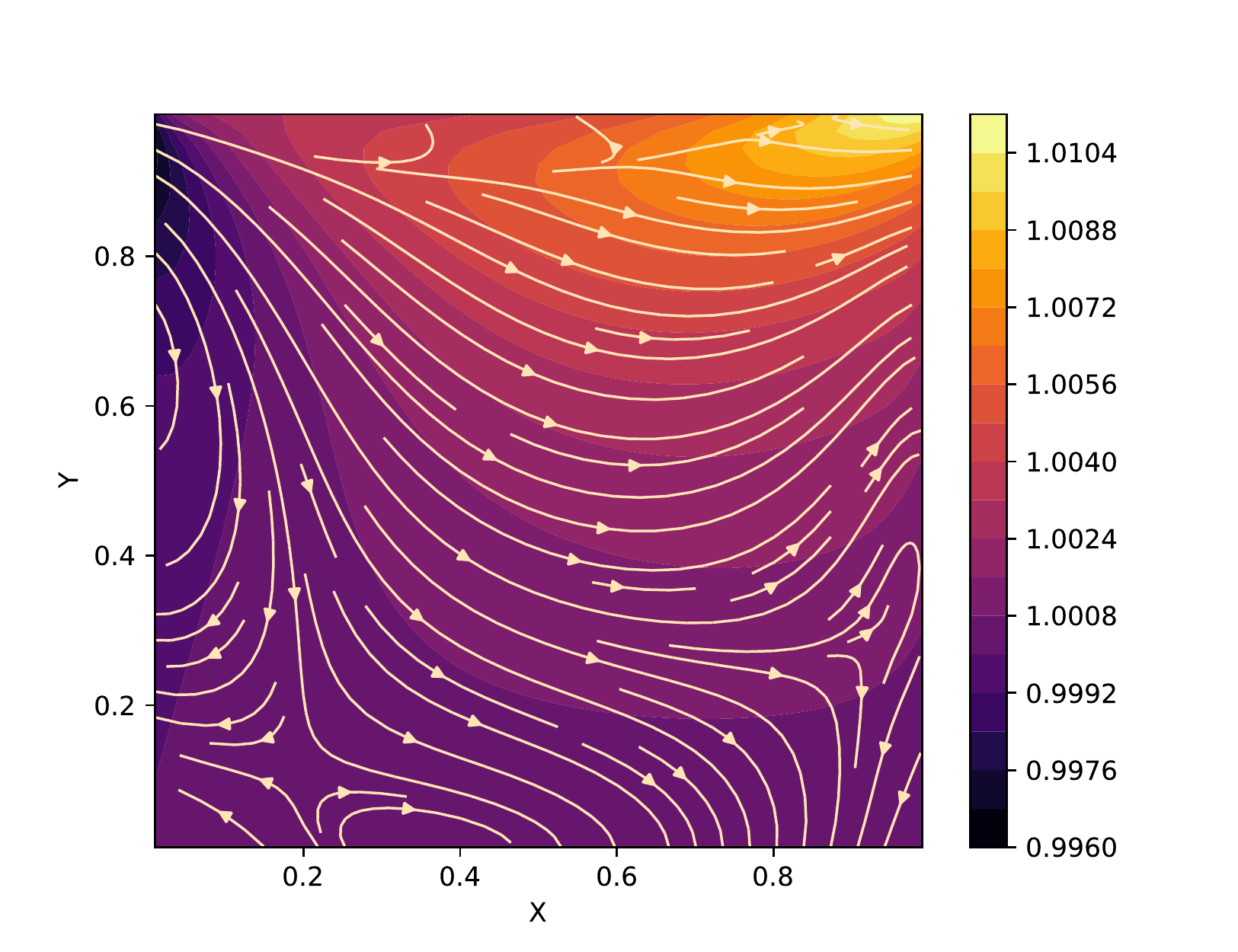}%
    }\hfill
    \subfloat[$\rm Kn=0.5$]{%
        \includegraphics[width=.33\linewidth]{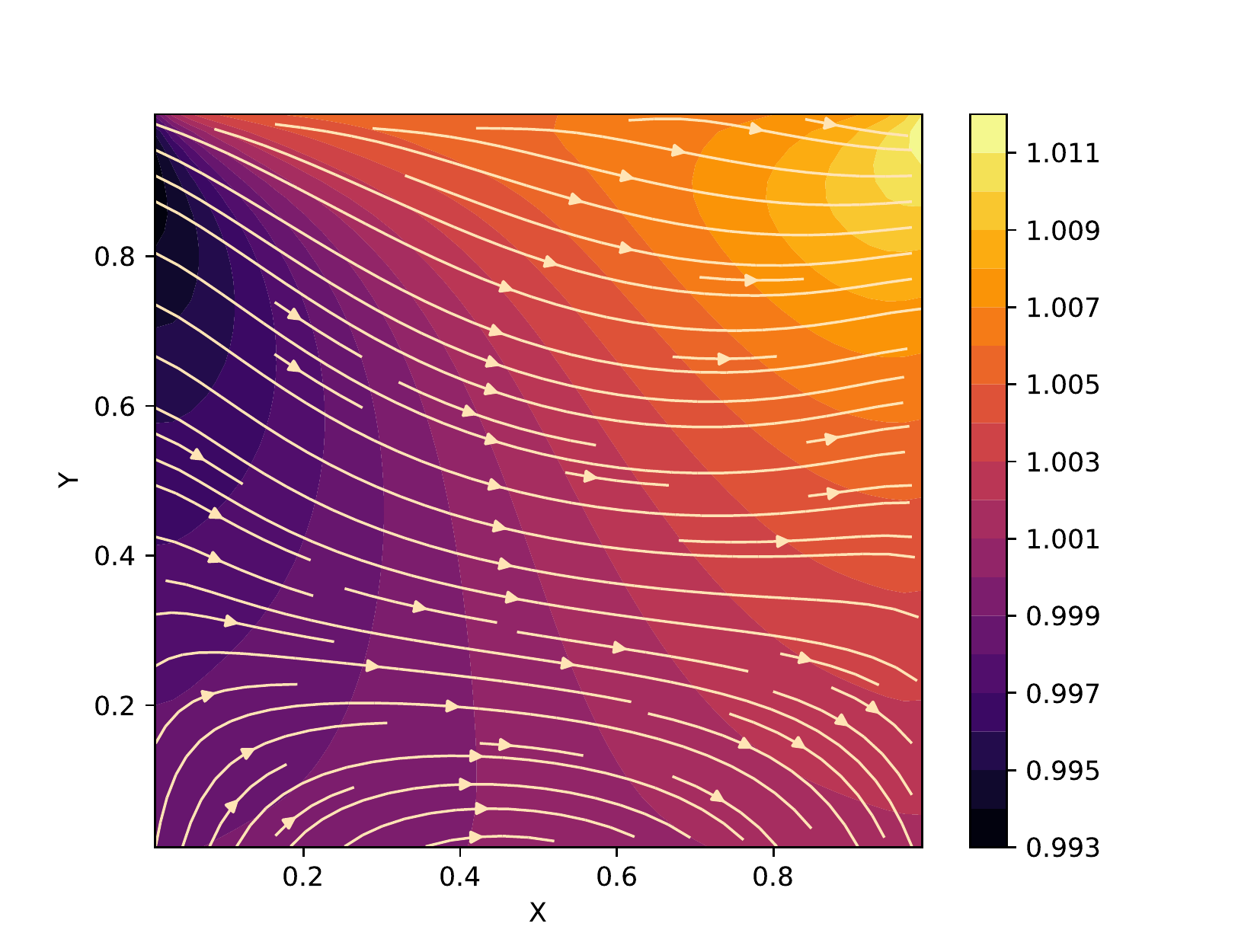}%
    }\hfill
    \subfloat[$\rm Kn=0.001$]{%
        \includegraphics[width=.33\linewidth]{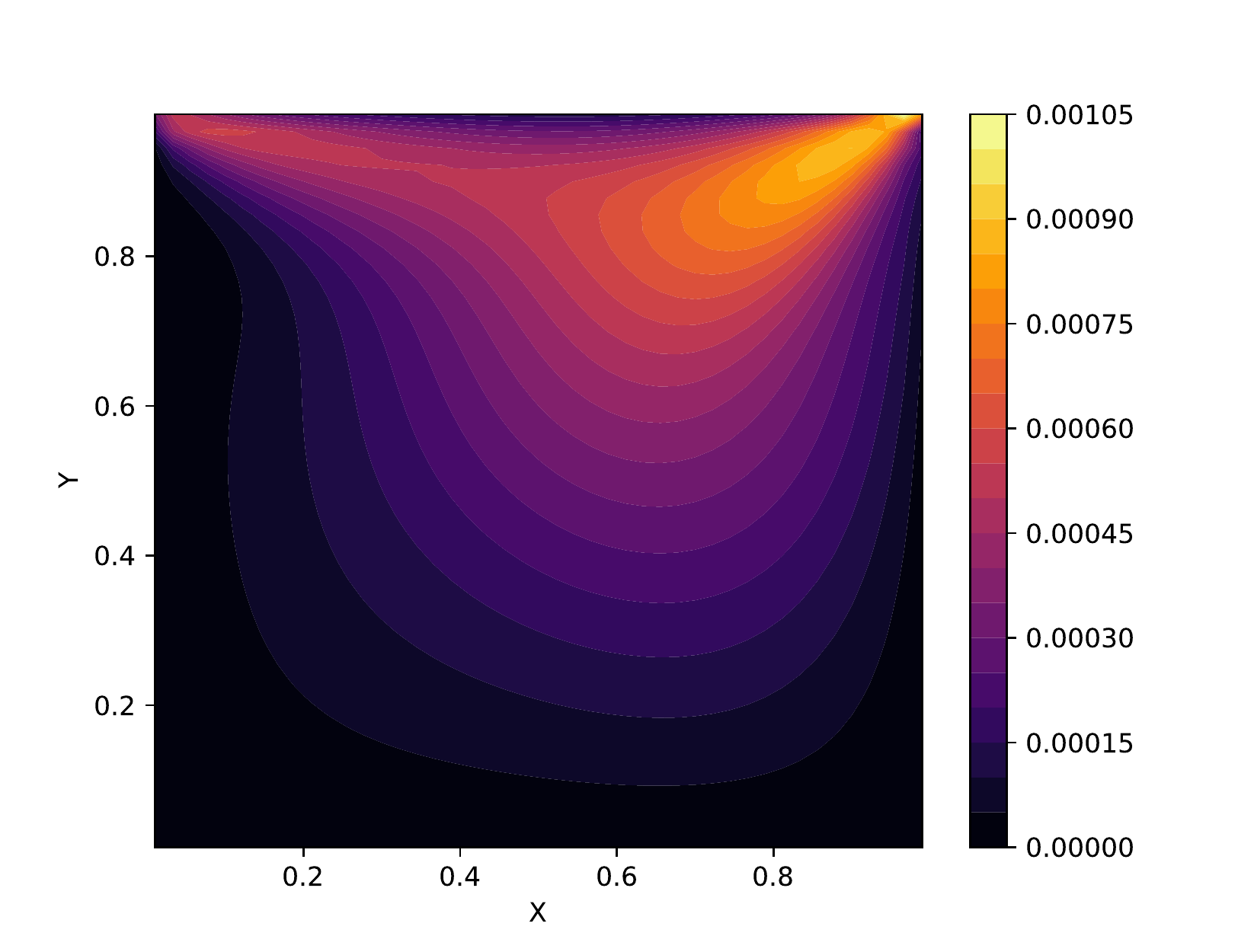}%
    }\hfill
    \subfloat[$\rm Kn=0.075$]{%
        \includegraphics[width=.33\linewidth]{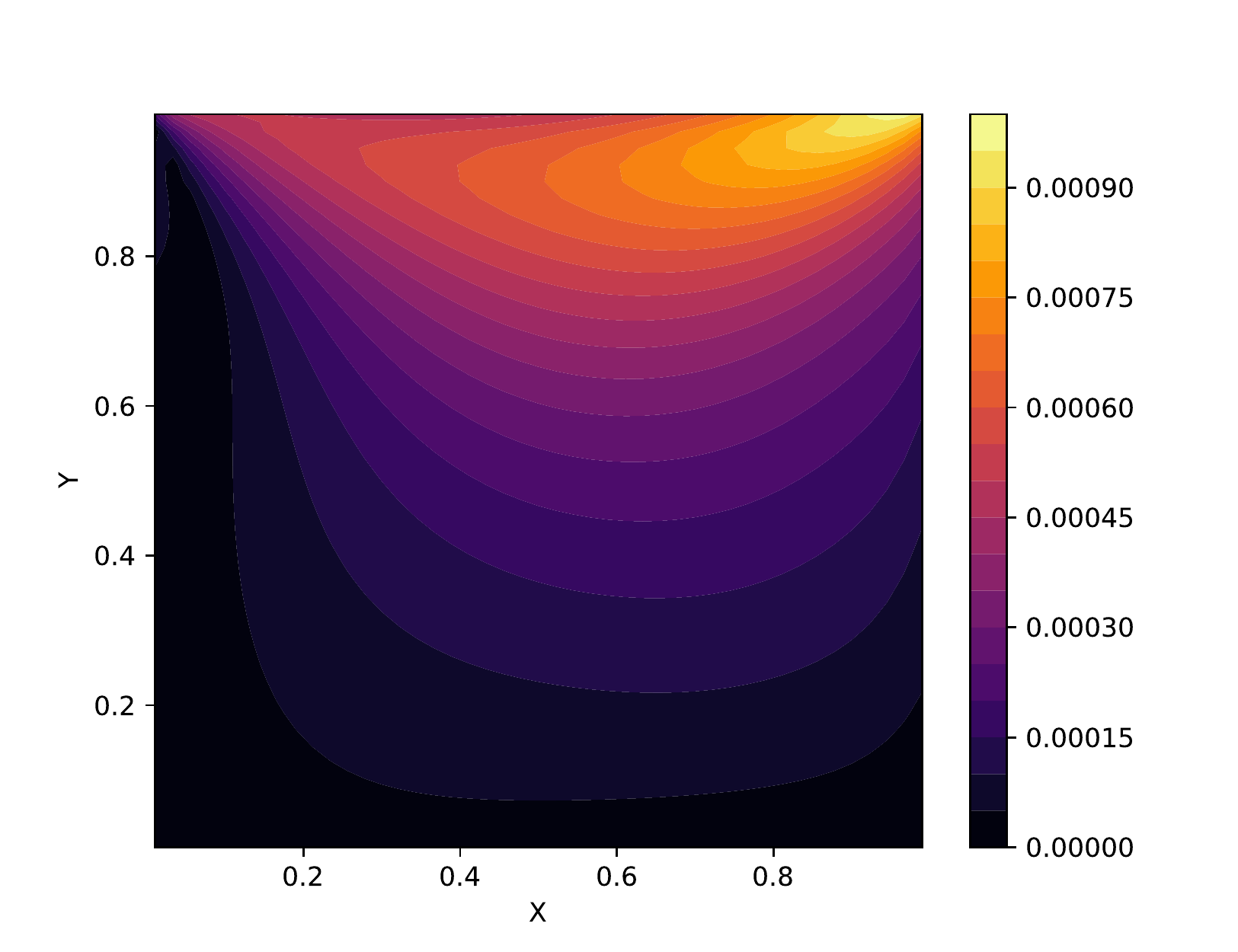}%
    }\hfill
    \subfloat[$\rm Kn=0.5$]{%
        \includegraphics[width=.33\linewidth]{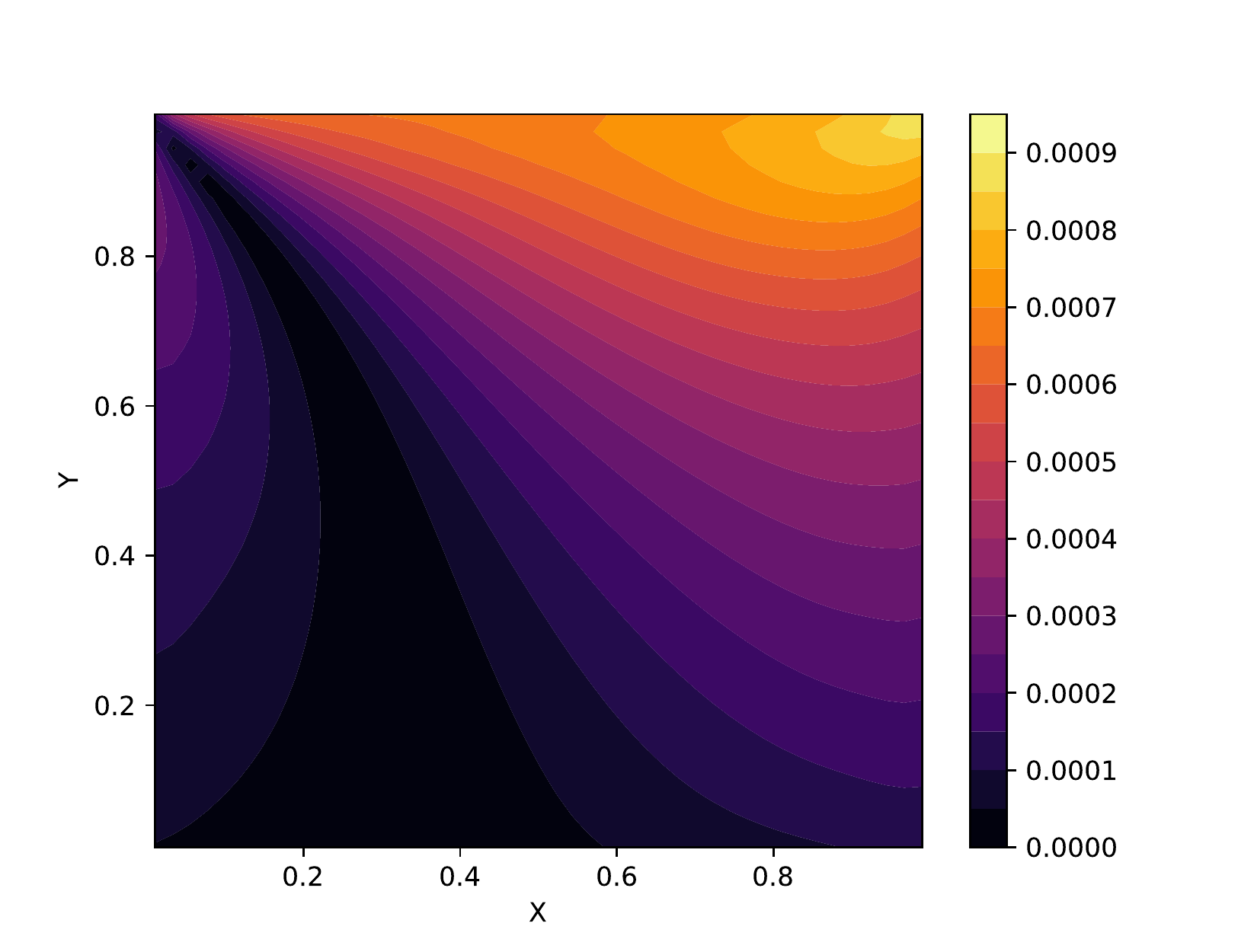}%
    }
\caption{Expectation values (first row) and standard deviations (second row) of temperature (contour) and heat flux (vector) at different reference Knudsen numbers in the lid-driven cavity.}
\label{pic:cavity heat}
\end{figure}

The heat transfer inside cavity is crossly coupled with flow transport.
Fig.\ref{pic:cavity heat} shows the expectation values and standard deviations of temperature contours along with heat flux vectors.
As demonstrated, the viscous heating at the top right corner contribute one local maximum of temperature.
Meanwhile, the loose coupling among particles leads to an expansion cooling effect around the top left, and results in a minimum in the transition regime.
With the increasing Knudsen number, the heat flux gradually deviates from the Fourier's law, and the heat transports from the cold to hot region.
Such an anti-gradient heat flux is induced by viscous shearing, which is a typical non-equilibrium phenomenon in the cavity flow.

\begin{figure}
    \subfloat[$\rm Kn=0.001$]{%
        \includegraphics[width=.33\linewidth]{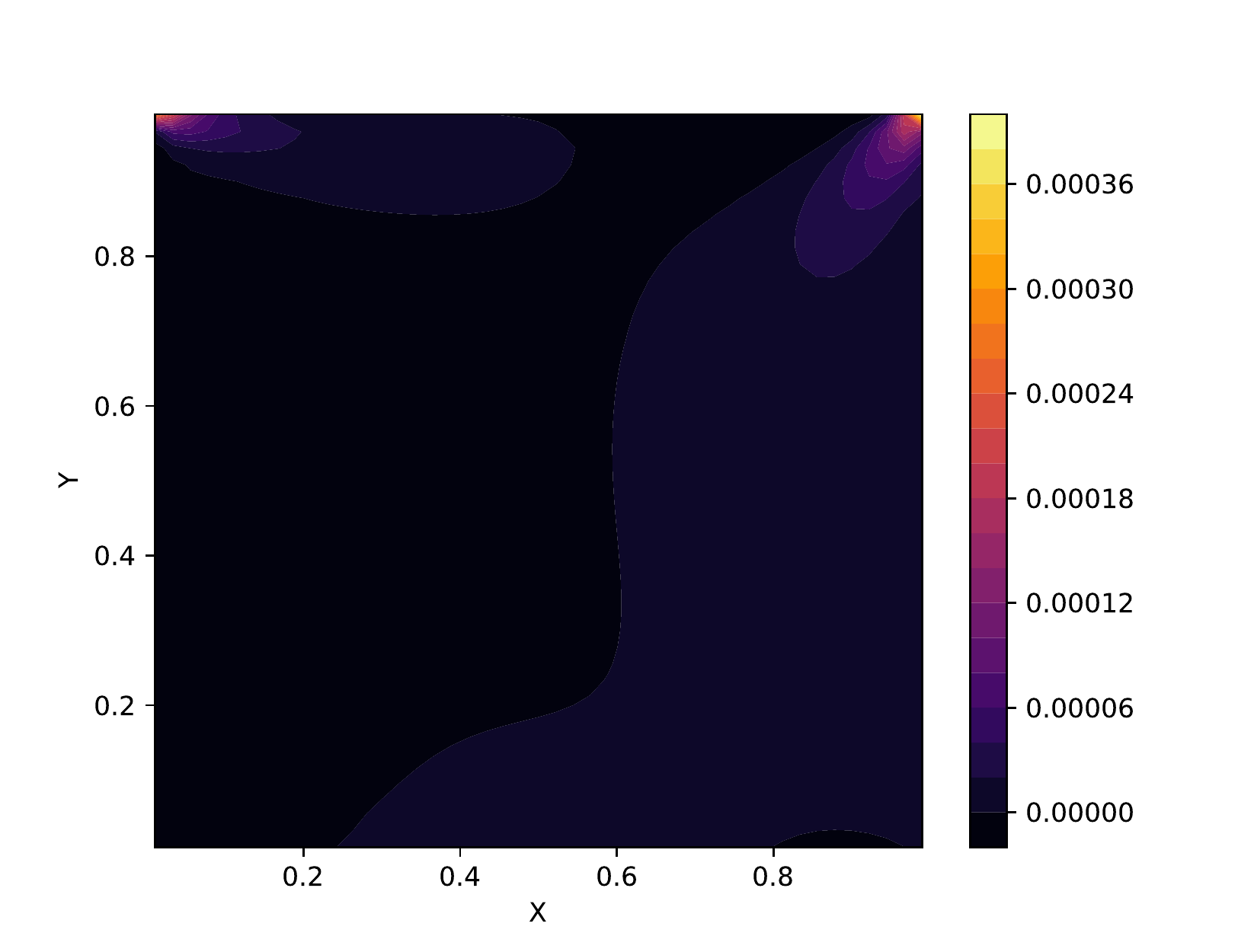}%
    }\hfill
    \subfloat[$\rm Kn=0.075$]{%
        \includegraphics[width=.33\linewidth]{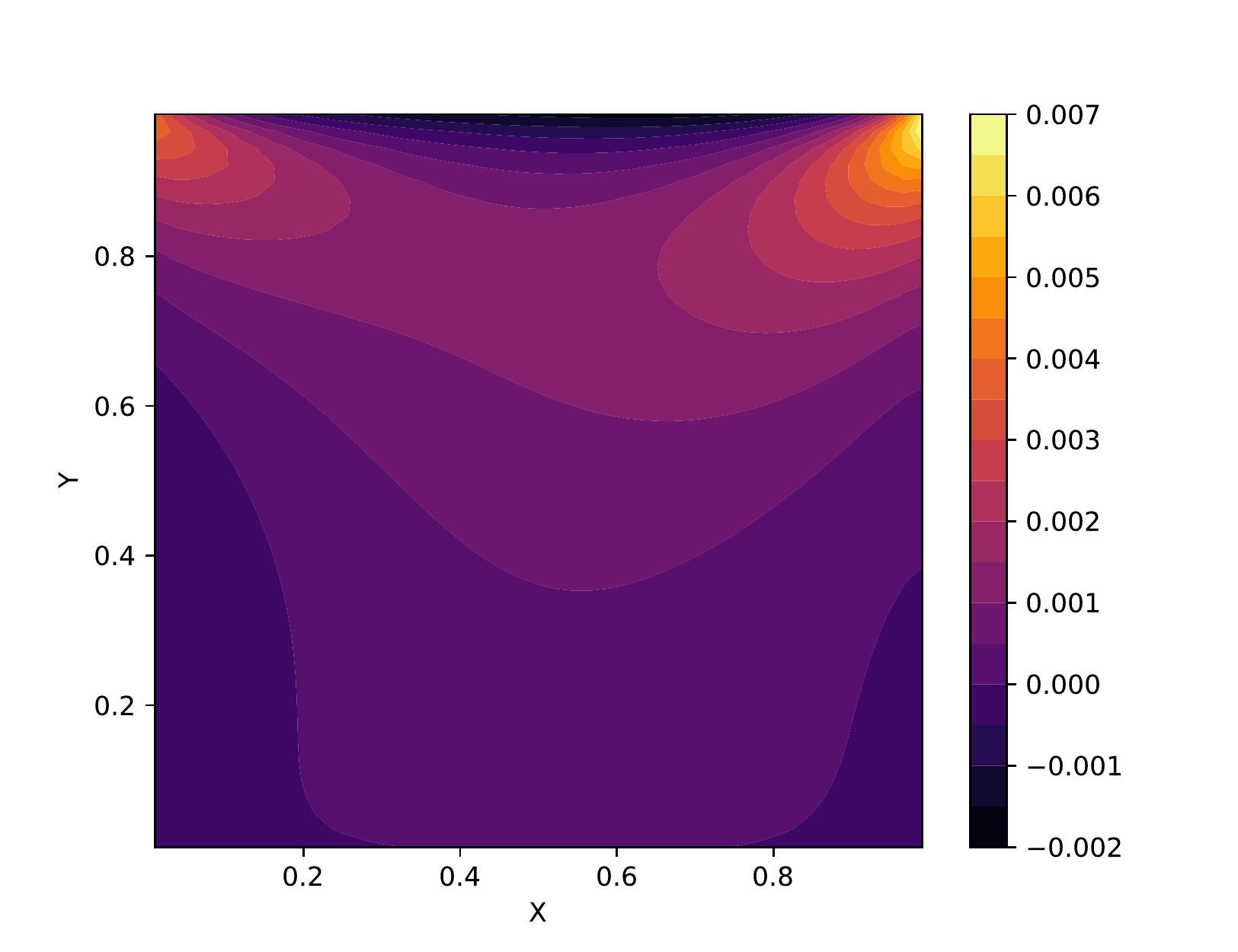}%
    }\hfill
    \subfloat[$\rm Kn=0.5$]{%
        \includegraphics[width=.33\linewidth]{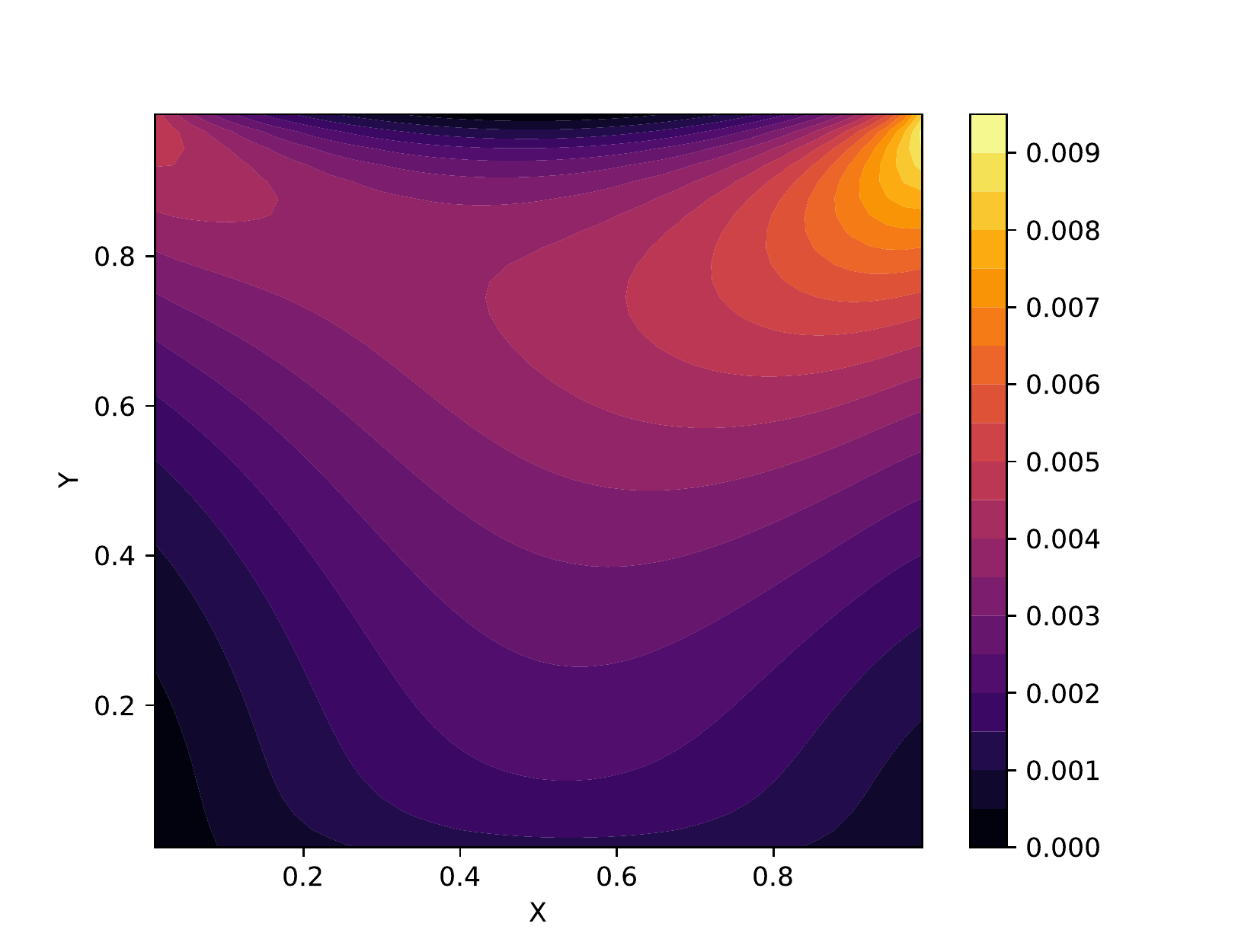}%
    }\hfill
    \subfloat[$\rm Kn=0.001$]{%
        \includegraphics[width=.33\linewidth]{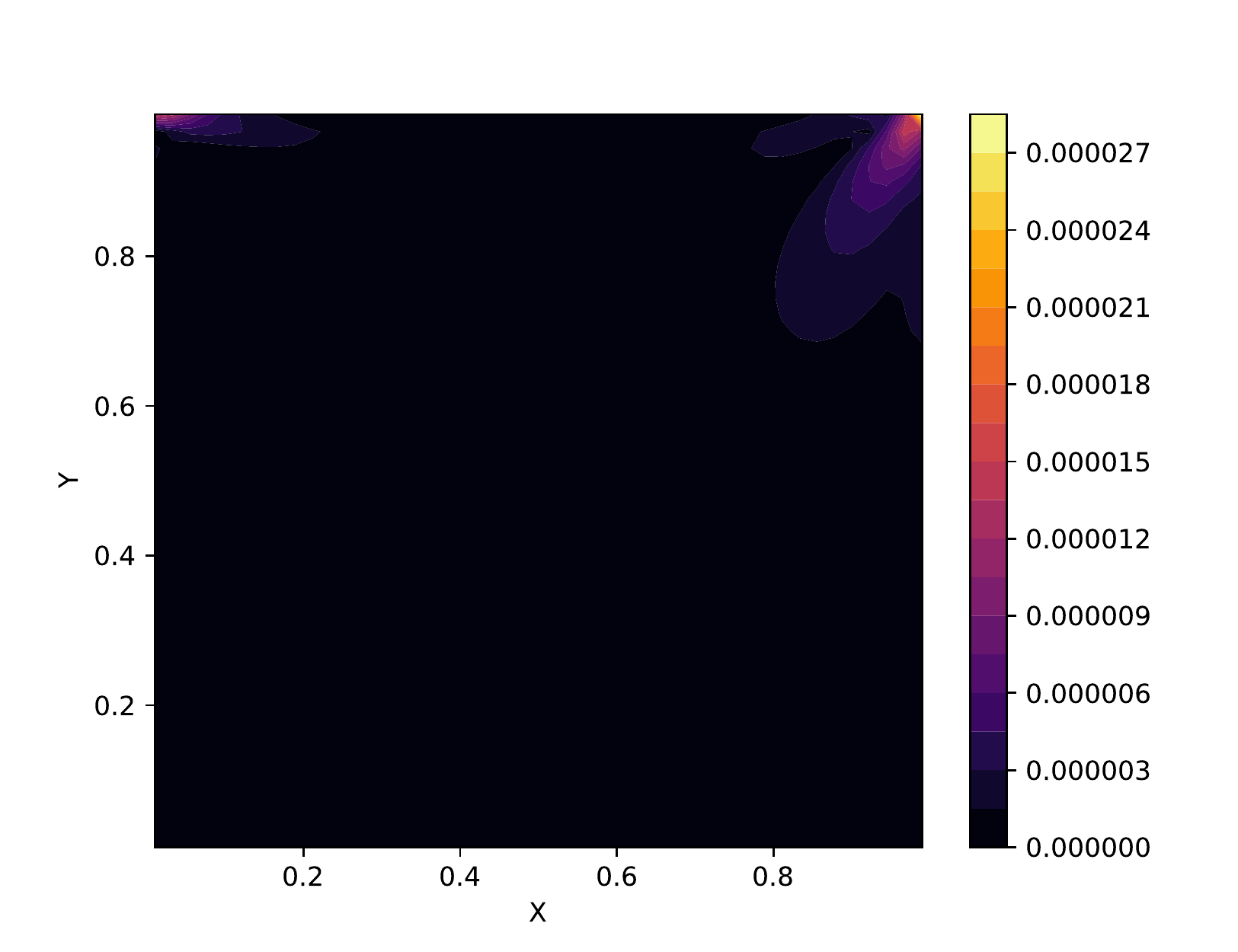}%
    }\hfill
    \subfloat[$\rm Kn=0.075$]{%
        \includegraphics[width=.33\linewidth]{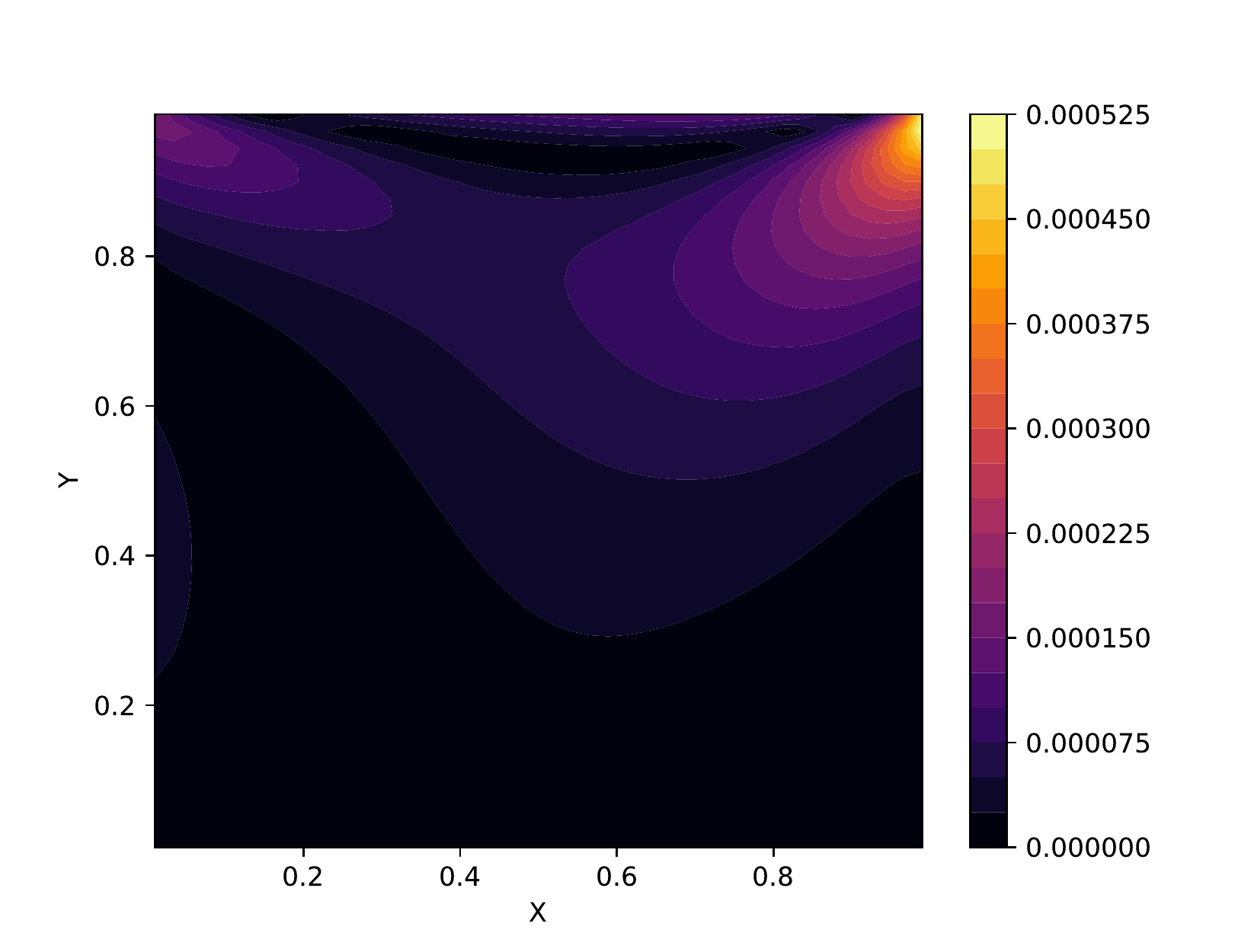}%
    }\hfill
    \subfloat[$\rm Kn=0.5$]{%
        \includegraphics[width=.33\linewidth]{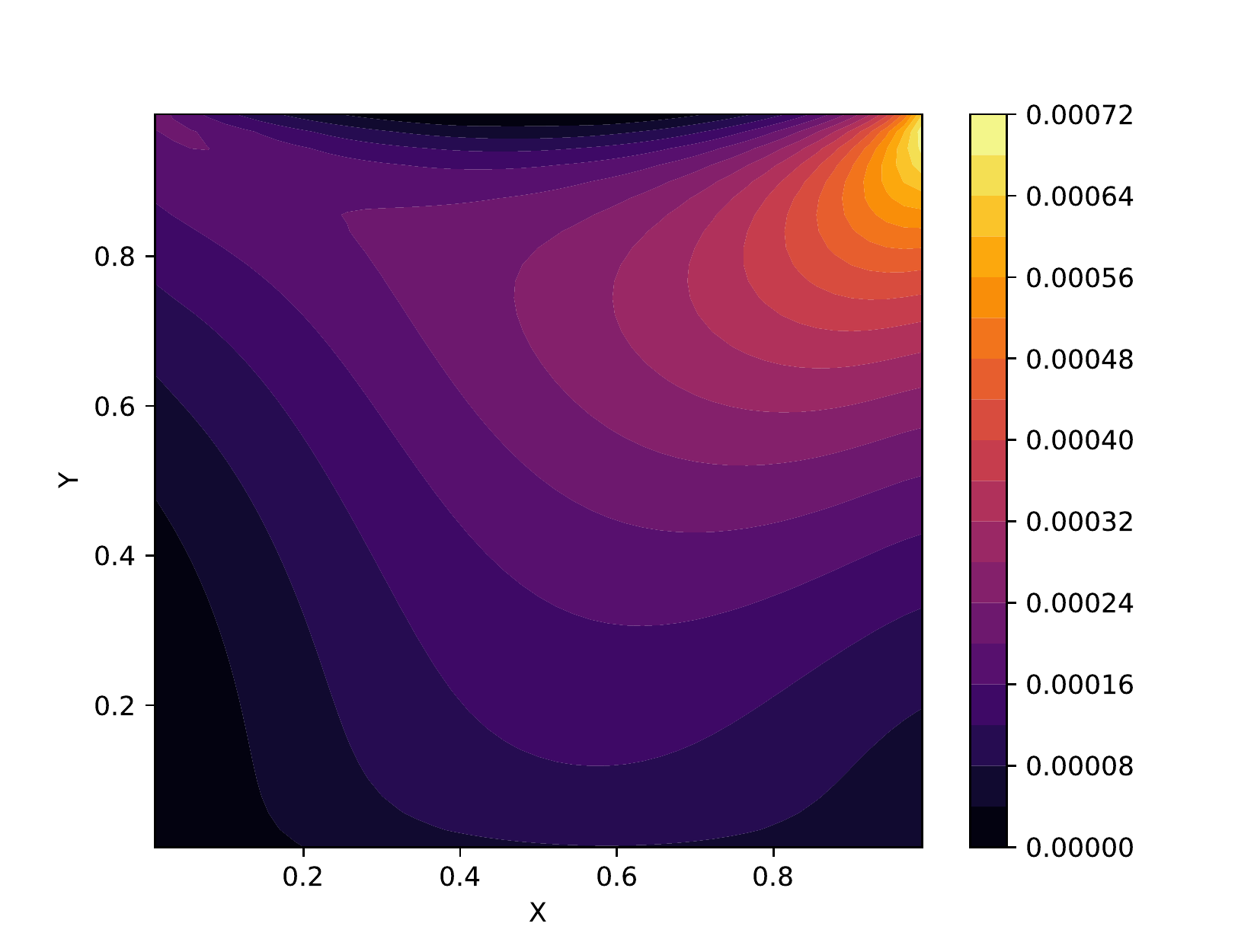}%
    }\hfill
    \subfloat[$\rm Kn=0.001$]{%
        \includegraphics[width=.33\linewidth]{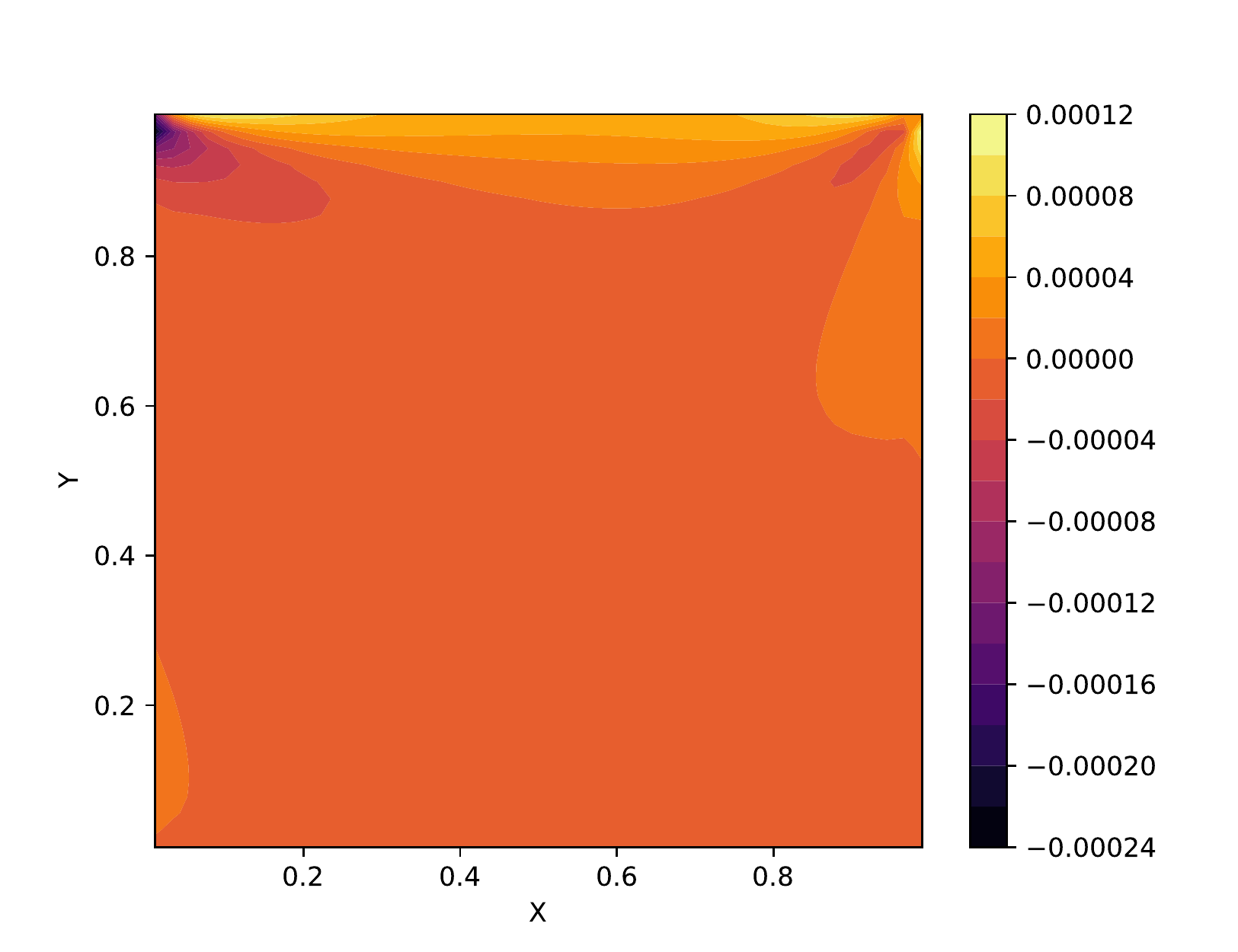}%
    }\hfill
    \subfloat[$\rm Kn=0.075$]{%
        \includegraphics[width=.33\linewidth]{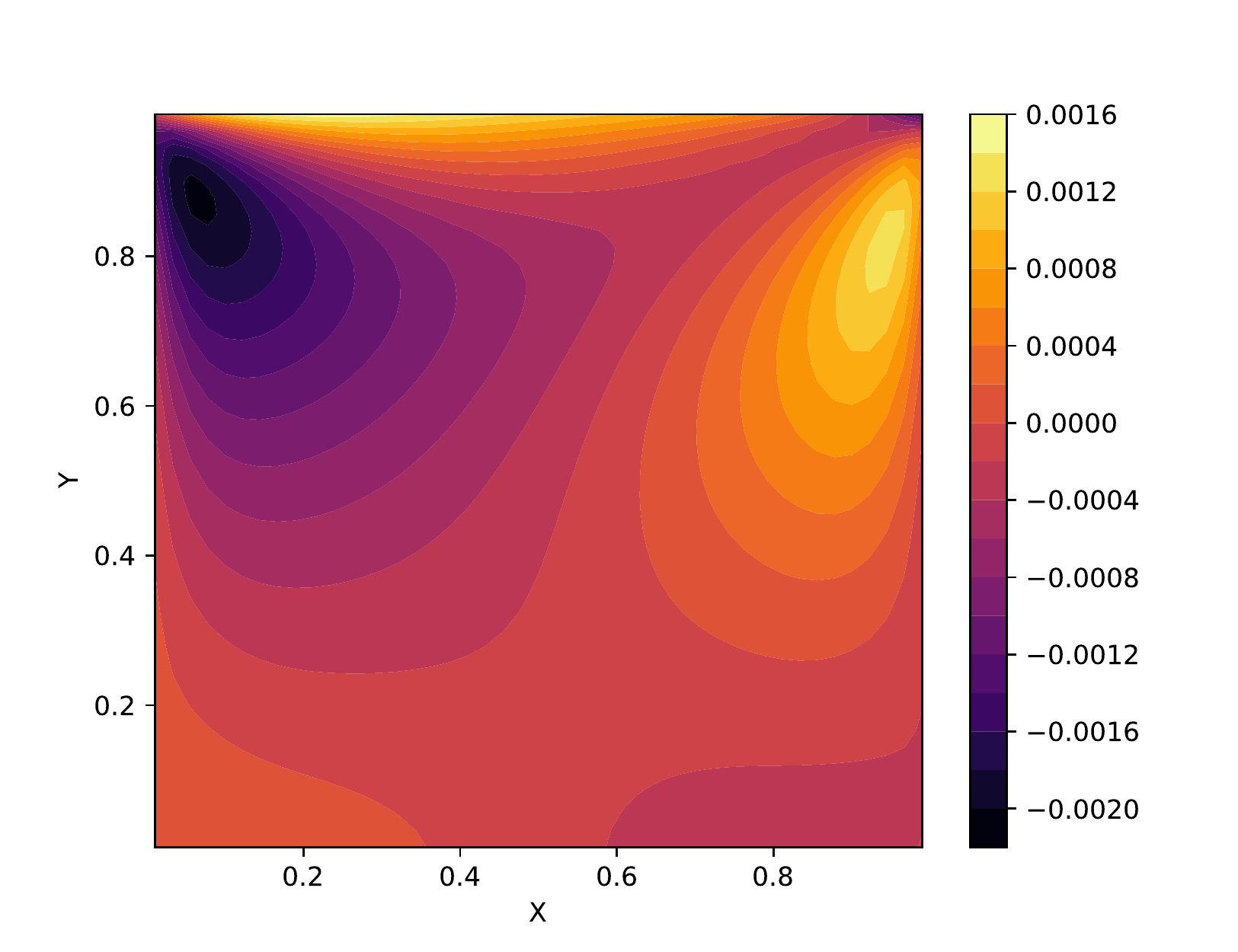}%
    }\hfill
    \subfloat[$\rm Kn=0.5$]{%
        \includegraphics[width=.33\linewidth]{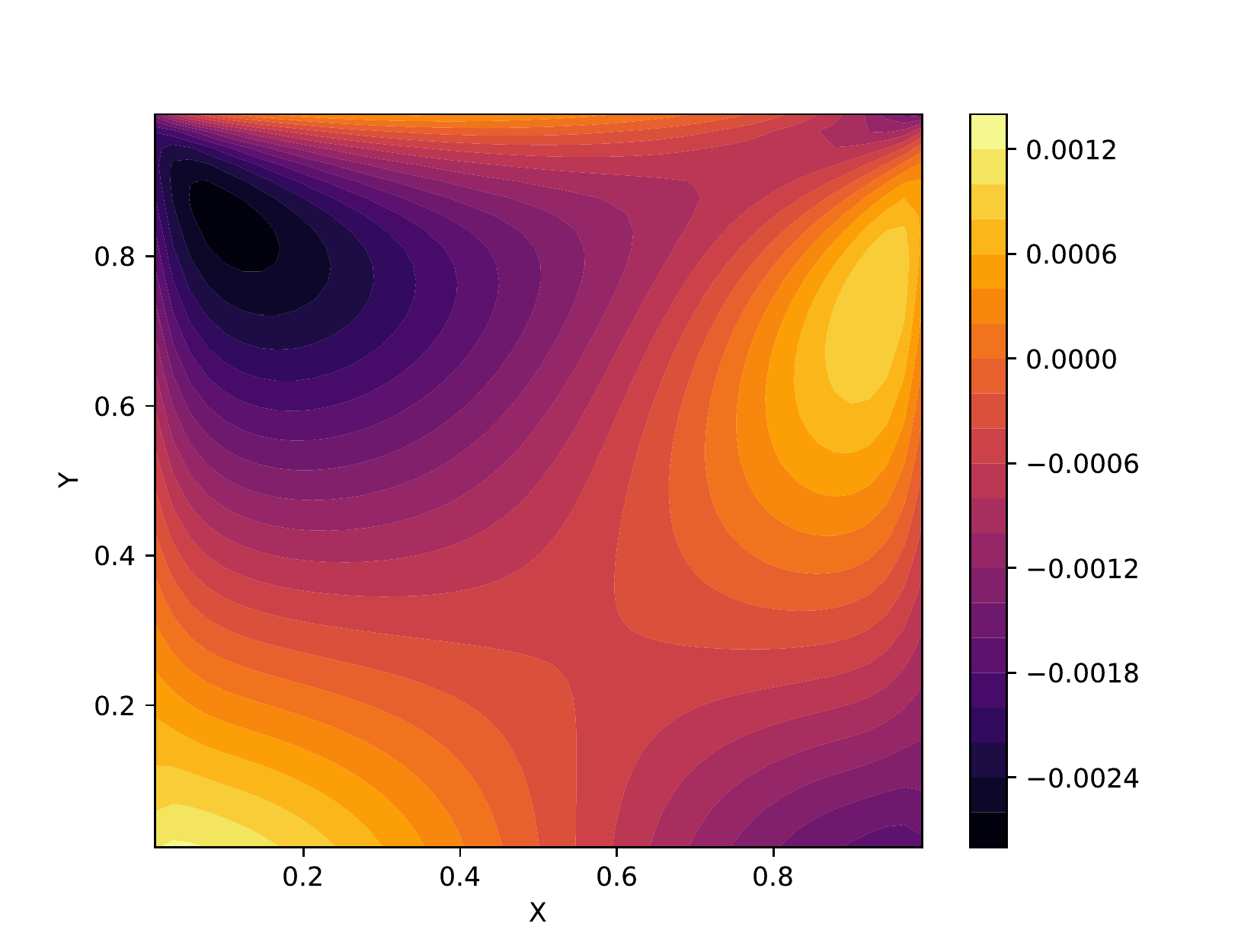}%
    }\hfill
    \subfloat[$\rm Kn=0.001$]{%
        \includegraphics[width=.33\linewidth]{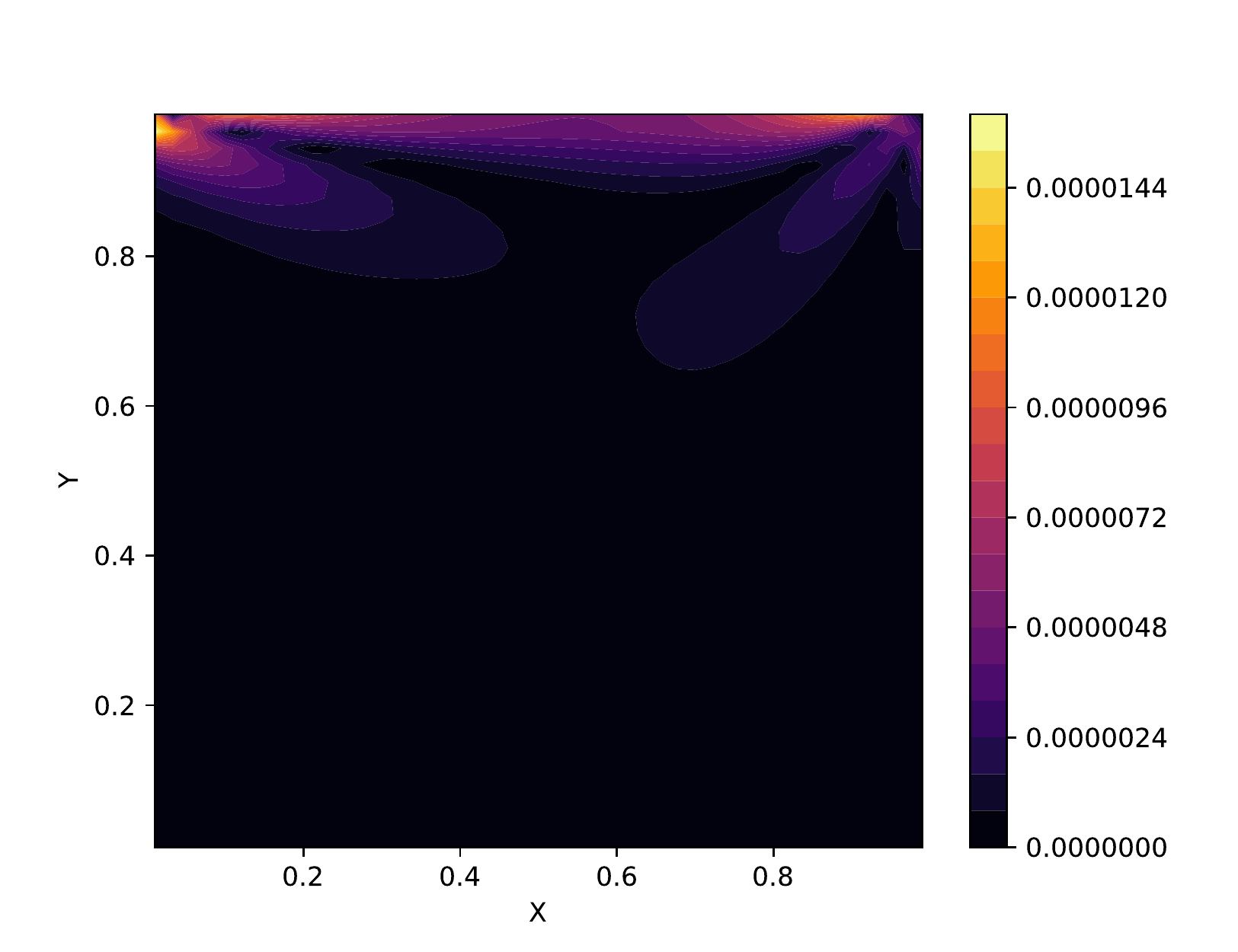}%
    }\hfill
    \subfloat[$\rm Kn=0.075$]{%
        \includegraphics[width=.33\linewidth]{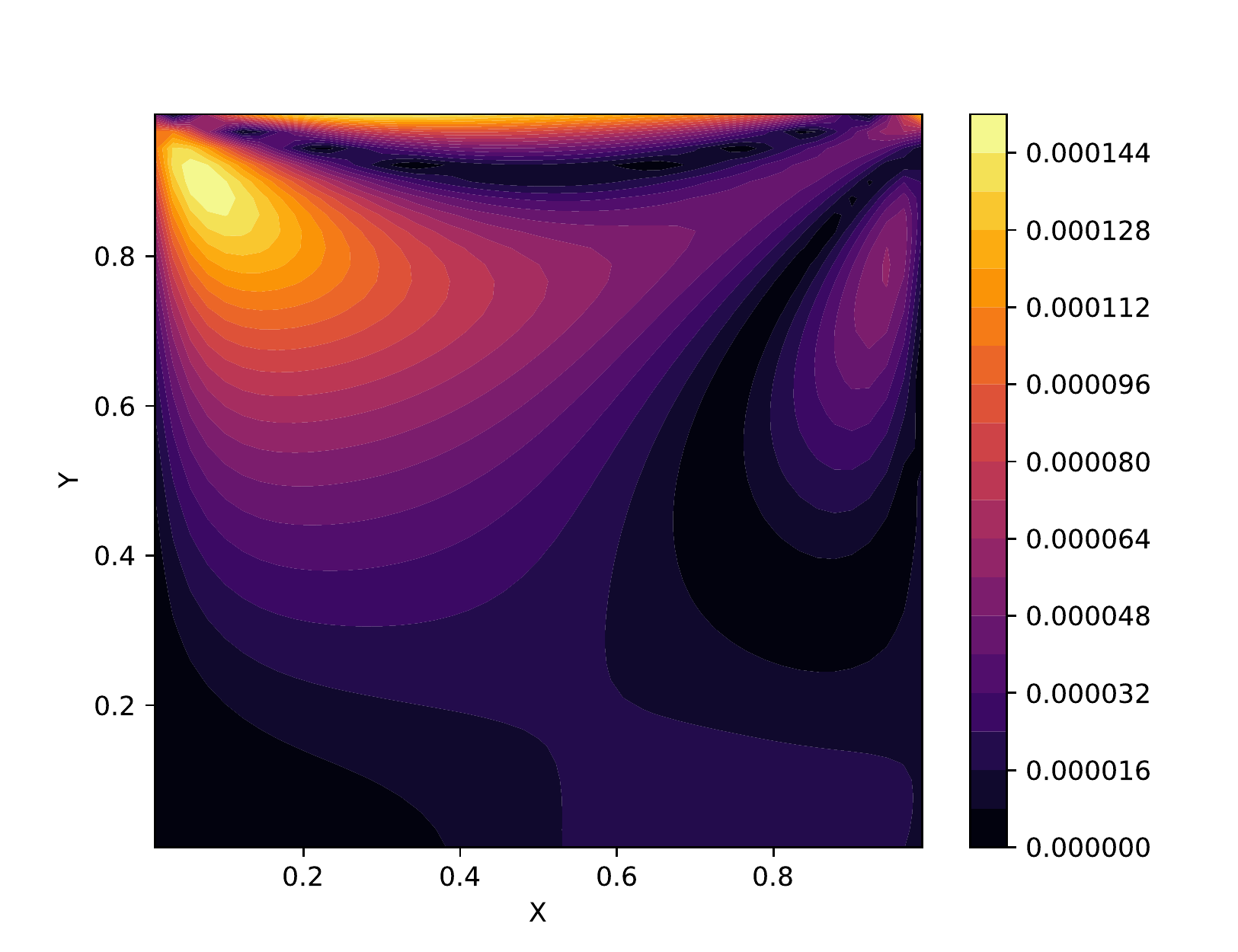}%
    }\hfill
    \subfloat[$\rm Kn=0.5$]{%
        \includegraphics[width=.33\linewidth]{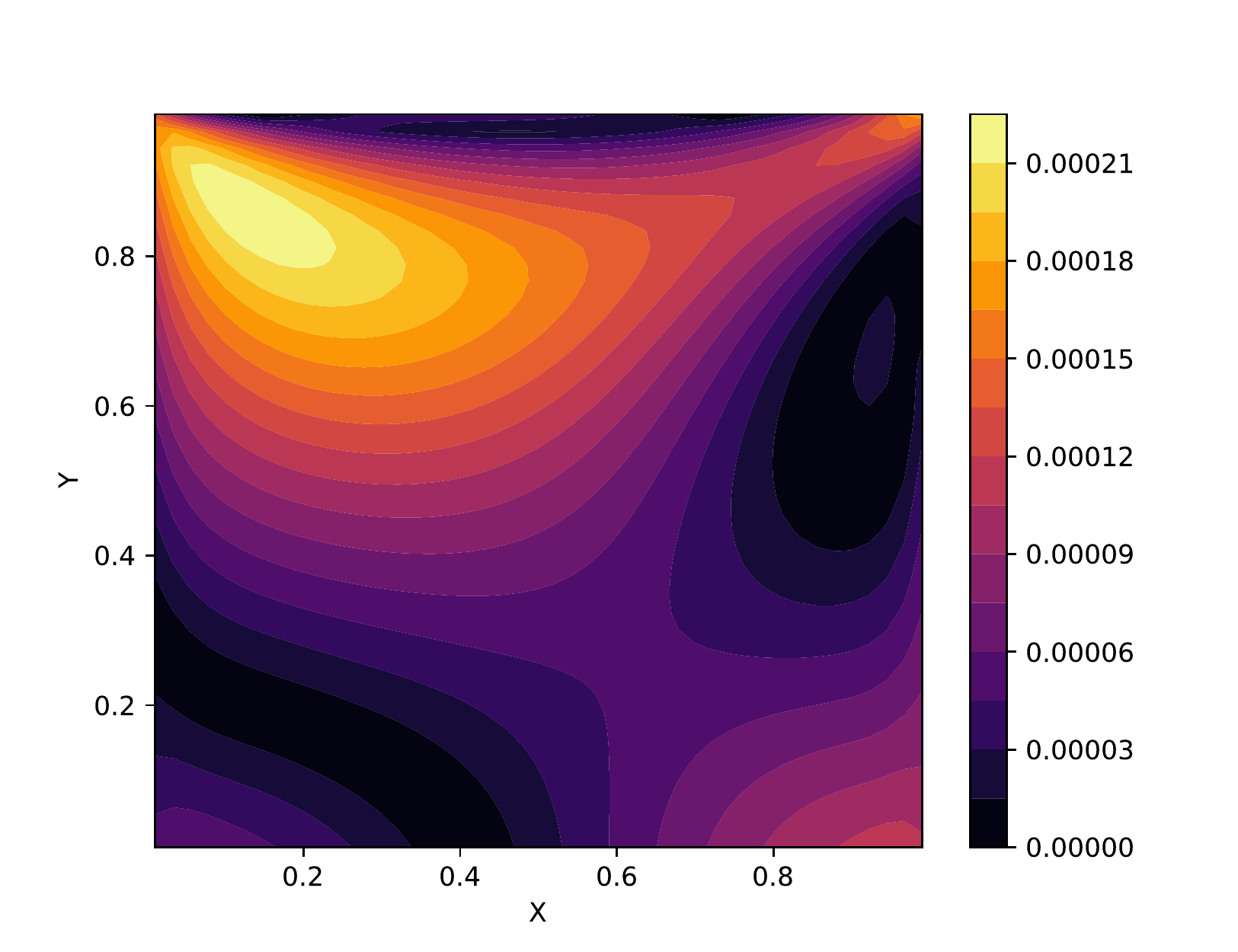}%
    }
\caption{Expectation values and standard deviations of heat flux at different reference Knudsen numbers in the lid-driven cavity. The four rows are $\mathbb E(q_x)$, $\mathbb S(q_x)$, $\mathbb E(q_y)$, and $\mathbb S(q_y)$ respectively.}
\label{pic:cavity heatflux}
\end{figure}

Fig.\ref{pic:cavity heat}(d)-(e) shows the standard deviations of temperature.
Compared with the mean fields, the distribution of temperature variances performs a semblable pattern.
As $\rm Kn$ increases, the loose coupling of particles enhances the freedom of transport phenomena, and the corner-effect zones from two vertices enlarge downwards.
The influence of main eddy decreases with increasing dissipation and the energy transport inside cavity presents the tendency with enhanced horizontal characteristics.
Fig.\ref{pic:cavity heatflux} presents the distribution of components of heat flux and their standard deviations.
In spite of the positive correlation between the mean field and its variance, it can also be noticed that such a correspondence doesn't necessarily happen one by one.
For example, at $\rm Kn=0.5$ the upper half of right side wall with high value of $\mathbb E(q_y)$ doesn't hold $\mathbb S(q_y)$ maximum routinely as it is at $\rm Kn=0.075$.
The standard deviation doesn't vary monotonically with respect to the Knudsen number, but could hold a maximum at a certain point in the transition regime.
It can be inferred that peculiarity exists for higher-order velocity moments of particle distribution function, which is more sensitive to the slight change of distribution function and its variance.

\section{Conclusion}

As the computational fluid dynamics plays a more important role in the study on flow mechanism and spacecraft design, deterministic theoretical and numerical solutions may not be taken for granted.
In this paper, a general methodology of modeling and simulating multi-scale flow dynamics is proposed in conjunction with uncertain quantification.
The Boltzmann model equation is reformulated with the stochastic Galerkin method, and theoretical analysis is presented quantitatively in both kinetic regime and its upscaling macroscopic moments system.
A newly developed stochastic kinetic scheme is employed for numerical investigations with full validations.
Different numerical experiments, including homogeneous relaxation of particle distribution function, normal shock wave structure, transient shear layer and lid-driven cavity in different flow regimes, are studied subject to different kinds of uncertainties from initial status, boundary conditions and intermolecular collision kernels.
Favorable agreements are achieved between theoretical analysis and numerical results.
New physical phenomena, such as the consistent propagating patterns of mean fields and uncertainties from continuum to rarefied regimes, are observed and analyzed systematically. 
The current method provides an innovative tool for sensitivity analysis, flow diagnoses, and optimization for the study of computational fluid dynamics, especially on non-equilibrium flow dynamics.
Confined to the computational resources, multi-dimensional uncertainties in probabilistic space will be further considered in future work.

\begin{acknowledgments}
The current research is funded by the Alexander von Humboldt Foundation.
\end{acknowledgments}

\section*{Data Availability Statement}
The data that support the findings of this study are available from the corresponding author upon reasonable request.

%\nocite{*}
\bibliography{v1}% Produces the bibliography via BibTeX.

\providecommand{\noopsort}[1]{}\providecommand{\singleletter}[1]{#1}%
\begin{thebibliography}{24}%
\makeatletter
\providecommand \@ifxundefined [1]{%
 \@ifx{#1\undefined}
}%
\providecommand \@ifnum [1]{%
 \ifnum #1\expandafter \@firstoftwo
 \else \expandafter \@secondoftwo
 \fi
}%
\providecommand \@ifx [1]{%
 \ifx #1\expandafter \@firstoftwo
 \else \expandafter \@secondoftwo
 \fi
}%
\providecommand \natexlab [1]{#1}%
\providecommand \enquote  [1]{``#1''}%
\providecommand \bibnamefont  [1]{#1}%
\providecommand \bibfnamefont [1]{#1}%
\providecommand \citenamefont [1]{#1}%
\providecommand \href@noop [0]{\@secondoftwo}%
\providecommand \href [0]{\begingroup \@sanitize@url \@href}%
\providecommand \@href[1]{\@@startlink{#1}\@@href}%
\providecommand \@@href[1]{\endgroup#1\@@endlink}%
\providecommand \@sanitize@url [0]{\catcode `\\12\catcode `\$12\catcode
  `\&12\catcode `\#12\catcode `\^12\catcode `\_12\catcode `\%12\relax}%
\providecommand \@@startlink[1]{}%
\providecommand \@@endlink[0]{}%
\providecommand \url  [0]{\begingroup\@sanitize@url \@url }%
\providecommand \@url [1]{\endgroup\@href {#1}{\urlprefix }}%
\providecommand \urlprefix  [0]{URL }%
\providecommand \Eprint [0]{\href }%
\providecommand \doibase [0]{http://dx.doi.org/}%
\providecommand \selectlanguage [0]{\@gobble}%
\providecommand \bibinfo  [0]{\@secondoftwo}%
\providecommand \bibfield  [0]{\@secondoftwo}%
\providecommand \translation [1]{[#1]}%
\providecommand \BibitemOpen [0]{}%
\providecommand \bibitemStop [0]{}%
\providecommand \bibitemNoStop [0]{.\EOS\space}%
\providecommand \EOS [0]{\spacefactor3000\relax}%
\providecommand \BibitemShut  [1]{\csname bibitem#1\endcsname}%
\let\auto@bib@innerbib\@empty
%</preamble>
\bibitem [{\citenamefont {Hilbert}(1902)}]{hilbert1902mathematical}%
  \BibitemOpen
  \bibfield  {author} {\bibinfo {author} {\bibfnamefont {D.}~\bibnamefont
  {Hilbert}},\ }\bibfield  {title} {\enquote {\bibinfo {title} {Mathematical
  problems},}\ }\href@noop {} {\bibfield  {journal} {\bibinfo  {journal}
  {Bulletin of the American Mathematical Society}\ }\textbf {\bibinfo {volume}
  {8}},\ \bibinfo {pages} {437--479} (\bibinfo {year} {1902})}\BibitemShut
  {NoStop}%
\bibitem [{\citenamefont {Chapman}\ and\ \citenamefont
  {Cowling}(1970)}]{chapman1970mathematical}%
  \BibitemOpen
  \bibfield  {author} {\bibinfo {author} {\bibfnamefont {S.}~\bibnamefont
  {Chapman}}\ and\ \bibinfo {author} {\bibfnamefont {T.~G.}\ \bibnamefont
  {Cowling}},\ }\href@noop {} {\emph {\bibinfo {title} {The mathematical theory
  of non-uniform gases: an account of the kinetic theory of viscosity, thermal
  conduction and diffusion in gases}}}\ (\bibinfo  {publisher} {Cambridge
  University Press},\ \bibinfo {year} {1970})\BibitemShut {NoStop}%
\bibitem [{\citenamefont {Grad}(1949)}]{grad1949kinetic}%
  \BibitemOpen
  \bibfield  {author} {\bibinfo {author} {\bibfnamefont {H.}~\bibnamefont
  {Grad}},\ }\bibfield  {title} {\enquote {\bibinfo {title} {On the kinetic
  theory of rarefied gases},}\ }\href@noop {} {\bibfield  {journal} {\bibinfo
  {journal} {Communications on pure and applied mathematics}\ }\textbf
  {\bibinfo {volume} {2}},\ \bibinfo {pages} {331--407} (\bibinfo {year}
  {1949})}\BibitemShut {NoStop}%
\bibitem [{\citenamefont {Lennard-Jones}(1924)}]{lennard1924determination}%
  \BibitemOpen
  \bibfield  {author} {\bibinfo {author} {\bibfnamefont {J.~E.}\ \bibnamefont
  {Lennard-Jones}},\ }\bibfield  {title} {\enquote {\bibinfo {title} {On the
  determination of molecular fields. ii. from the equation of state of gas},}\
  }\href@noop {} {\bibfield  {journal} {\bibinfo  {journal} {Proc. Roy. Soc.
  A}\ }\textbf {\bibinfo {volume} {106}},\ \bibinfo {pages} {463--477}
  (\bibinfo {year} {1924})}\BibitemShut {NoStop}%
\bibitem [{\citenamefont {Cacuci}(2003)}]{cacuci2003sensitivity}%
  \BibitemOpen
  \bibfield  {author} {\bibinfo {author} {\bibfnamefont {D.~G.}\ \bibnamefont
  {Cacuci}},\ }\href@noop {} {\emph {\bibinfo {title} {Sensitivity and
  Uncertainty Analysis, Volume I: Theory}}}\ (\bibinfo  {publisher} {Boca
  Raton, FL: Chapman \& Hall/CRC},\ \bibinfo {year} {2003})\BibitemShut
  {NoStop}%
\bibitem [{\citenamefont {Saltelli}, \citenamefont {Tarantola},\ and\
  \citenamefont {Chan}(1999)}]{saltelli1999quantitative}%
  \BibitemOpen
  \bibfield  {author} {\bibinfo {author} {\bibfnamefont {A.}~\bibnamefont
  {Saltelli}}, \bibinfo {author} {\bibfnamefont {S.}~\bibnamefont {Tarantola}},
  \ and\ \bibinfo {author} {\bibfnamefont {K.-S.}\ \bibnamefont {Chan}},\
  }\bibfield  {title} {\enquote {\bibinfo {title} {A quantitative
  model-independent method for global sensitivity analysis of model output},}\
  }\href@noop {} {\bibfield  {journal} {\bibinfo  {journal} {Technometrics}\
  }\textbf {\bibinfo {volume} {41}},\ \bibinfo {pages} {39--56} (\bibinfo
  {year} {1999})}\BibitemShut {NoStop}%
\bibitem [{\citenamefont {Xiu}\ and\ \citenamefont
  {Karniadakis}(2003)}]{xiu2003modeling}%
  \BibitemOpen
  \bibfield  {author} {\bibinfo {author} {\bibfnamefont {D.}~\bibnamefont
  {Xiu}}\ and\ \bibinfo {author} {\bibfnamefont {G.~E.}\ \bibnamefont
  {Karniadakis}},\ }\bibfield  {title} {\enquote {\bibinfo {title} {Modeling
  uncertainty in flow simulations via generalized polynomial chaos},}\
  }\href@noop {} {\bibfield  {journal} {\bibinfo  {journal} {Journal of
  computational physics}\ }\textbf {\bibinfo {volume} {187}},\ \bibinfo {pages}
  {137--167} (\bibinfo {year} {2003})}\BibitemShut {NoStop}%
\bibitem [{\citenamefont {Walters}\ and\ \citenamefont
  {Huyse}(2002)}]{walters2002uncertainty}%
  \BibitemOpen
  \bibfield  {author} {\bibinfo {author} {\bibfnamefont {R.~W.}\ \bibnamefont
  {Walters}}\ and\ \bibinfo {author} {\bibfnamefont {L.}~\bibnamefont
  {Huyse}},\ }\href@noop {} {\enquote {\bibinfo {title} {Uncertainty analysis
  for fluid mechanics with applications},}\ }\bibinfo {type} {Tech. Rep.}\
  (\bibinfo  {institution} {National Aeronautics and Space Administration
  Hampton VA Langley Research Center},\ \bibinfo {year} {2002})\BibitemShut
  {NoStop}%
\bibitem [{\citenamefont {Najm}(2009)}]{najm2009uncertainty}%
  \BibitemOpen
  \bibfield  {author} {\bibinfo {author} {\bibfnamefont {H.~N.}\ \bibnamefont
  {Najm}},\ }\bibfield  {title} {\enquote {\bibinfo {title} {Uncertainty
  quantification and polynomial chaos techniques in computational fluid
  dynamics},}\ }\href@noop {} {\bibfield  {journal} {\bibinfo  {journal}
  {Annual review of fluid mechanics}\ }\textbf {\bibinfo {volume} {41}},\
  \bibinfo {pages} {35--52} (\bibinfo {year} {2009})}\BibitemShut {NoStop}%
\bibitem [{\citenamefont {Bourgat}, \citenamefont {Le~Tallec},\ and\
  \citenamefont {Tidriri}(1995)}]{bourgat1995coupling}%
  \BibitemOpen
  \bibfield  {author} {\bibinfo {author} {\bibfnamefont {J.-F.}\ \bibnamefont
  {Bourgat}}, \bibinfo {author} {\bibfnamefont {P.}~\bibnamefont {Le~Tallec}},
  \ and\ \bibinfo {author} {\bibfnamefont {M.~D.}\ \bibnamefont {Tidriri}},\
  }\bibfield  {title} {\enquote {\bibinfo {title} {Coupling boltzmann and
  navier-stokes equations by friction},}\ }\href@noop {} {\  (\bibinfo {year}
  {1995})}\BibitemShut {NoStop}%
\bibitem [{\citenamefont {Sun}, \citenamefont {Boyd},\ and\ \citenamefont
  {Candler}(2004)}]{sun2004hybrid}%
  \BibitemOpen
  \bibfield  {author} {\bibinfo {author} {\bibfnamefont {Q.}~\bibnamefont
  {Sun}}, \bibinfo {author} {\bibfnamefont {I.~D.}\ \bibnamefont {Boyd}}, \
  and\ \bibinfo {author} {\bibfnamefont {G.~V.}\ \bibnamefont {Candler}},\
  }\bibfield  {title} {\enquote {\bibinfo {title} {A hybrid continuum/particle
  approach for modeling subsonic, rarefied gas flows},}\ }\href@noop {}
  {\bibfield  {journal} {\bibinfo  {journal} {Journal of Computational
  Physics}\ }\textbf {\bibinfo {volume} {194}},\ \bibinfo {pages} {256--277}
  (\bibinfo {year} {2004})}\BibitemShut {NoStop}%
\bibitem [{\citenamefont {Wijesinghe}\ \emph {et~al.}(2004)\citenamefont
  {Wijesinghe}, \citenamefont {Hornung}, \citenamefont {Garcia},\ and\
  \citenamefont {Hadjiconstantinou}}]{wijesinghe2004three}%
  \BibitemOpen
  \bibfield  {author} {\bibinfo {author} {\bibfnamefont {H.}~\bibnamefont
  {Wijesinghe}}, \bibinfo {author} {\bibfnamefont {R.}~\bibnamefont {Hornung}},
  \bibinfo {author} {\bibfnamefont {A.}~\bibnamefont {Garcia}}, \ and\ \bibinfo
  {author} {\bibfnamefont {N.}~\bibnamefont {Hadjiconstantinou}},\ }\bibfield
  {title} {\enquote {\bibinfo {title} {Three-dimensional hybrid
  continuum-atomistic simulations for multiscale hydrodynamics},}\ }\href@noop
  {} {\bibfield  {journal} {\bibinfo  {journal} {J. Fluids Eng.}\ }\textbf
  {\bibinfo {volume} {126}},\ \bibinfo {pages} {768--777} (\bibinfo {year}
  {2004})}\BibitemShut {NoStop}%
\bibitem [{\citenamefont {Degond}, \citenamefont {Dimarco},\ and\ \citenamefont
  {Mieussens}(2007)}]{degond2007moving}%
  \BibitemOpen
  \bibfield  {author} {\bibinfo {author} {\bibfnamefont {P.}~\bibnamefont
  {Degond}}, \bibinfo {author} {\bibfnamefont {G.}~\bibnamefont {Dimarco}}, \
  and\ \bibinfo {author} {\bibfnamefont {L.}~\bibnamefont {Mieussens}},\
  }\bibfield  {title} {\enquote {\bibinfo {title} {A moving interface method
  for dynamic kinetic--fluid coupling},}\ }\href@noop {} {\bibfield  {journal}
  {\bibinfo  {journal} {Journal of Computational Physics}\ }\textbf {\bibinfo
  {volume} {227}},\ \bibinfo {pages} {1176--1208} (\bibinfo {year}
  {2007})}\BibitemShut {NoStop}%
\bibitem [{\citenamefont {Lemou}\ and\ \citenamefont
  {Mieussens}(2008)}]{lemou2008new}%
  \BibitemOpen
  \bibfield  {author} {\bibinfo {author} {\bibfnamefont {M.}~\bibnamefont
  {Lemou}}\ and\ \bibinfo {author} {\bibfnamefont {L.}~\bibnamefont
  {Mieussens}},\ }\bibfield  {title} {\enquote {\bibinfo {title} {A new
  asymptotic preserving scheme based on micro-macro formulation for linear
  kinetic equations in the diffusion limit},}\ }\href@noop {} {\bibfield
  {journal} {\bibinfo  {journal} {SIAM Journal on Scientific Computing}\
  }\textbf {\bibinfo {volume} {31}},\ \bibinfo {pages} {334--368} (\bibinfo
  {year} {2008})}\BibitemShut {NoStop}%
\bibitem [{\citenamefont {Filbet}\ and\ \citenamefont
  {Jin}(2010)}]{filbet2010class}%
  \BibitemOpen
  \bibfield  {author} {\bibinfo {author} {\bibfnamefont {F.}~\bibnamefont
  {Filbet}}\ and\ \bibinfo {author} {\bibfnamefont {S.}~\bibnamefont {Jin}},\
  }\bibfield  {title} {\enquote {\bibinfo {title} {A class of
  asymptotic-preserving schemes for kinetic equations and related problems with
  stiff sources},}\ }\href@noop {} {\bibfield  {journal} {\bibinfo  {journal}
  {Journal of Computational Physics}\ }\textbf {\bibinfo {volume} {229}},\
  \bibinfo {pages} {7625--7648} (\bibinfo {year} {2010})}\BibitemShut {NoStop}%
\bibitem [{\citenamefont {Xu}\ and\ \citenamefont
  {Huang}(2010)}]{xu2010unified}%
  \BibitemOpen
  \bibfield  {author} {\bibinfo {author} {\bibfnamefont {K.}~\bibnamefont
  {Xu}}\ and\ \bibinfo {author} {\bibfnamefont {J.-C.}\ \bibnamefont {Huang}},\
  }\bibfield  {title} {\enquote {\bibinfo {title} {A unified gas-kinetic scheme
  for continuum and rarefied flows},}\ }\href@noop {} {\bibfield  {journal}
  {\bibinfo  {journal} {Journal of Computational Physics}\ }\textbf {\bibinfo
  {volume} {229}},\ \bibinfo {pages} {7747--7764} (\bibinfo {year}
  {2010})}\BibitemShut {NoStop}%
\bibitem [{\citenamefont {Xiao}, \citenamefont {Cai},\ and\ \citenamefont
  {Xu}(2017)}]{xiao2017well}%
  \BibitemOpen
  \bibfield  {author} {\bibinfo {author} {\bibfnamefont {T.}~\bibnamefont
  {Xiao}}, \bibinfo {author} {\bibfnamefont {Q.}~\bibnamefont {Cai}}, \ and\
  \bibinfo {author} {\bibfnamefont {K.}~\bibnamefont {Xu}},\ }\bibfield
  {title} {\enquote {\bibinfo {title} {A well-balanced unified gas-kinetic
  scheme for multiscale flow transport under gravitational field},}\ }\href
  {\doibase http://dx.doi.org/10.1016/j.jcp.2016.12.022} {\bibfield  {journal}
  {\bibinfo  {journal} {Journal of Computational Physics}\ }\textbf {\bibinfo
  {volume} {332}},\ \bibinfo {pages} {475 -- 491} (\bibinfo {year}
  {2017})}\BibitemShut {NoStop}%
\bibitem [{\citenamefont {Hu}\ and\ \citenamefont
  {Jin}(2017)}]{hu2017uncertainty}%
  \BibitemOpen
  \bibfield  {author} {\bibinfo {author} {\bibfnamefont {J.}~\bibnamefont
  {Hu}}\ and\ \bibinfo {author} {\bibfnamefont {S.}~\bibnamefont {Jin}},\
  }\bibfield  {title} {\enquote {\bibinfo {title} {Uncertainty quantification
  for kinetic equations},}\ }in\ \href@noop {} {\emph {\bibinfo {booktitle}
  {Uncertainty Quantification for Hyperbolic and Kinetic Equations}}}\
  (\bibinfo  {publisher} {Springer},\ \bibinfo {year} {2017})\ pp.\ \bibinfo
  {pages} {193--229}\BibitemShut {NoStop}%
\bibitem [{\citenamefont {Dimarco}\ and\ \citenamefont
  {Pareschi}(2019)}]{dimarco2019multi}%
  \BibitemOpen
  \bibfield  {author} {\bibinfo {author} {\bibfnamefont {G.}~\bibnamefont
  {Dimarco}}\ and\ \bibinfo {author} {\bibfnamefont {L.}~\bibnamefont
  {Pareschi}},\ }\bibfield  {title} {\enquote {\bibinfo {title} {Multi-scale
  control variate methods for uncertainty quantification in kinetic
  equations},}\ }\href@noop {} {\bibfield  {journal} {\bibinfo  {journal}
  {Journal of Computational Physics}\ }\textbf {\bibinfo {volume} {388}},\
  \bibinfo {pages} {63--89} (\bibinfo {year} {2019})}\BibitemShut {NoStop}%
\bibitem [{\citenamefont {Xiao}\ and\ \citenamefont
  {Frank}(2020{\natexlab{a}})}]{xiao2020stochasticflow}%
  \BibitemOpen
  \bibfield  {author} {\bibinfo {author} {\bibfnamefont {T.}~\bibnamefont
  {Xiao}}\ and\ \bibinfo {author} {\bibfnamefont {M.}~\bibnamefont {Frank}},\
  }\bibfield  {title} {\enquote {\bibinfo {title} {A stochastic kinetic scheme
  for multi-scale flow transport with uncertainty quantification},}\
  }\href@noop {} {\bibfield  {journal} {\bibinfo  {journal} {arXiv preprint
  arXiv:2002.00277}\ } (\bibinfo {year} {2020}{\natexlab{a}})}\BibitemShut
  {NoStop}%
\bibitem [{\citenamefont {Wu}\ \emph {et~al.}(2013)\citenamefont {Wu},
  \citenamefont {White}, \citenamefont {Scanlon}, \citenamefont {Reese},\ and\
  \citenamefont {Zhang}}]{wu2013deterministic}%
  \BibitemOpen
  \bibfield  {author} {\bibinfo {author} {\bibfnamefont {L.}~\bibnamefont
  {Wu}}, \bibinfo {author} {\bibfnamefont {C.}~\bibnamefont {White}}, \bibinfo
  {author} {\bibfnamefont {T.~J.}\ \bibnamefont {Scanlon}}, \bibinfo {author}
  {\bibfnamefont {J.~M.}\ \bibnamefont {Reese}}, \ and\ \bibinfo {author}
  {\bibfnamefont {Y.}~\bibnamefont {Zhang}},\ }\bibfield  {title} {\enquote
  {\bibinfo {title} {Deterministic numerical solutions of the boltzmann
  equation using the fast spectral method},}\ }\href@noop {} {\bibfield
  {journal} {\bibinfo  {journal} {Journal of Computational Physics}\ }\textbf
  {\bibinfo {volume} {250}},\ \bibinfo {pages} {27--52} (\bibinfo {year}
  {2013})}\BibitemShut {NoStop}%
\bibitem [{\citenamefont {Xiao}\ and\ \citenamefont
  {Frank}(2020{\natexlab{b}})}]{xiao2020stochasticplasma}%
  \BibitemOpen
  \bibfield  {author} {\bibinfo {author} {\bibfnamefont {T.}~\bibnamefont
  {Xiao}}\ and\ \bibinfo {author} {\bibfnamefont {M.}~\bibnamefont {Frank}},\
  }\bibfield  {title} {\enquote {\bibinfo {title} {A stochastic kinetic scheme
  for multi-scale plasma transport with uncertainty quantification},}\
  }\href@noop {} {\bibfield  {journal} {\bibinfo  {journal} {arXiv preprint
  arXiv:2006.03477}\ } (\bibinfo {year} {2020}{\natexlab{b}})}\BibitemShut
  {NoStop}%
\bibitem [{\citenamefont {Xiao}\ \emph {et~al.}(2020)\citenamefont {Xiao},
  \citenamefont {Liu}, \citenamefont {Xu},\ and\ \citenamefont
  {Cai}}]{xiao2020velocity}%
  \BibitemOpen
  \bibfield  {author} {\bibinfo {author} {\bibfnamefont {T.}~\bibnamefont
  {Xiao}}, \bibinfo {author} {\bibfnamefont {C.}~\bibnamefont {Liu}}, \bibinfo
  {author} {\bibfnamefont {K.}~\bibnamefont {Xu}}, \ and\ \bibinfo {author}
  {\bibfnamefont {Q.}~\bibnamefont {Cai}},\ }\bibfield  {title} {\enquote
  {\bibinfo {title} {A velocity-space adaptive unified gas kinetic scheme for
  continuum and rarefied flows},}\ }\href@noop {} {\bibfield  {journal}
  {\bibinfo  {journal} {Journal of Computational Physics}\ ,\ \bibinfo {pages}
  {109535}} (\bibinfo {year} {2020})}\BibitemShut {NoStop}%
\bibitem [{\citenamefont {Xiao}\ \emph {et~al.}(2018)\citenamefont {Xiao},
  \citenamefont {Xu}, \citenamefont {Cai},\ and\ \citenamefont
  {Qian}}]{xiao2018investigation}%
  \BibitemOpen
  \bibfield  {author} {\bibinfo {author} {\bibfnamefont {T.}~\bibnamefont
  {Xiao}}, \bibinfo {author} {\bibfnamefont {K.}~\bibnamefont {Xu}}, \bibinfo
  {author} {\bibfnamefont {Q.}~\bibnamefont {Cai}}, \ and\ \bibinfo {author}
  {\bibfnamefont {T.}~\bibnamefont {Qian}},\ }\bibfield  {title} {\enquote
  {\bibinfo {title} {An investigation of non-equilibrium heat transport in a
  gas system under external force field},}\ }\href@noop {} {\bibfield
  {journal} {\bibinfo  {journal} {International Journal of Heat and Mass
  Transfer}\ }\textbf {\bibinfo {volume} {126}},\ \bibinfo {pages} {362--379}
  (\bibinfo {year} {2018})}\BibitemShut {NoStop}%
\end{thebibliography}%

\onecolumngrid

\end{document}